\documentclass[conference]{IEEEtran}
\IEEEoverridecommandlockouts

\usepackage{amsmath,amssymb,amsfonts}
\usepackage{xspace}
\usepackage{xcolor}
\usepackage{graphicx}
\usepackage[normalem]{ulem}
\usepackage{enumitem}

\usepackage{hyperref}
\hypersetup{linkcolor=black,citecolor=black,anchorcolor=black,filecolor=black,menucolor=black,runcolor=black,urlcolor=black,hidelinks}
\usepackage{breakurl}

\usepackage{booktabs}
\usepackage{multirow}
\usepackage{makecell}
\usepackage{ragged2e}

\usepackage{caption}
\usepackage{subcaption}
\usepackage{placeins}

\usepackage{listings}

\usepackage{tikz}
\usetikzlibrary{calc}
\usetikzlibrary{shapes.geometric}
\usetikzlibrary{decorations.pathreplacing}
\usetikzlibrary{positioning}

\usepackage{flushend}
\usepackage{datetime}

\usepackage{cite}

\makeatletter
\newcommand{\DefMacro}{\@ifstar\@DefMacroAllowRedefine\@DefMacro}
\newcommand{\@DefMacro}[2]{\expandafter\newcommand\csname rmk-#1\endcsname{#2}}
\newcommand{\@DefMacroAllowRedefine}[2]{\expandafter\providecommand\csname rmk-#1\endcsname{} \expandafter\renewcommand\csname rmk-#1\endcsname{#2}}
\makeatother
\newcommand{\UseMacro}[1]{\csname rmk-#1\endcsname}

\newcommand{\XSpace}[1]{}
\newcommand{\XComment}[1]{}

\newcommand{\MyPara}[1]{\vspace{2pt}\noindent\textbf{#1}.}
\newcommand{\MyParaOnly}[1]{\noindent\textbf{#1}}

\newcommand{\Code}[1]{{\ifmmode{\mathtt{#1}}\else$\mathtt{#1}$\fi}}
\newcommand{\CodeIn}[1]{{\ifmmode{\mathtt{#1}}\else$\mathtt{#1}$\fi}}
\newcommand{\CodeDash}{\text{\texttt{-}}}

\newcolumntype{R}[1]{>{\RaggedLeft\arraybackslash}p{#1}}
\newcolumntype{L}[1]{>{\RaggedRight\arraybackslash}p{#1}}

\definecolor{gray}{RGB}{211,211,211}
\newcommand{\jbasicstyle}{\small\ttfamily}

\newcommand{\jnumberstyle}{\scriptsize}

\lstdefinelanguage{pseudo}
{
morekeywords={},
keywordstyle=\bfseries,
lineskip=-0.1em,
numbers=left,
numberstyle=\jnumberstyle,
numbersep=4pt,
basicstyle=\jbasicstyle,
breaklines=true,
breakautoindent=true,
tabsize=2,
columns=fullflexible,
morecomment=*[l][\textsl]{//},
mathescape=true,
xleftmargin=10pt,
}

\lstdefinelanguage{todo-comment}
{
morekeywords={},
keywordstyle=\bfseries,
lineskip=-0.1em,
numbers=none,
basicstyle=\scriptsize\ttfamily,
breaklines=true,
breakautoindent=true,
tabsize=2,
columns=fullflexible,
morecomment=*[l][\textsl]{//},
mathescape=true,
xleftmargin=0pt,
}

\definecolor{keywordcolor}{rgb}{0,0,1}
\definecolor{modifiercolor}{rgb}{0.5,0,0.5}
\definecolor{datatypecolor}{rgb}{0.82,0.16,0.46}
\definecolor{methodcolor}{rgb}{0.25,0.5,0.35}
\definecolor{byzantine}{rgb}{0.74, 0.2, 0.64}
\definecolor{cadetblue}{rgb}{0.37, 0.62, 0.63}
\definecolor{cadet}{rgb}{0.0, 0.42, 0.24}
\definecolor{brown(web)}{rgb}{0.65, 0.16, 0.16}
\definecolor{bluegray}{rgb}{0.2, 0.2, 0.6}

\lstdefinelanguage{java-pretty}
{
language=java,
numbers=left,
basicstyle=\scriptsize\ttfamily,
numberstyle=\scriptsize,
aboveskip=3pt,
belowskip=1pt,
breaklines=true,
columns=fullflexible,
xleftmargin=14pt,
tabsize=2,
showstringspaces=false,
deletekeywords={public, private, protected, static, final, class, interface, abstract, implements, extends, if, else, while, do, for, switch, case, default, break, continue, return, int, long, double, float, boolean, char, void, String,this},
morekeywords=[1]{if, else, while, do, for, switch, case, default, break, continue, return},
keywordstyle=[1]\color{byzantine}\bfseries,
morekeywords=[2]{public, private, protected, static, final, class, interface, abstract, implements, extends},
keywordstyle=[2]\color{bluegray}\bfseries,
morekeywords=[3]{int, long, double, float, boolean, char, void, String, @Override, @Test},
keywordstyle=[3]\color{cadet}\bfseries,
morekeywords=[4]{class, interface, extends, implements, new, super, throw, throws, try, catch, finally},
keywordstyle=[4]\color{methodcolor},
morecomment=[l]{//},
commentstyle=\color{cadet},
stringstyle=\color{brown(web)},
}

\definecolor{gitgreen}{rgb}{0.0, 0.5, 0.0}
\definecolor{gitred}{rgb}{0.7, 0.0, 0.0}
\definecolor{gitgray}{rgb}{0.5, 0.5, 0.5}

\lstdefinelanguage{java-diff}
{
language=java-pretty,
morecomment=[l]{+},
showstringspaces=false,
keepspaces=true,
breaklines=true,
}

\newcommand{\Title}{Retrofitting Code Using LLMs to Support Exceptional Behavior}
\newcommand{\eg}{e.g.,\xspace}

\newcommand{\exampleRepo}{sagiegurari/fax4j\xspace}

\newcommand{\exampleException}{\CodeIn{FaxException}\xspace}
\newcommand{\exampleMethod}{\CodeIn{initialize}\xspace}
\newcommand{\exampleClass}{\CodeIn{AbstractService}\xspace}

\newcommand{\EBTs}{EBTs\xspace}
\newcommand{\EBT}{EBT\xspace}

\newcommand{\LLM}{LLM\xspace}
\newcommand{\LLMs}{LLMs\xspace}
\newcommand{\TDD}{TDD\xspace}
\newcommand{\Tool}{\textsc{ExCoder}\xspace}
\newcommand{\cond}{exceptional condition check\xspace}

\newcommand{\conds}{exceptional condition checks\xspace}
\newcommand{\Conds}{Exceptional condition checks\xspace}
\newcommand{\ts}{\CodeIn{throw} statement\xspace}

\newcommand{\tss}{\CodeIn{throw} statements\xspace}

\newcommand{\ERC}{ERC\xspace}
\newcommand{\ERCs}{ERC\xspace}
\newcommand{\task}{\ERC{} retrofitting\xspace}

\newcommand{\nEBTs}{non-EBTs\xspace}

\newcommand{\NEBTs}{Non-EBTs\xspace}

\newcommand{\trycatch}{\CodeIn{try/catch} block\xspace}
\newcommand{\trycatchs}{\CodeIn{try/catch} blocks\xspace}
\newcommand{\ifstate}{\CodeIn{if} statement\xspace}
\newcommand{\ifstates}{\CodeIn{if} statements\xspace}

\newcommand{\switchstates}{\CodeIn{switch} statements\xspace}

\newcommand{\ContextAS}{Available symbols\xspace}

\newcommand{\contextAS}{available symbols\xspace}
\newcommand{\ShortASText}{AvSym\xspace}
\newcommand{\FormulaAS}{c_\text{\ShortASText}}
\newcommand{\ShortAS}{$\FormulaAS$\xspace}

\newcommand{\ContextEC}{Exception constructors\xspace}

\newcommand{\contextEC}{exception constructors\xspace}
\newcommand{\ShortECText}{ExCons\xspace}
\newcommand{\FormulaEC}{c_\text{\ShortECText}}
\newcommand{\ShortEC}{$\FormulaEC$\xspace}

\newcommand{\ContextNEBT}{\NEBTs}
\newcommand{\contextNEBT}{\nEBTs}
\newcommand{\ShortNEBTText}{nEBTs\xspace}
\newcommand{\FormulaNEBT}{c_\text{\ShortNEBTText}}
\newcommand{\ShortNEBT}{$\FormulaNEBT$\xspace}

\newcommand{\ContextLC}{Line coverage\xspace}

\newcommand{\contextLC}{line coverage\xspace}
\newcommand{\ShortLCText}{LCov\xspace}
\newcommand{\FormulaLC}{c_\text{\ShortLCText}}
\newcommand{\ShortLC}{$\FormulaLC$\xspace}

\newcommand{\ContextTE}{Thrown exception\xspace}

\newcommand{\contextTE}{thrown exception\xspace}
\newcommand{\ShortTEText}{ThrExc\xspace}
\newcommand{\FormulaTE}{c_\text{\ShortTEText}}
\newcommand{\ShortTE}{$\FormulaTE$\xspace}

\newcommand{\CateTL}{Too Lenient\xspace}

\newcommand{\catetl}{too lenient\xspace}

\newcommand{\CateTS}{Too Strict\xspace}

\newcommand{\CateWH}{Wrong Handling\xspace}

\newcommand{\CateCD}{Destroyed Code\xspace}

\newcommand{\java}{Java\xspace}
\newcommand{\Base}{Base\xspace}

\newcommand{\tarmethod}{target method\xspace}
\newcommand{\tarmethods}{target methods\xspace}

\newcommand{\conEng}{context engineering\xspace}

\DefMacro{stat-num-projects-all-data}{118}
\DefMacro{stat-num-methods-all-data}{518}
\DefMacro{stat-etest-sum-all-data}{934}
\DefMacro{stat-exception-type-all-data}{100}
\DefMacro{stat-catch-throw-all-data}{49}
\DefMacro{stat-if-throw-all-data}{364}
\DefMacro{stat-switch-throw-all-data}{12}
\DefMacro{stat-rest-throw-all-data}{133}
\DefMacro{stat-total-throw-all-data}{558}

\DefMacro{stat-num-projects-real-non-direct-mega-test-data}{150}
\DefMacro{stat-num-methods-real-non-direct-mega-test-data}{1,099}

\DefMacro{stat-num-methods-real-mega-test-data}{546}

\DefMacro{res-real-mega-test-data-with-exception-with-project-with-gold-with-throw-run-ebts-pass-at-k-compiled-at-1-llama_cpp-phi4:14b-q8_0-tuctn-all-info-multi_ebt}{84.18}
\DefMacro{res-real-mega-test-data-with-exception-with-project-with-gold-with-throw-run-ebts-pass-at-k-compiled-at-10-llama_cpp-phi4:14b-q8_0-tuctn-all-info-multi_ebt}{84.87}
\DefMacro{res-real-mega-test-data-with-exception-with-project-with-gold-with-throw-run-ebts-pass-at-k-compiled-at-5-llama_cpp-phi4:14b-q8_0-tuctn-all-info-multi_ebt}{84.54}
\DefMacro{res-real-mega-test-data-with-exception-with-project-with-gold-with-throw-run-ebts-pass-at-k-pass-at-1-llama_cpp-phi4:14b-q8_0-tuctn-all-info-multi_ebt}{74.61}
\DefMacro{res-real-mega-test-data-with-exception-with-project-with-gold-with-throw-run-ebts-pass-at-k-pass-at-10-llama_cpp-phi4:14b-q8_0-tuctn-all-info-multi_ebt}{75.33}
\DefMacro{res-real-mega-test-data-with-exception-with-project-with-gold-with-throw-run-ebts-pass-at-k-pass-at-5-llama_cpp-phi4:14b-q8_0-tuctn-all-info-multi_ebt}{75.00}
\DefMacro{res-real-mega-test-data-with-exception-with-project-with-gold-with-throw-run-all-pass-at-k-pass-at-1-llama_cpp-phi4:14b-q8_0-tuctn-all-info-multi_ebt}{74.28}
\DefMacro{res-real-mega-test-data-with-exception-with-project-with-gold-with-throw-run-all-pass-at-k-pass-at-10-llama_cpp-phi4:14b-q8_0-tuctn-all-info-multi_ebt}{75.00}
\DefMacro{res-real-mega-test-data-with-exception-with-project-with-gold-with-throw-run-all-pass-at-k-pass-at-5-llama_cpp-phi4:14b-q8_0-tuctn-all-info-multi_ebt}{74.67}
\DefMacro{res-real-mega-test-data-with-exception-with-project-with-gold-with-throw-run-all-with-tools-pass-at-k-pass-at-1-llama_cpp-phi4:14b-q8_0-tuctn-all-info-multi_ebt}{61.48}
\DefMacro{res-real-mega-test-data-with-exception-with-project-with-gold-with-throw-run-all-with-tools-pass-at-k-pass-at-10-llama_cpp-phi4:14b-q8_0-tuctn-all-info-multi_ebt}{62.17}
\DefMacro{res-real-mega-test-data-with-exception-with-project-with-gold-with-throw-run-all-with-tools-pass-at-k-pass-at-5-llama_cpp-phi4:14b-q8_0-tuctn-all-info-multi_ebt}{61.84}

\DefMacro{OverQwenLargeRepairPassAllFive}{11.75\xspace}
\DefMacro{OverQwenLargeRepairPassAllToolsFive}{7.41\xspace}

\DefMacro{res-real-mega-test-data-with-exception-with-project-with-gold-with-throw-run-ebts-pass-at-k-compiled-at-1-llama_cpp-llama3.1:8b-instruct-q8_0-base-multi_ebt}{68.09}
\DefMacro{res-real-mega-test-data-with-exception-with-project-with-gold-with-throw-run-ebts-pass-at-k-compiled-at-10-llama_cpp-llama3.1:8b-instruct-q8_0-base-multi_ebt}{70.07}
\DefMacro{res-real-mega-test-data-with-exception-with-project-with-gold-with-throw-run-ebts-pass-at-k-compiled-at-5-llama_cpp-llama3.1:8b-instruct-q8_0-base-multi_ebt}{69.08}
\DefMacro{res-real-mega-test-data-with-exception-with-project-with-gold-with-throw-run-ebts-pass-at-k-pass-at-1-llama_cpp-llama3.1:8b-instruct-q8_0-base-multi_ebt}{55.72}
\DefMacro{res-real-mega-test-data-with-exception-with-project-with-gold-with-throw-run-ebts-pass-at-k-pass-at-10-llama_cpp-llama3.1:8b-instruct-q8_0-base-multi_ebt}{58.22}
\DefMacro{res-real-mega-test-data-with-exception-with-project-with-gold-with-throw-run-ebts-pass-at-k-pass-at-5-llama_cpp-llama3.1:8b-instruct-q8_0-base-multi_ebt}{57.07}
\DefMacro{res-real-mega-test-data-with-exception-with-project-with-gold-with-throw-run-all-pass-at-k-pass-at-1-llama_cpp-llama3.1:8b-instruct-q8_0-base-multi_ebt}{55.39}
\DefMacro{res-real-mega-test-data-with-exception-with-project-with-gold-with-throw-run-all-pass-at-k-pass-at-10-llama_cpp-llama3.1:8b-instruct-q8_0-base-multi_ebt}{57.89}
\DefMacro{res-real-mega-test-data-with-exception-with-project-with-gold-with-throw-run-all-pass-at-k-pass-at-5-llama_cpp-llama3.1:8b-instruct-q8_0-base-multi_ebt}{56.74}
\DefMacro{res-real-mega-test-data-with-exception-with-project-with-gold-with-throw-run-all-with-tools-pass-at-k-pass-at-1-llama_cpp-llama3.1:8b-instruct-q8_0-base-multi_ebt}{41.28}
\DefMacro{res-real-mega-test-data-with-exception-with-project-with-gold-with-throw-run-all-with-tools-pass-at-k-pass-at-10-llama_cpp-llama3.1:8b-instruct-q8_0-base-multi_ebt}{43.42}
\DefMacro{res-real-mega-test-data-with-exception-with-project-with-gold-with-throw-run-all-with-tools-pass-at-k-pass-at-5-llama_cpp-llama3.1:8b-instruct-q8_0-base-multi_ebt}{42.43}

\DefMacro{res-real-mega-test-data-with-exception-with-project-with-gold-with-throw-run-ebts-pass-at-k-compiled-at-1-llama_cpp-qwen2.5-coder:32b-instruct-q8_0-cmtu-multi_ebt}{92.14}
\DefMacro{res-real-mega-test-data-with-exception-with-project-with-gold-with-throw-run-ebts-pass-at-k-compiled-at-10-llama_cpp-qwen2.5-coder:32b-instruct-q8_0-cmtu-multi_ebt}{92.43}
\DefMacro{res-real-mega-test-data-with-exception-with-project-with-gold-with-throw-run-ebts-pass-at-k-compiled-at-5-llama_cpp-qwen2.5-coder:32b-instruct-q8_0-cmtu-multi_ebt}{92.27}
\DefMacro{res-real-mega-test-data-with-exception-with-project-with-gold-with-throw-run-ebts-pass-at-k-pass-at-1-llama_cpp-qwen2.5-coder:32b-instruct-q8_0-cmtu-multi_ebt}{78.29}
\DefMacro{res-real-mega-test-data-with-exception-with-project-with-gold-with-throw-run-ebts-pass-at-k-pass-at-10-llama_cpp-qwen2.5-coder:32b-instruct-q8_0-cmtu-multi_ebt}{78.62}
\DefMacro{res-real-mega-test-data-with-exception-with-project-with-gold-with-throw-run-ebts-pass-at-k-pass-at-5-llama_cpp-qwen2.5-coder:32b-instruct-q8_0-cmtu-multi_ebt}{78.45}
\DefMacro{res-real-mega-test-data-with-exception-with-project-with-gold-with-throw-run-all-pass-at-k-pass-at-1-llama_cpp-qwen2.5-coder:32b-instruct-q8_0-cmtu-multi_ebt}{78.29}
\DefMacro{res-real-mega-test-data-with-exception-with-project-with-gold-with-throw-run-all-pass-at-k-pass-at-10-llama_cpp-qwen2.5-coder:32b-instruct-q8_0-cmtu-multi_ebt}{78.62}
\DefMacro{res-real-mega-test-data-with-exception-with-project-with-gold-with-throw-run-all-pass-at-k-pass-at-5-llama_cpp-qwen2.5-coder:32b-instruct-q8_0-cmtu-multi_ebt}{78.45}
\DefMacro{res-real-mega-test-data-with-exception-with-project-with-gold-with-throw-run-all-with-tools-pass-at-k-pass-at-1-llama_cpp-qwen2.5-coder:32b-instruct-q8_0-cmtu-multi_ebt}{68.72}
\DefMacro{res-real-mega-test-data-with-exception-with-project-with-gold-with-throw-run-all-with-tools-pass-at-k-pass-at-10-llama_cpp-qwen2.5-coder:32b-instruct-q8_0-cmtu-multi_ebt}{69.41}
\DefMacro{res-real-mega-test-data-with-exception-with-project-with-gold-with-throw-run-all-with-tools-pass-at-k-pass-at-5-llama_cpp-qwen2.5-coder:32b-instruct-q8_0-cmtu-multi_ebt}{69.08}

\DefMacro{OverQwenLargeCompFive}{9.87\xspace}
\DefMacro{OverQwenLargePassEBTFive}{13.15\xspace}
\DefMacro{OverQwenLargePassAllOne}{12.56\xspace}
\DefMacro{OverQwenLargePassAllFive}{12.82\xspace}
\DefMacro{OverQwenLargePassAllTen}{13.15\xspace}
\DefMacro{OverQwenLargePassAllToolsFive}{12.17\xspace}
\DefMacro{OverGPTCompFive}{7.42\xspace}
\DefMacro{OverGPTPassEBTFive}{8.27\xspace}
\DefMacro{OverGPTPassAllOne}{12.30\xspace}
\DefMacro{OverGPTPassAllFive}{8.27\xspace}
\DefMacro{OverGPTPassAllTen}{6.91\xspace}
\DefMacro{OverGPTPassAllToolsFive}{8.03\xspace}

\DefMacro{res-real-mega-test-data-with-exception-with-project-with-gold-with-throw-run-ebts-pass-at-k-compiled-at-1-llama_cpp-qwen2.5-coder:7b-instruct-q8_0-base-multi_ebt}{84.14}
\DefMacro{res-real-mega-test-data-with-exception-with-project-with-gold-with-throw-run-ebts-pass-at-k-compiled-at-10-llama_cpp-qwen2.5-coder:7b-instruct-q8_0-base-multi_ebt}{84.54}
\DefMacro{res-real-mega-test-data-with-exception-with-project-with-gold-with-throw-run-ebts-pass-at-k-compiled-at-5-llama_cpp-qwen2.5-coder:7b-instruct-q8_0-base-multi_ebt}{84.38}
\DefMacro{res-real-mega-test-data-with-exception-with-project-with-gold-with-throw-run-ebts-pass-at-k-pass-at-1-llama_cpp-qwen2.5-coder:7b-instruct-q8_0-base-multi_ebt}{64.54}
\DefMacro{res-real-mega-test-data-with-exception-with-project-with-gold-with-throw-run-ebts-pass-at-k-pass-at-10-llama_cpp-qwen2.5-coder:7b-instruct-q8_0-base-multi_ebt}{65.13}
\DefMacro{res-real-mega-test-data-with-exception-with-project-with-gold-with-throw-run-ebts-pass-at-k-pass-at-5-llama_cpp-qwen2.5-coder:7b-instruct-q8_0-base-multi_ebt}{64.80}
\DefMacro{res-real-mega-test-data-with-exception-with-project-with-gold-with-throw-run-all-pass-at-k-pass-at-1-llama_cpp-qwen2.5-coder:7b-instruct-q8_0-base-multi_ebt}{64.54}
\DefMacro{res-real-mega-test-data-with-exception-with-project-with-gold-with-throw-run-all-pass-at-k-pass-at-10-llama_cpp-qwen2.5-coder:7b-instruct-q8_0-base-multi_ebt}{65.13}
\DefMacro{res-real-mega-test-data-with-exception-with-project-with-gold-with-throw-run-all-pass-at-k-pass-at-5-llama_cpp-qwen2.5-coder:7b-instruct-q8_0-base-multi_ebt}{64.80}
\DefMacro{res-real-mega-test-data-with-exception-with-project-with-gold-with-throw-run-all-with-tools-pass-at-k-pass-at-1-llama_cpp-qwen2.5-coder:7b-instruct-q8_0-base-multi_ebt}{54.05}
\DefMacro{res-real-mega-test-data-with-exception-with-project-with-gold-with-throw-run-all-with-tools-pass-at-k-pass-at-10-llama_cpp-qwen2.5-coder:7b-instruct-q8_0-base-multi_ebt}{54.93}
\DefMacro{res-real-mega-test-data-with-exception-with-project-with-gold-with-throw-run-all-with-tools-pass-at-k-pass-at-5-llama_cpp-qwen2.5-coder:7b-instruct-q8_0-base-multi_ebt}{54.44}

\DefMacro{res-real-mega-test-data-with-exception-with-project-with-gold-with-throw-run-ebts-pass-at-k-compiled-at-1-llama_cpp-phi4:14b-q8_0-base-multi_ebt}{75.72}
\DefMacro{res-real-mega-test-data-with-exception-with-project-with-gold-with-throw-run-ebts-pass-at-k-compiled-at-10-llama_cpp-phi4:14b-q8_0-base-multi_ebt}{76.97}
\DefMacro{res-real-mega-test-data-with-exception-with-project-with-gold-with-throw-run-ebts-pass-at-k-compiled-at-5-llama_cpp-phi4:14b-q8_0-base-multi_ebt}{76.32}
\DefMacro{res-real-mega-test-data-with-exception-with-project-with-gold-with-throw-run-ebts-pass-at-k-pass-at-1-llama_cpp-phi4:14b-q8_0-base-multi_ebt}{61.55}
\DefMacro{res-real-mega-test-data-with-exception-with-project-with-gold-with-throw-run-ebts-pass-at-k-pass-at-10-llama_cpp-phi4:14b-q8_0-base-multi_ebt}{62.50}
\DefMacro{res-real-mega-test-data-with-exception-with-project-with-gold-with-throw-run-ebts-pass-at-k-pass-at-5-llama_cpp-phi4:14b-q8_0-base-multi_ebt}{62.01}
\DefMacro{res-real-mega-test-data-with-exception-with-project-with-gold-with-throw-run-all-pass-at-k-pass-at-1-llama_cpp-phi4:14b-q8_0-base-multi_ebt}{61.55}
\DefMacro{res-real-mega-test-data-with-exception-with-project-with-gold-with-throw-run-all-pass-at-k-pass-at-10-llama_cpp-phi4:14b-q8_0-base-multi_ebt}{62.50}
\DefMacro{res-real-mega-test-data-with-exception-with-project-with-gold-with-throw-run-all-pass-at-k-pass-at-5-llama_cpp-phi4:14b-q8_0-base-multi_ebt}{62.01}
\DefMacro{res-real-mega-test-data-with-exception-with-project-with-gold-with-throw-run-all-with-tools-pass-at-k-pass-at-1-llama_cpp-phi4:14b-q8_0-base-multi_ebt}{48.42}
\DefMacro{res-real-mega-test-data-with-exception-with-project-with-gold-with-throw-run-all-with-tools-pass-at-k-pass-at-10-llama_cpp-phi4:14b-q8_0-base-multi_ebt}{50.00}
\DefMacro{res-real-mega-test-data-with-exception-with-project-with-gold-with-throw-run-all-with-tools-pass-at-k-pass-at-5-llama_cpp-phi4:14b-q8_0-base-multi_ebt}{49.18}

\DefMacro{stat-num-projects-real-mega-test-data-with-exception-with-project-with-gold-with-throw}{75}
\DefMacro{stat-num-methods-real-mega-test-data-with-exception-with-project-with-gold-with-throw}{304}
\DefMacro{stat-etest-sum-real-mega-test-data-with-exception-with-project-with-gold-with-throw}{640}
\DefMacro{stat-exception-type-real-mega-test-data-with-exception-with-project-with-gold-with-throw}{60}
\DefMacro{stat-tool-tests-avg-real-mega-test-data-with-exception-with-project-with-gold-with-throw}{77.03}
\DefMacro{stat-catch-throw-real-mega-test-data-with-exception-with-project-with-gold-with-throw}{15}
\DefMacro{stat-if-throw-real-mega-test-data-with-exception-with-project-with-gold-with-throw}{217}
\DefMacro{stat-switch-throw-real-mega-test-data-with-exception-with-project-with-gold-with-throw}{6}
\DefMacro{stat-rest-throw-real-mega-test-data-with-exception-with-project-with-gold-with-throw}{93}
\DefMacro{stat-total-throw-real-mega-test-data-with-exception-with-project-with-gold-with-throw}{331}

\DefMacro{res-real-mega-test-data-with-exception-with-project-with-gold-with-throw-run-ebts-pass-at-k-compiled-at-1-llama_cpp-qwen2.5-coder:32b-instruct-q8_0-only-avsym-multi_ebt}{90.00}
\DefMacro{res-real-mega-test-data-with-exception-with-project-with-gold-with-throw-run-ebts-pass-at-k-compiled-at-10-llama_cpp-qwen2.5-coder:32b-instruct-q8_0-only-avsym-multi_ebt}{90.13}
\DefMacro{res-real-mega-test-data-with-exception-with-project-with-gold-with-throw-run-ebts-pass-at-k-compiled-at-5-llama_cpp-qwen2.5-coder:32b-instruct-q8_0-only-avsym-multi_ebt}{90.13}
\DefMacro{res-real-mega-test-data-with-exception-with-project-with-gold-with-throw-run-ebts-pass-at-k-pass-at-1-llama_cpp-qwen2.5-coder:32b-instruct-q8_0-only-avsym-multi_ebt}{81.45}
\DefMacro{res-real-mega-test-data-with-exception-with-project-with-gold-with-throw-run-ebts-pass-at-k-pass-at-10-llama_cpp-qwen2.5-coder:32b-instruct-q8_0-only-avsym-multi_ebt}{82.57}
\DefMacro{res-real-mega-test-data-with-exception-with-project-with-gold-with-throw-run-ebts-pass-at-k-pass-at-5-llama_cpp-qwen2.5-coder:32b-instruct-q8_0-only-avsym-multi_ebt}{82.07}
\DefMacro{res-real-mega-test-data-with-exception-with-project-with-gold-with-throw-run-all-pass-at-k-pass-at-1-llama_cpp-qwen2.5-coder:32b-instruct-q8_0-only-avsym-multi_ebt}{81.45}
\DefMacro{res-real-mega-test-data-with-exception-with-project-with-gold-with-throw-run-all-pass-at-k-pass-at-10-llama_cpp-qwen2.5-coder:32b-instruct-q8_0-only-avsym-multi_ebt}{82.57}
\DefMacro{res-real-mega-test-data-with-exception-with-project-with-gold-with-throw-run-all-pass-at-k-pass-at-5-llama_cpp-qwen2.5-coder:32b-instruct-q8_0-only-avsym-multi_ebt}{82.07}
\DefMacro{res-real-mega-test-data-with-exception-with-project-with-gold-with-throw-run-all-with-tools-pass-at-k-pass-at-1-llama_cpp-qwen2.5-coder:32b-instruct-q8_0-only-avsym-multi_ebt}{72.27}
\DefMacro{res-real-mega-test-data-with-exception-with-project-with-gold-with-throw-run-all-with-tools-pass-at-k-pass-at-10-llama_cpp-qwen2.5-coder:32b-instruct-q8_0-only-avsym-multi_ebt}{74.01}
\DefMacro{res-real-mega-test-data-with-exception-with-project-with-gold-with-throw-run-all-with-tools-pass-at-k-pass-at-5-llama_cpp-qwen2.5-coder:32b-instruct-q8_0-only-avsym-multi_ebt}{73.27}

\DefMacro{res-real-mega-test-data-with-exception-with-project-with-gold-with-throw-run-ebts-pass-at-k-compiled-at-1-llama_cpp-qwen2.5-coder:32b-instruct-q8_0-only-lcov-multi_ebt}{84.80}
\DefMacro{res-real-mega-test-data-with-exception-with-project-with-gold-with-throw-run-ebts-pass-at-k-compiled-at-10-llama_cpp-qwen2.5-coder:32b-instruct-q8_0-only-lcov-multi_ebt}{85.86}
\DefMacro{res-real-mega-test-data-with-exception-with-project-with-gold-with-throw-run-ebts-pass-at-k-compiled-at-5-llama_cpp-qwen2.5-coder:32b-instruct-q8_0-only-lcov-multi_ebt}{85.45}
\DefMacro{res-real-mega-test-data-with-exception-with-project-with-gold-with-throw-run-ebts-pass-at-k-pass-at-1-llama_cpp-qwen2.5-coder:32b-instruct-q8_0-only-lcov-multi_ebt}{72.11}
\DefMacro{res-real-mega-test-data-with-exception-with-project-with-gold-with-throw-run-ebts-pass-at-k-pass-at-10-llama_cpp-qwen2.5-coder:32b-instruct-q8_0-only-lcov-multi_ebt}{73.68}
\DefMacro{res-real-mega-test-data-with-exception-with-project-with-gold-with-throw-run-ebts-pass-at-k-pass-at-5-llama_cpp-qwen2.5-coder:32b-instruct-q8_0-only-lcov-multi_ebt}{73.04}
\DefMacro{res-real-mega-test-data-with-exception-with-project-with-gold-with-throw-run-all-pass-at-k-pass-at-1-llama_cpp-qwen2.5-coder:32b-instruct-q8_0-only-lcov-multi_ebt}{72.11}
\DefMacro{res-real-mega-test-data-with-exception-with-project-with-gold-with-throw-run-all-pass-at-k-pass-at-10-llama_cpp-qwen2.5-coder:32b-instruct-q8_0-only-lcov-multi_ebt}{73.68}
\DefMacro{res-real-mega-test-data-with-exception-with-project-with-gold-with-throw-run-all-pass-at-k-pass-at-5-llama_cpp-qwen2.5-coder:32b-instruct-q8_0-only-lcov-multi_ebt}{73.04}
\DefMacro{res-real-mega-test-data-with-exception-with-project-with-gold-with-throw-run-all-with-tools-pass-at-k-pass-at-1-llama_cpp-qwen2.5-coder:32b-instruct-q8_0-only-lcov-multi_ebt}{62.60}
\DefMacro{res-real-mega-test-data-with-exception-with-project-with-gold-with-throw-run-all-with-tools-pass-at-k-pass-at-10-llama_cpp-qwen2.5-coder:32b-instruct-q8_0-only-lcov-multi_ebt}{63.49}
\DefMacro{res-real-mega-test-data-with-exception-with-project-with-gold-with-throw-run-all-with-tools-pass-at-k-pass-at-5-llama_cpp-qwen2.5-coder:32b-instruct-q8_0-only-lcov-multi_ebt}{63.16}

\DefMacro{res-real-mega-test-data-with-exception-with-project-with-gold-with-throw-run-ebts-pass-at-k-compiled-at-1-azure-gpt-5-mini-tuctn-all-info-multi_ebt}{97.86}
\DefMacro{res-real-mega-test-data-with-exception-with-project-with-gold-with-throw-run-ebts-pass-at-k-compiled-at-10-azure-gpt-5-mini-tuctn-all-info-multi_ebt}{99.67}
\DefMacro{res-real-mega-test-data-with-exception-with-project-with-gold-with-throw-run-ebts-pass-at-k-compiled-at-5-azure-gpt-5-mini-tuctn-all-info-multi_ebt}{99.50}
\DefMacro{res-real-mega-test-data-with-exception-with-project-with-gold-with-throw-run-ebts-pass-at-k-pass-at-1-azure-gpt-5-mini-tuctn-all-info-multi_ebt}{95.03}
\DefMacro{res-real-mega-test-data-with-exception-with-project-with-gold-with-throw-run-ebts-pass-at-k-pass-at-10-azure-gpt-5-mini-tuctn-all-info-multi_ebt}{98.36}
\DefMacro{res-real-mega-test-data-with-exception-with-project-with-gold-with-throw-run-ebts-pass-at-k-pass-at-5-azure-gpt-5-mini-tuctn-all-info-multi_ebt}{97.92}
\DefMacro{res-real-mega-test-data-with-exception-with-project-with-gold-with-throw-run-all-pass-at-k-pass-at-1-azure-gpt-5-mini-tuctn-all-info-multi_ebt}{95.03}
\DefMacro{res-real-mega-test-data-with-exception-with-project-with-gold-with-throw-run-all-pass-at-k-pass-at-10-azure-gpt-5-mini-tuctn-all-info-multi_ebt}{98.36}
\DefMacro{res-real-mega-test-data-with-exception-with-project-with-gold-with-throw-run-all-pass-at-k-pass-at-5-azure-gpt-5-mini-tuctn-all-info-multi_ebt}{97.92}
\DefMacro{res-real-mega-test-data-with-exception-with-project-with-gold-with-throw-run-all-with-tools-pass-at-k-pass-at-1-azure-gpt-5-mini-tuctn-all-info-multi_ebt}{77.30}
\DefMacro{res-real-mega-test-data-with-exception-with-project-with-gold-with-throw-run-all-with-tools-pass-at-k-pass-at-10-azure-gpt-5-mini-tuctn-all-info-multi_ebt}{84.87}
\DefMacro{res-real-mega-test-data-with-exception-with-project-with-gold-with-throw-run-all-with-tools-pass-at-k-pass-at-5-azure-gpt-5-mini-tuctn-all-info-multi_ebt}{83.82}

\DefMacro{res-real-mega-test-data-with-exception-with-project-with-gold-with-throw-run-ebts-pass-at-k-compiled-at-1-azure-gpt-5-mini-base-multi_ebt}{85.59}
\DefMacro{res-real-mega-test-data-with-exception-with-project-with-gold-with-throw-run-ebts-pass-at-k-compiled-at-10-azure-gpt-5-mini-base-multi_ebt}{93.75}
\DefMacro{res-real-mega-test-data-with-exception-with-project-with-gold-with-throw-run-ebts-pass-at-k-compiled-at-5-azure-gpt-5-mini-base-multi_ebt}{92.08}
\DefMacro{res-real-mega-test-data-with-exception-with-project-with-gold-with-throw-run-ebts-pass-at-k-pass-at-1-azure-gpt-5-mini-base-multi_ebt}{82.83}
\DefMacro{res-real-mega-test-data-with-exception-with-project-with-gold-with-throw-run-ebts-pass-at-k-pass-at-10-azure-gpt-5-mini-base-multi_ebt}{91.45}
\DefMacro{res-real-mega-test-data-with-exception-with-project-with-gold-with-throw-run-ebts-pass-at-k-pass-at-5-azure-gpt-5-mini-base-multi_ebt}{89.65}
\DefMacro{res-real-mega-test-data-with-exception-with-project-with-gold-with-throw-run-all-pass-at-k-pass-at-1-azure-gpt-5-mini-base-multi_ebt}{82.73}
\DefMacro{res-real-mega-test-data-with-exception-with-project-with-gold-with-throw-run-all-pass-at-k-pass-at-10-azure-gpt-5-mini-base-multi_ebt}{91.45}
\DefMacro{res-real-mega-test-data-with-exception-with-project-with-gold-with-throw-run-all-pass-at-k-pass-at-5-azure-gpt-5-mini-base-multi_ebt}{89.65}
\DefMacro{res-real-mega-test-data-with-exception-with-project-with-gold-with-throw-run-all-with-tools-pass-at-k-pass-at-1-azure-gpt-5-mini-base-multi_ebt}{65.49}
\DefMacro{res-real-mega-test-data-with-exception-with-project-with-gold-with-throw-run-all-with-tools-pass-at-k-pass-at-10-azure-gpt-5-mini-base-multi_ebt}{79.61}
\DefMacro{res-real-mega-test-data-with-exception-with-project-with-gold-with-throw-run-all-with-tools-pass-at-k-pass-at-5-azure-gpt-5-mini-base-multi_ebt}{75.79}

\DefMacro{res-real-mega-test-data-with-exception-with-project-with-gold-with-throw-run-ebts-pass-at-k-compiled-at-1-llama_cpp-qwen2.5-coder:32b-instruct-q8_0-only-threxc-multi_ebt}{84.18}
\DefMacro{res-real-mega-test-data-with-exception-with-project-with-gold-with-throw-run-ebts-pass-at-k-compiled-at-10-llama_cpp-qwen2.5-coder:32b-instruct-q8_0-only-threxc-multi_ebt}{85.53}
\DefMacro{res-real-mega-test-data-with-exception-with-project-with-gold-with-throw-run-ebts-pass-at-k-compiled-at-5-llama_cpp-qwen2.5-coder:32b-instruct-q8_0-only-threxc-multi_ebt}{85.37}
\DefMacro{res-real-mega-test-data-with-exception-with-project-with-gold-with-throw-run-ebts-pass-at-k-pass-at-1-llama_cpp-qwen2.5-coder:32b-instruct-q8_0-only-threxc-multi_ebt}{72.57}
\DefMacro{res-real-mega-test-data-with-exception-with-project-with-gold-with-throw-run-ebts-pass-at-k-pass-at-10-llama_cpp-qwen2.5-coder:32b-instruct-q8_0-only-threxc-multi_ebt}{73.68}
\DefMacro{res-real-mega-test-data-with-exception-with-project-with-gold-with-throw-run-ebts-pass-at-k-pass-at-5-llama_cpp-qwen2.5-coder:32b-instruct-q8_0-only-threxc-multi_ebt}{73.48}
\DefMacro{res-real-mega-test-data-with-exception-with-project-with-gold-with-throw-run-all-pass-at-k-pass-at-1-llama_cpp-qwen2.5-coder:32b-instruct-q8_0-only-threxc-multi_ebt}{72.53}
\DefMacro{res-real-mega-test-data-with-exception-with-project-with-gold-with-throw-run-all-pass-at-k-pass-at-10-llama_cpp-qwen2.5-coder:32b-instruct-q8_0-only-threxc-multi_ebt}{73.68}
\DefMacro{res-real-mega-test-data-with-exception-with-project-with-gold-with-throw-run-all-pass-at-k-pass-at-5-llama_cpp-qwen2.5-coder:32b-instruct-q8_0-only-threxc-multi_ebt}{73.48}
\DefMacro{res-real-mega-test-data-with-exception-with-project-with-gold-with-throw-run-all-with-tools-pass-at-k-pass-at-1-llama_cpp-qwen2.5-coder:32b-instruct-q8_0-only-threxc-multi_ebt}{62.89}
\DefMacro{res-real-mega-test-data-with-exception-with-project-with-gold-with-throw-run-all-with-tools-pass-at-k-pass-at-10-llama_cpp-qwen2.5-coder:32b-instruct-q8_0-only-threxc-multi_ebt}{64.14}
\DefMacro{res-real-mega-test-data-with-exception-with-project-with-gold-with-throw-run-all-with-tools-pass-at-k-pass-at-5-llama_cpp-qwen2.5-coder:32b-instruct-q8_0-only-threxc-multi_ebt}{63.85}

\DefMacro{res-real-mega-test-data-with-exception-with-project-with-gold-with-throw-run-ebts-pass-at-k-compiled-at-1-llama_cpp-qwen2.5-coder:32b-instruct-q8_0-base-multi_ebt}{85.20}
\DefMacro{res-real-mega-test-data-with-exception-with-project-with-gold-with-throw-run-ebts-pass-at-k-compiled-at-10-llama_cpp-qwen2.5-coder:32b-instruct-q8_0-base-multi_ebt}{85.53}
\DefMacro{res-real-mega-test-data-with-exception-with-project-with-gold-with-throw-run-ebts-pass-at-k-compiled-at-5-llama_cpp-qwen2.5-coder:32b-instruct-q8_0-base-multi_ebt}{85.36}
\DefMacro{res-real-mega-test-data-with-exception-with-project-with-gold-with-throw-run-ebts-pass-at-k-pass-at-1-llama_cpp-qwen2.5-coder:32b-instruct-q8_0-base-multi_ebt}{73.36}
\DefMacro{res-real-mega-test-data-with-exception-with-project-with-gold-with-throw-run-ebts-pass-at-k-pass-at-10-llama_cpp-qwen2.5-coder:32b-instruct-q8_0-base-multi_ebt}{73.36}
\DefMacro{res-real-mega-test-data-with-exception-with-project-with-gold-with-throw-run-ebts-pass-at-k-pass-at-5-llama_cpp-qwen2.5-coder:32b-instruct-q8_0-base-multi_ebt}{73.36}
\DefMacro{res-real-mega-test-data-with-exception-with-project-with-gold-with-throw-run-all-pass-at-k-pass-at-1-llama_cpp-qwen2.5-coder:32b-instruct-q8_0-base-multi_ebt}{73.36}
\DefMacro{res-real-mega-test-data-with-exception-with-project-with-gold-with-throw-run-all-pass-at-k-pass-at-10-llama_cpp-qwen2.5-coder:32b-instruct-q8_0-base-multi_ebt}{73.36}
\DefMacro{res-real-mega-test-data-with-exception-with-project-with-gold-with-throw-run-all-pass-at-k-pass-at-5-llama_cpp-qwen2.5-coder:32b-instruct-q8_0-base-multi_ebt}{73.36}
\DefMacro{res-real-mega-test-data-with-exception-with-project-with-gold-with-throw-run-all-with-tools-pass-at-k-pass-at-1-llama_cpp-qwen2.5-coder:32b-instruct-q8_0-base-multi_ebt}{63.55}
\DefMacro{res-real-mega-test-data-with-exception-with-project-with-gold-with-throw-run-all-with-tools-pass-at-k-pass-at-10-llama_cpp-qwen2.5-coder:32b-instruct-q8_0-base-multi_ebt}{64.14}
\DefMacro{res-real-mega-test-data-with-exception-with-project-with-gold-with-throw-run-all-with-tools-pass-at-k-pass-at-5-llama_cpp-qwen2.5-coder:32b-instruct-q8_0-base-multi_ebt}{63.82}

\DefMacro{overlap-real-mega-test-data-with-exception-with-project-with-gold-with-throw-run-all-llama_cpp-qwen2.5-coder:32b-instruct-q8_0-multi_ebt-only-base}{8}
\DefMacro{overlap-real-mega-test-data-with-exception-with-project-with-gold-with-throw-run-all-llama_cpp-qwen2.5-coder:32b-instruct-q8_0-multi_ebt-only-tuctn-all-info}{48}

\DefMacro{res-real-mega-test-data-with-exception-with-project-with-gold-with-throw-run-ebts-pass-at-k-compiled-at-1-llama_cpp-qwen2.5-coder:7b-instruct-q8_0-tuctn-all-info-multi_ebt}{90.53}
\DefMacro{res-real-mega-test-data-with-exception-with-project-with-gold-with-throw-run-ebts-pass-at-k-compiled-at-10-llama_cpp-qwen2.5-coder:7b-instruct-q8_0-tuctn-all-info-multi_ebt}{91.45}
\DefMacro{res-real-mega-test-data-with-exception-with-project-with-gold-with-throw-run-ebts-pass-at-k-compiled-at-5-llama_cpp-qwen2.5-coder:7b-instruct-q8_0-tuctn-all-info-multi_ebt}{90.95}
\DefMacro{res-real-mega-test-data-with-exception-with-project-with-gold-with-throw-run-ebts-pass-at-k-pass-at-1-llama_cpp-qwen2.5-coder:7b-instruct-q8_0-tuctn-all-info-multi_ebt}{76.97}
\DefMacro{res-real-mega-test-data-with-exception-with-project-with-gold-with-throw-run-ebts-pass-at-k-pass-at-10-llama_cpp-qwen2.5-coder:7b-instruct-q8_0-tuctn-all-info-multi_ebt}{77.63}
\DefMacro{res-real-mega-test-data-with-exception-with-project-with-gold-with-throw-run-ebts-pass-at-k-pass-at-5-llama_cpp-qwen2.5-coder:7b-instruct-q8_0-tuctn-all-info-multi_ebt}{77.30}
\DefMacro{res-real-mega-test-data-with-exception-with-project-with-gold-with-throw-run-all-pass-at-k-pass-at-1-llama_cpp-qwen2.5-coder:7b-instruct-q8_0-tuctn-all-info-multi_ebt}{76.32}
\DefMacro{res-real-mega-test-data-with-exception-with-project-with-gold-with-throw-run-all-pass-at-k-pass-at-10-llama_cpp-qwen2.5-coder:7b-instruct-q8_0-tuctn-all-info-multi_ebt}{76.97}
\DefMacro{res-real-mega-test-data-with-exception-with-project-with-gold-with-throw-run-all-pass-at-k-pass-at-5-llama_cpp-qwen2.5-coder:7b-instruct-q8_0-tuctn-all-info-multi_ebt}{76.64}
\DefMacro{res-real-mega-test-data-with-exception-with-project-with-gold-with-throw-run-all-with-tools-pass-at-k-pass-at-1-llama_cpp-qwen2.5-coder:7b-instruct-q8_0-tuctn-all-info-multi_ebt}{65.82}
\DefMacro{res-real-mega-test-data-with-exception-with-project-with-gold-with-throw-run-all-with-tools-pass-at-k-pass-at-10-llama_cpp-qwen2.5-coder:7b-instruct-q8_0-tuctn-all-info-multi_ebt}{67.11}
\DefMacro{res-real-mega-test-data-with-exception-with-project-with-gold-with-throw-run-all-with-tools-pass-at-k-pass-at-5-llama_cpp-qwen2.5-coder:7b-instruct-q8_0-tuctn-all-info-multi_ebt}{66.45}

\DefMacro{res-real-mega-test-data-with-exception-with-project-with-gold-with-throw-run-ebts-pass-at-k-compiled-at-1-llama_cpp-qwen2.5-coder:32b-instruct-q8_0-base-repair@4-repair}{90.07}
\DefMacro{res-real-mega-test-data-with-exception-with-project-with-gold-with-throw-run-ebts-pass-at-k-compiled-at-10-llama_cpp-qwen2.5-coder:32b-instruct-q8_0-base-repair@4-repair}{90.79}
\DefMacro{res-real-mega-test-data-with-exception-with-project-with-gold-with-throw-run-ebts-pass-at-k-compiled-at-5-llama_cpp-qwen2.5-coder:32b-instruct-q8_0-base-repair@4-repair}{90.62}
\DefMacro{res-real-mega-test-data-with-exception-with-project-with-gold-with-throw-run-ebts-pass-at-k-pass-at-1-llama_cpp-qwen2.5-coder:32b-instruct-q8_0-base-repair@4-repair}{83.78}
\DefMacro{res-real-mega-test-data-with-exception-with-project-with-gold-with-throw-run-ebts-pass-at-k-pass-at-10-llama_cpp-qwen2.5-coder:32b-instruct-q8_0-base-repair@4-repair}{84.54}
\DefMacro{res-real-mega-test-data-with-exception-with-project-with-gold-with-throw-run-ebts-pass-at-k-pass-at-5-llama_cpp-qwen2.5-coder:32b-instruct-q8_0-base-repair@4-repair}{84.47}
\DefMacro{res-real-mega-test-data-with-exception-with-project-with-gold-with-throw-run-all-pass-at-k-pass-at-1-llama_cpp-qwen2.5-coder:32b-instruct-q8_0-base-repair@4-repair}{83.78}
\DefMacro{res-real-mega-test-data-with-exception-with-project-with-gold-with-throw-run-all-pass-at-k-pass-at-10-llama_cpp-qwen2.5-coder:32b-instruct-q8_0-base-repair@4-repair}{84.54}
\DefMacro{res-real-mega-test-data-with-exception-with-project-with-gold-with-throw-run-all-pass-at-k-pass-at-5-llama_cpp-qwen2.5-coder:32b-instruct-q8_0-base-repair@4-repair}{84.47}
\DefMacro{res-real-mega-test-data-with-exception-with-project-with-gold-with-throw-run-all-with-tools-pass-at-k-pass-at-1-llama_cpp-qwen2.5-coder:32b-instruct-q8_0-base-repair@4-repair}{71.12}
\DefMacro{res-real-mega-test-data-with-exception-with-project-with-gold-with-throw-run-all-with-tools-pass-at-k-pass-at-10-llama_cpp-qwen2.5-coder:32b-instruct-q8_0-base-repair@4-repair}{72.37}
\DefMacro{res-real-mega-test-data-with-exception-with-project-with-gold-with-throw-run-all-with-tools-pass-at-k-pass-at-5-llama_cpp-qwen2.5-coder:32b-instruct-q8_0-base-repair@4-repair}{71.87}

\DefMacro{res-real-mega-test-data-with-exception-with-project-with-gold-with-throw-run-ebts-pass-at-k-compiled-at-1-llama_cpp-qwen2.5-coder:32b-instruct-q8_0-tuctn-all-info-repair@4-repair}{97.34}
\DefMacro{res-real-mega-test-data-with-exception-with-project-with-gold-with-throw-run-ebts-pass-at-k-compiled-at-10-llama_cpp-qwen2.5-coder:32b-instruct-q8_0-tuctn-all-info-repair@4-repair}{97.37}
\DefMacro{res-real-mega-test-data-with-exception-with-project-with-gold-with-throw-run-ebts-pass-at-k-compiled-at-5-llama_cpp-qwen2.5-coder:32b-instruct-q8_0-tuctn-all-info-repair@4-repair}{97.37}
\DefMacro{res-real-mega-test-data-with-exception-with-project-with-gold-with-throw-run-ebts-pass-at-k-pass-at-1-llama_cpp-qwen2.5-coder:32b-instruct-q8_0-tuctn-all-info-repair@4-repair}{96.02}
\DefMacro{res-real-mega-test-data-with-exception-with-project-with-gold-with-throw-run-ebts-pass-at-k-pass-at-10-llama_cpp-qwen2.5-coder:32b-instruct-q8_0-tuctn-all-info-repair@4-repair}{96.38}
\DefMacro{res-real-mega-test-data-with-exception-with-project-with-gold-with-throw-run-ebts-pass-at-k-pass-at-5-llama_cpp-qwen2.5-coder:32b-instruct-q8_0-tuctn-all-info-repair@4-repair}{96.22}
\DefMacro{res-real-mega-test-data-with-exception-with-project-with-gold-with-throw-run-all-pass-at-k-pass-at-1-llama_cpp-qwen2.5-coder:32b-instruct-q8_0-tuctn-all-info-repair@4-repair}{96.02}
\DefMacro{res-real-mega-test-data-with-exception-with-project-with-gold-with-throw-run-all-pass-at-k-pass-at-10-llama_cpp-qwen2.5-coder:32b-instruct-q8_0-tuctn-all-info-repair@4-repair}{96.38}
\DefMacro{res-real-mega-test-data-with-exception-with-project-with-gold-with-throw-run-all-pass-at-k-pass-at-5-llama_cpp-qwen2.5-coder:32b-instruct-q8_0-tuctn-all-info-repair@4-repair}{96.22}
\DefMacro{res-real-mega-test-data-with-exception-with-project-with-gold-with-throw-run-all-with-tools-pass-at-k-pass-at-1-llama_cpp-qwen2.5-coder:32b-instruct-q8_0-tuctn-all-info-repair@4-repair}{78.95}
\DefMacro{res-real-mega-test-data-with-exception-with-project-with-gold-with-throw-run-all-with-tools-pass-at-k-pass-at-10-llama_cpp-qwen2.5-coder:32b-instruct-q8_0-tuctn-all-info-repair@4-repair}{79.61}
\DefMacro{res-real-mega-test-data-with-exception-with-project-with-gold-with-throw-run-all-with-tools-pass-at-k-pass-at-5-llama_cpp-qwen2.5-coder:32b-instruct-q8_0-tuctn-all-info-repair@4-repair}{79.28}

\DefMacro{stat-direct-throw-percentage}{49.7}
\DefMacro{stat-rm-comp-fail-percentage}{12.5}

\DefMacro{res-real-mega-test-data-with-exception-with-project-with-gold-with-throw-run-ebts-pass-at-k-compiled-at-1-llama_cpp-qwen2.5-coder:32b-instruct-q8_0-tuctn-all-info-multi_ebt}{95.10}
\DefMacro{res-real-mega-test-data-with-exception-with-project-with-gold-with-throw-run-ebts-pass-at-k-compiled-at-10-llama_cpp-qwen2.5-coder:32b-instruct-q8_0-tuctn-all-info-multi_ebt}{95.39}
\DefMacro{res-real-mega-test-data-with-exception-with-project-with-gold-with-throw-run-ebts-pass-at-k-compiled-at-5-llama_cpp-qwen2.5-coder:32b-instruct-q8_0-tuctn-all-info-multi_ebt}{95.23}
\DefMacro{res-real-mega-test-data-with-exception-with-project-with-gold-with-throw-run-ebts-pass-at-k-pass-at-1-llama_cpp-qwen2.5-coder:32b-instruct-q8_0-tuctn-all-info-multi_ebt}{86.25}
\DefMacro{res-real-mega-test-data-with-exception-with-project-with-gold-with-throw-run-ebts-pass-at-k-pass-at-10-llama_cpp-qwen2.5-coder:32b-instruct-q8_0-tuctn-all-info-multi_ebt}{86.84}
\DefMacro{res-real-mega-test-data-with-exception-with-project-with-gold-with-throw-run-ebts-pass-at-k-pass-at-5-llama_cpp-qwen2.5-coder:32b-instruct-q8_0-tuctn-all-info-multi_ebt}{86.51}
\DefMacro{res-real-mega-test-data-with-exception-with-project-with-gold-with-throw-run-all-pass-at-k-pass-at-1-llama_cpp-qwen2.5-coder:32b-instruct-q8_0-tuctn-all-info-multi_ebt}{85.92}
\DefMacro{res-real-mega-test-data-with-exception-with-project-with-gold-with-throw-run-all-pass-at-k-pass-at-10-llama_cpp-qwen2.5-coder:32b-instruct-q8_0-tuctn-all-info-multi_ebt}{86.51}
\DefMacro{res-real-mega-test-data-with-exception-with-project-with-gold-with-throw-run-all-pass-at-k-pass-at-5-llama_cpp-qwen2.5-coder:32b-instruct-q8_0-tuctn-all-info-multi_ebt}{86.18}
\DefMacro{res-real-mega-test-data-with-exception-with-project-with-gold-with-throw-run-all-with-tools-pass-at-k-pass-at-1-llama_cpp-qwen2.5-coder:32b-instruct-q8_0-tuctn-all-info-multi_ebt}{75.72}
\DefMacro{res-real-mega-test-data-with-exception-with-project-with-gold-with-throw-run-all-with-tools-pass-at-k-pass-at-10-llama_cpp-qwen2.5-coder:32b-instruct-q8_0-tuctn-all-info-multi_ebt}{76.32}
\DefMacro{res-real-mega-test-data-with-exception-with-project-with-gold-with-throw-run-all-with-tools-pass-at-k-pass-at-5-llama_cpp-qwen2.5-coder:32b-instruct-q8_0-tuctn-all-info-multi_ebt}{75.99}

\DefMacro{ComplexityLocMax}{83\xspace}
\DefMacro{ComplexityLocClosedMax}{14\xspace}
\DefMacro{ComplexityLocOpenLabel}{15+\xspace}
\DefMacro{ComplexityLocOpenPercent}{19.7\%\xspace}
\DefMacro{ComplexityLocMidRangeLabel}{6 to 14\xspace}
\DefMacro{ComplexityCcMax}{19\xspace}
\DefMacro{ComplexityCcOpenMin}{5\xspace}
\DefMacro{ComplexityCcOpenLabel}{5+\xspace}
\DefMacro{ComplexityCcOpenPercent}{14.1\%\xspace}
\DefMacro{ComplexityCcOpenGapValue}{12.8\xspace}

\DefMacro{res-real-mega-test-data-with-exception-with-project-with-gold-with-throw-run-ebts-pass-at-k-compiled-at-1-llama_cpp-llama3.1:8b-instruct-q8_0-tuctn-all-info-multi_ebt}{77.99}
\DefMacro{res-real-mega-test-data-with-exception-with-project-with-gold-with-throw-run-ebts-pass-at-k-compiled-at-10-llama_cpp-llama3.1:8b-instruct-q8_0-tuctn-all-info-multi_ebt}{80.92}
\DefMacro{res-real-mega-test-data-with-exception-with-project-with-gold-with-throw-run-ebts-pass-at-k-compiled-at-5-llama_cpp-llama3.1:8b-instruct-q8_0-tuctn-all-info-multi_ebt}{79.44}
\DefMacro{res-real-mega-test-data-with-exception-with-project-with-gold-with-throw-run-ebts-pass-at-k-pass-at-1-llama_cpp-llama3.1:8b-instruct-q8_0-tuctn-all-info-multi_ebt}{60.53}
\DefMacro{res-real-mega-test-data-with-exception-with-project-with-gold-with-throw-run-ebts-pass-at-k-pass-at-10-llama_cpp-llama3.1:8b-instruct-q8_0-tuctn-all-info-multi_ebt}{64.14}
\DefMacro{res-real-mega-test-data-with-exception-with-project-with-gold-with-throw-run-ebts-pass-at-k-pass-at-5-llama_cpp-llama3.1:8b-instruct-q8_0-tuctn-all-info-multi_ebt}{62.34}
\DefMacro{res-real-mega-test-data-with-exception-with-project-with-gold-with-throw-run-all-pass-at-k-pass-at-1-llama_cpp-llama3.1:8b-instruct-q8_0-tuctn-all-info-multi_ebt}{60.20}
\DefMacro{res-real-mega-test-data-with-exception-with-project-with-gold-with-throw-run-all-pass-at-k-pass-at-10-llama_cpp-llama3.1:8b-instruct-q8_0-tuctn-all-info-multi_ebt}{63.82}
\DefMacro{res-real-mega-test-data-with-exception-with-project-with-gold-with-throw-run-all-pass-at-k-pass-at-5-llama_cpp-llama3.1:8b-instruct-q8_0-tuctn-all-info-multi_ebt}{62.01}
\DefMacro{res-real-mega-test-data-with-exception-with-project-with-gold-with-throw-run-all-with-tools-pass-at-k-pass-at-1-llama_cpp-llama3.1:8b-instruct-q8_0-tuctn-all-info-multi_ebt}{45.99}
\DefMacro{res-real-mega-test-data-with-exception-with-project-with-gold-with-throw-run-all-with-tools-pass-at-k-pass-at-10-llama_cpp-llama3.1:8b-instruct-q8_0-tuctn-all-info-multi_ebt}{49.34}
\DefMacro{res-real-mega-test-data-with-exception-with-project-with-gold-with-throw-run-all-with-tools-pass-at-k-pass-at-5-llama_cpp-llama3.1:8b-instruct-q8_0-tuctn-all-info-multi_ebt}{47.70}

\DefMacro{res-real-mega-test-data-with-exception-with-project-with-gold-with-throw-run-ebts-pass-at-k-compiled-at-1-llama_cpp-qwen2.5-coder:32b-instruct-q8_0-only-nebt-multi_ebt}{85.10}
\DefMacro{res-real-mega-test-data-with-exception-with-project-with-gold-with-throw-run-ebts-pass-at-k-compiled-at-10-llama_cpp-qwen2.5-coder:32b-instruct-q8_0-only-nebt-multi_ebt}{86.18}
\DefMacro{res-real-mega-test-data-with-exception-with-project-with-gold-with-throw-run-ebts-pass-at-k-compiled-at-5-llama_cpp-qwen2.5-coder:32b-instruct-q8_0-only-nebt-multi_ebt}{85.99}
\DefMacro{res-real-mega-test-data-with-exception-with-project-with-gold-with-throw-run-ebts-pass-at-k-pass-at-1-llama_cpp-qwen2.5-coder:32b-instruct-q8_0-only-nebt-multi_ebt}{73.06}
\DefMacro{res-real-mega-test-data-with-exception-with-project-with-gold-with-throw-run-ebts-pass-at-k-pass-at-10-llama_cpp-qwen2.5-coder:32b-instruct-q8_0-only-nebt-multi_ebt}{74.34}
\DefMacro{res-real-mega-test-data-with-exception-with-project-with-gold-with-throw-run-ebts-pass-at-k-pass-at-5-llama_cpp-qwen2.5-coder:32b-instruct-q8_0-only-nebt-multi_ebt}{74.00}
\DefMacro{res-real-mega-test-data-with-exception-with-project-with-gold-with-throw-run-all-pass-at-k-pass-at-1-llama_cpp-qwen2.5-coder:32b-instruct-q8_0-only-nebt-multi_ebt}{73.06}
\DefMacro{res-real-mega-test-data-with-exception-with-project-with-gold-with-throw-run-all-pass-at-k-pass-at-10-llama_cpp-qwen2.5-coder:32b-instruct-q8_0-only-nebt-multi_ebt}{74.34}
\DefMacro{res-real-mega-test-data-with-exception-with-project-with-gold-with-throw-run-all-pass-at-k-pass-at-5-llama_cpp-qwen2.5-coder:32b-instruct-q8_0-only-nebt-multi_ebt}{74.00}
\DefMacro{res-real-mega-test-data-with-exception-with-project-with-gold-with-throw-run-all-with-tools-pass-at-k-pass-at-1-llama_cpp-qwen2.5-coder:32b-instruct-q8_0-only-nebt-multi_ebt}{63.62}
\DefMacro{res-real-mega-test-data-with-exception-with-project-with-gold-with-throw-run-all-with-tools-pass-at-k-pass-at-10-llama_cpp-qwen2.5-coder:32b-instruct-q8_0-only-nebt-multi_ebt}{64.80}
\DefMacro{res-real-mega-test-data-with-exception-with-project-with-gold-with-throw-run-all-with-tools-pass-at-k-pass-at-5-llama_cpp-qwen2.5-coder:32b-instruct-q8_0-only-nebt-multi_ebt}{64.54}

\DefMacro{stat-num-methods-real-mega-test-data-with-exception}{525}

\DefMacro{stat-num-methods-real-mega-test-data-with-exception-with-project-with-gold}{449}
\DefMacro{stat-unreported-exception-real-mega-test-data-with-exception-with-project-with-gold}{4}
\DefMacro{stat-missing-return-only-real-mega-test-data-with-exception-with-project-with-gold}{34}
\DefMacro{stat-total-rm-comp-fail-real-mega-test-data-with-exception-with-project-with-gold}{38}

\DefMacro{stat-num-projects-mega-val-data-with-exception-with-project-with-gold-with-throw}{43}
\DefMacro{stat-num-methods-mega-val-data-with-exception-with-project-with-gold-with-throw}{214}
\DefMacro{stat-etest-sum-mega-val-data-with-exception-with-project-with-gold-with-throw}{294}
\DefMacro{stat-exception-type-mega-val-data-with-exception-with-project-with-gold-with-throw}{48}
\DefMacro{stat-catch-throw-mega-val-data-with-exception-with-project-with-gold-with-throw}{34}
\DefMacro{stat-if-throw-mega-val-data-with-exception-with-project-with-gold-with-throw}{147}
\DefMacro{stat-switch-throw-mega-val-data-with-exception-with-project-with-gold-with-throw}{6}
\DefMacro{stat-rest-throw-mega-val-data-with-exception-with-project-with-gold-with-throw}{40}
\DefMacro{stat-total-throw-mega-val-data-with-exception-with-project-with-gold-with-throw}{227}

\DefMacro{qual-base-total-samples-count}{223}
\DefMacro{qual-base-semantic-match-yes-count}{166}
\DefMacro{qual-base-semantic-match-no-of-total-pct}{25.56}
\DefMacro{qual-base-semantic-match-yes-of-methods-pct}{54.61}
\DefMacro{qual-base-too-lenient-pct}{64.91}
\DefMacro{qual-base-too-strict-pct}{10.53}
\DefMacro{qual-base-wrong-handling-pct}{12.28}
\DefMacro{qual-base-destroyed-code-pct}{12.28}
\DefMacro{qual-tuctn-all-info-total-samples-count}{263}
\DefMacro{qual-tuctn-all-info-semantic-match-yes-count}{198}
\DefMacro{qual-tuctn-all-info-semantic-match-no-of-total-pct}{24.71}
\DefMacro{qual-tuctn-all-info-semantic-match-yes-of-methods-pct}{65.13}
\DefMacro{qual-tuctn-all-info-too-lenient-pct}{63.08}
\DefMacro{qual-tuctn-all-info-too-strict-pct}{12.31}
\DefMacro{qual-tuctn-all-info-wrong-handling-pct}{13.85}
\DefMacro{qual-tuctn-all-info-destroyed-code-pct}{10.77}
\DefMacro{qual-fp-rate-gap-pct}{0.8}
\DefMacro{qual-max-share-gap-pct}{1.8}

\DefMacro{qual-human-rater-count}{two}
\DefMacro{qual-agent-model}{Claude Code}
\DefMacro{qual-agent-harness-version}{2.1.239}
\DefMacro{qual-agent-model-version}{Claude Opus 5}

\DefMacro{agree-kappa-min}{0.77}
\DefMacro{agree-kappa-max}{0.87}

\newcommand{\tableCapVspace}{\vspace{-1pt}}
\DefMacro{TCap-dataset-stats}{Statistics of our dataset.\tableCapVspace}
\DefMacro{TCap-dataset-throw-count}{Statistics of \ERC types.\tableCapVspace}
\DefMacro{TCap-res-prompt-comp-repair}{Comparison of \Tool and the baseline prompt after 4 iterations of \LLM self-repair on Qwen 2.5 Coder 32b. Each iteration returns the failing samples to the \LLM with its original context and the corresponding compiler or test errors.\tableCapVspace}
\DefMacro{TCap-res-prompt-comp-multi-ebt}{Effect of each context component, on Qwen 2.5 Coder 32b: (1)~\textbf{\ShortAS}, methods and variables usable inside the \tarmethod; (2)~\textbf{\ShortEC}, constructors of the exceptions specified by the \EBTs; (3)~\textbf{\ShortTE}, exceptions thrown inside the \tarmethod when running the \EBTs; (4)~\textbf{\ShortLC}, line coverage of each \EBT; (5)~\textbf{\ShortNEBT}, \nEBTs defined for the \tarmethod.\tableCapVspace}
\DefMacro{TCap-res-model-comp-multi-ebt}{Performance of \Tool compared with the baseline prompt across 5 \LLMs of different sizes and architectures. We sample 10 outputs per task and compute the metrics as described in Section~\ref{sec:eval-metrics}.\tableCapVspace}

\DefMacro{TCap-qualitative}{Classification of false positives over the \tarmethods for which each prompt generates a passing \ERC (\UseMacro{qual-base-total-samples-count} for the baseline and \UseMacro{qual-tuctn-all-info-total-samples-count} for \Tool), using \UseMacro{THead-qwen2.5-coder:32b-instruct-q8-0}. The first row is the share of passing \ERC that is a false positive; the remaining rows are the shares of those false positives per category.\tableCapVspace}

\DefMacro{THead-qwen2.5-coder:32b-instruct-q8-0}{Qwen 2.5 Coder 32b}
\DefMacro{THead-qwen2.5-coder:7b-instruct-q8-0}{Qwen 2.5 Coder 7b}
\DefMacro{THead-llama3.1:8b-instruct-q8-0}{Llama3.1 8b}
\DefMacro{THead-phi4:14b-q8-0}{Phi4 14b}
\DefMacro{THead-gpt-5-mini}{GPT-5 Mini}

\DefMacro{THead-model}{Models}
\DefMacro{THead-prompt}{Prompt}
\DefMacro{THead-compiled-at-k}{compiled@k}
\DefMacro{THead-ebts-pass-at-k}{pass@k (\EBTs)}
\DefMacro{THead-all-pass-at-k}{pass@k (All User)}
\DefMacro{THead-allntools-pass-at-k}{pass@k (All User+Tools)}
\DefMacro{THead-run-all-with-tools-pass-at-k-pass-at-1-small}{k=1}
\DefMacro{THead-run-all-with-tools-pass-at-k-pass-at-5-small}{k=5}
\DefMacro{THead-run-all-with-tools-pass-at-k-pass-at-10-small}{k=10}
\DefMacro{THead-run-all-with-tools-pass-at-k-pass-at-5}{pass@5 (All User+Tools)}
\DefMacro{THead-run-ebts-pass-at-k-compiled-at-10-small}{k=10}
\DefMacro{THead-run-ebts-pass-at-k-pass-at-10-small}{k=10}
\DefMacro{THead-run-all-pass-at-k-pass-at-10-small}{k=10}
\DefMacro{THead-run-ebts-pass-at-k-compiled-at-5-small}{k=5}
\DefMacro{THead-run-ebts-pass-at-k-pass-at-5-small}{k=5}
\DefMacro{THead-run-all-pass-at-k-pass-at-5-small}{k=5}
\DefMacro{THead-run-ebts-pass-at-k-compiled-at-1-small}{k=1}
\DefMacro{THead-run-ebts-pass-at-k-pass-at-1-small}{k=1}
\DefMacro{THead-run-all-pass-at-k-pass-at-1-small}{k=1}
\DefMacro{THead-run-ebts-pass-at-k-compiled-at-5}{compiled@5}
\DefMacro{THead-run-ebts-pass-at-k-pass-at-5}{pass@5 (\EBTs)}
\DefMacro{THead-run-all-pass-at-k-pass-at-5}{pass@5 (All User)}
\DefMacro{THead-run-all-pass-at-k-pass-at-10}{pass@10 (All User)}

\DefMacro{THead-base}{\Base}
\DefMacro{THead-base-repair@4}{Iter-\Base}
\DefMacro{THead-only-avsym}{Only w/ \ShortAS}
\DefMacro{THead-only-lcov}{Only w/ \ShortLC}
\DefMacro{THead-only-nebt}{Only w/ \ShortNEBT}
\DefMacro{THead-only-threxc}{Only w/ \ShortTE}
\DefMacro{THead-cmtu}{Only w/ \ShortEC}
\DefMacro{THead-tuctn-all-info}{\Tool}
\DefMacro{THead-tuctn-all-info-repair@4}{Iter-\Tool}

\DefMacro{THead-data-stats-test-data}{Eval}
\DefMacro{THead-data-stats-real-mega-test-data-with-exception-with-project-with-gold-with-throw}{Eval}
\DefMacro{THead-data-stats-val-no-switch-clean-v3}{Valid}
\DefMacro{THead-data-stats-mega-val-data-with-exception-with-project-with-gold-with-throw}{Valid}
\DefMacro{THead-data-stats-all-data}{All}
\DefMacro{THead-data-stats-num-projects}{\# Projects}
\DefMacro{THead-data-stats-num-methods}{\# Methods}
\DefMacro{THead-data-stats-etest-sum}{\# EBTs}
\DefMacro{THead-data-stats-exception-type}{\# Exception Types}
\DefMacro{THead-data-stats-catch-throw}{\#\CodeIn{try\CodeDash catch}}
\DefMacro{THead-data-stats-if-throw}{\#\CodeIn{if}}
\DefMacro{THead-data-stats-switch-throw}{\#\CodeIn{switch}}
\DefMacro{THead-data-stats-rest-throw}{\# Other}
\DefMacro{THead-data-stats-total-throw}{\# All}

\DefMacro{tool-test-cap}{400\xspace}
\DefMacro{eval-num-samples}{10\xspace}
\DefMacro{eval-temperature}{0.8\xspace}

\DefMacro{THead-qual-too-lenient}{\CateTL}
\DefMacro{THead-qual-too-strict}{\CateTS}
\DefMacro{THead-qual-wrong-handling}{\CateWH}
\DefMacro{THead-qual-destroyed-code}{\CateCD}
\DefMacro{THead-qual-semantic-match-no}{Total}

\newcommand{\NumCollectedProjects}{\UseMacro{stat-num-projects-real-non-direct-mega-test-data}\xspace}
\newcommand{\NumCollectedMethods}{\UseMacro{stat-num-methods-real-non-direct-mega-test-data}\xspace}
\newcommand{\NumDirectThrowMethods}{\UseMacro{stat-num-methods-real-mega-test-data}\xspace}
\newcommand{\DirectThrowPercentage}{\UseMacro{stat-direct-throw-percentage}\%\xspace}
\newcommand{\NumRmCompFailMethods}{\UseMacro{stat-total-rm-comp-fail-real-mega-test-data-with-exception-with-project-with-gold}\xspace}
\newcommand{\RmCompFailPercentage}{\UseMacro{stat-rm-comp-fail-percentage}\%\xspace}
\newcommand{\NumRmMissingReturnMethods}{\UseMacro{stat-missing-return-only-real-mega-test-data-with-exception-with-project-with-gold}\xspace}
\newcommand{\NumRmUnreportedMethods}{\UseMacro{stat-unreported-exception-real-mega-test-data-with-exception-with-project-with-gold}\xspace}

\begin{document}

\title{\Title}

\makeatletter
\newcommand{\linebreakand}{%
\end{@IEEEauthorhalign}
\hfill\mbox{}\par
\mbox{}\hfill\begin{@IEEEauthorhalign}
}
\makeatother
\author{
\IEEEauthorblockN{Linghan Zhong}
\IEEEauthorblockA{\textit{The University of Texas at Austin, USA} \\
linghanz@cs.utexas.edu}
\and
\IEEEauthorblockN{Jiyang Zhang}
\IEEEauthorblockA{\textit{The University of Texas at Austin, USA} \\
jiyang.zhang@utexas.edu}
\and
\IEEEauthorblockN{Jayanth Srinivasa}
\IEEEauthorblockA{\textit{Cisco Systems, USA} \\
jasriniv@cisco.com}
\linebreakand
\IEEEauthorblockN{Junyi Jessy Li}
\IEEEauthorblockA{\textit{The University of Texas at Austin, USA} \\
jessy@austin.utexas.edu}
\and
\IEEEauthorblockN{Milos Gligoric}
\IEEEauthorblockA{\textit{The University of Texas at Austin, USA} \\
gligoric@utexas.edu}
}

\maketitle

\begin{abstract}

Exception Related Code (\ERC), which includes \tss, conditions
(\ifstates) that guard those throw statements, and \trycatchs, is an essential
component of software systems, allowing developers to detect and handle
exceptional states that deviate from the expected program behavior.
However, manually writing \ERCs across large codebases is tedious.
We propose a novel task: retrofitting existing code with \ERCs.
Namely, given code (without \ERCs) and Exceptional Behavior Tests
(\EBTs) (e.g., check if method throws InvalidArgumentException if null
is given as the value to the argument) we aim to automatically generate
missing \ERCs, such that the given tests pass.
We design and implement Exception Coder (\Tool) that performs \conEng
to help Large Language Models (\LLMs) tackle this task.
\Tool integrates static and dynamic program analysis with \LLMs by
providing the extracted contextual information to the \LLMs.
To evaluate \Tool, we build
a benchmark constructed from GitHub Java repositories, where we
systematically remove \ERCs in
\UseMacro{stat-num-methods-real-mega-test-data-with-exception-with-project-with-gold-with-throw}
methods from
\UseMacro{stat-num-projects-real-mega-test-data-with-exception-with-project-with-gold-with-throw}
projects.
Our results demonstrate that \Tool provides an effective, though imperfect,
solution to this problem in automated code generation, offering developers the first way
to implement \ERCs following test-driven development.
When combined with \UseMacro{THead-qwen2.5-coder:32b-instruct-q8-0},
\Tool achieves pass@1, 5, and 10 rates of
\UseMacro{res-real-mega-test-data-with-exception-with-project-with-gold-with-throw-run-all-pass-at-k-pass-at-1-llama_cpp-qwen2.5-coder:32b-instruct-q8_0-tuctn-all-info-multi_ebt}\%
(\UseMacro{OverQwenLargePassAllOne} percentage points over baseline),
\UseMacro{res-real-mega-test-data-with-exception-with-project-with-gold-with-throw-run-all-pass-at-k-pass-at-5-llama_cpp-qwen2.5-coder:32b-instruct-q8_0-tuctn-all-info-multi_ebt}\%
(\UseMacro{OverQwenLargePassAllFive} p.p. over baseline), and
\UseMacro{res-real-mega-test-data-with-exception-with-project-with-gold-with-throw-run-all-pass-at-k-pass-at-10-llama_cpp-qwen2.5-coder:32b-instruct-q8_0-tuctn-all-info-multi_ebt}\%
(\UseMacro{OverQwenLargePassAllTen} p.p. over baseline), respectively,
on developer-written test suites.
Our manual inspection of the generated code further reveals limitations of \Tool,
pointing to directions for future work.

\end{abstract}

\section{Introduction}

Exceptions are supported in many modern programming languages (\eg
C\#, Java, Python) and are widely used by developers to indicate that
exceptional behaviors, i.e., events outside the expected execution flow
of a program, have occurred.  Writing exception-related code (\ERC),
such as \tss, \trycatchs, and \conds (\ifstates) that guard the \tss,
is critical for developing reliable software systems, as explicit
error signaling prevents hard-to-diagnose failures from occurring
further downstream and enables callers to handle exceptions.

\begin{figure}[t]
\begin{subfigure}{\columnwidth}
\begin{lstlisting}[language=java-diff]
public abstract class AbstractService implements Service {
    public final synchronized void initialize(Map<String, String> configuration) {
+       if (this.initialized) { (*@\label{example:guard-expression}@*)
+          throw new FaxException("Service is already initialized."); (*@\label{example:throw-statement}@*)
+       }(*@\label{example:cond-end}@*)
        this.initialized = true;
        this.serviceConfiguration = new ConfigurationHolderImpl(configuration, this.propertyPart);
        LoggerManager loggerManager = LoggerManager.getInstance();
        this.serviceLogger = loggerManager.getLogger();
        this.initializeImpl();
    }
}
\end{lstlisting}
\caption{A \tarmethod with an \cond implemented using an \ifstate to guard the \ts.}
\label{fig:example}
\end{subfigure}
\begin{subfigure}{\columnwidth}
\begin{lstlisting}[language=java-pretty]
public class AbstractServiceTest {
    @Before
    public void setUp() throws Exception {
        ...
        this.service.initialize(this.configuration);
    }
    @Test(expected = FaxException.class)
    public void initializeAgainTest() throws Exception {
        this.service.initialize(this.configuration);
    }
}
\end{lstlisting}
\caption{An Exceptional Behavior Test (\EBT) and its setup.}
\label{fig:example-ebt}
\end{subfigure}
\caption{\label{fig:example-all}A \tarmethod and its \EBT, from \exampleClass in \CodeIn{\exampleRepo}.}
\end{figure}

Figure~\ref{fig:example} shows an example of \ERC from a \java project
\CodeIn{\exampleRepo}. The original implementation allows an \exampleClass
instance to be initialized multiple times. When reinitialization occurs,
important state information within the object could be reset, potentially
causing bugs in downstream operations. To mitigate this issue, developers can
add an \ifstate (line~\ref{example:guard-expression}) to the \tarmethod to
check for the exceptional condition and a \ts
(line~\ref{example:throw-statement}) to signal that an exception has occurred.
This approach notifies users of the multiple initialization and halts
execution, preventing more complex errors that could arise.
Note that \conds are not limited to \ifstates: they can also be
written in other forms, such as \switchstates whose
\CodeIn{default} case rejects an unsupported value, or \trycatchs that catch generic exceptions when they occur and throw more descriptive ones.

Despite its importance, writing \ERCs in appropriate locations is
laborious, particularly in large and complex codebases where
exceptional behaviors may arise across multiple contexts. To signal
exceptions effectively, developers must carefully reason about the
specific conditions to decide whether an exceptional behavior has occurred and
identify the control flow branches where exceptions need to be
signaled.

Large language models (\LLMs) have shown remarkable capabilities in code
generation and program understanding, making them a promising tool for \task.
Prior work on using \LLMs to generate \ERC~\cite{sunLLMRuntimeError2024,caiProgrammingAssistantException2024,renMisuseMasteryEnhancing2023, zhangLearningHandleExceptions2020} has primarily focused on
avoiding unhandled exceptions, which involves automatically detecting locations
where exceptions could be thrown and inserting error handling structures to
resolve them. However, since they mainly focus on ensuring robust program
execution by avoiding exceptions, these approaches are insufficient when
developers conversely have specific exceptional behaviors they wish to support
with \ERCs.

Furthermore, unlike these existing approaches that analyze \tarmethods for
potentially unhandled exceptions, retrofitting \ERCs requires knowing the conditions
under which developers want exceptions to be thrown. However, formulating the
intended \conds into natural language prompts can be as laborious for developers as
implementing the \ERCs. A natural solution to this problem is allowing
developers to express their intention through concrete test cases, following the
principles of Test-Driven Development
(TDD)~\cite{pancurImpactTestdrivenDevelopment2011,pancurEmpiricalEvaluationTestdriven2003,huangEmpiricalInvestigationEffectiveness2009}.
TDD, a methodology where automated tests that verify desired functionality are
written \emph{before} the implementation code, is already familiar to many
developers. This approach provides an intuitive way for developers to specify
exceptional behavior by simply providing example inputs and setups that should
trigger exceptions, rather than requiring them to explicitly formulate the
\conds for when exceptions should be thrown.

In this paper, we propose the novel task of retrofitting existing code
with \ERC given Exceptional Behavior Tests
(\EBTs). Figure~\ref{fig:example-ebt} illustrates an example \EBT and
its corresponding setup for the \tarmethod~\exampleMethod shown in
Figure~\ref{fig:example}. In this example, the
developer initializes the \exampleClass in the setup phase and again
within the \EBT, then verifies whether a \exampleException is thrown.

While our task is based on TDD principles, applying existing TDD-based
tools to retrofit \ERC presents unique challenges. Much prior
work~\cite{zhangAutoCodeRoverAutonomousProgram2024,
shinnReflexionLanguageAgents2023,
mathewsTestDrivenDevelopmentCode2024,
liCompetitionLevelCodeGeneration2022} in TDD has explored applying
TDD principles to LLM-based code generation (but not \ERC).
They often involve a generation and
evaluation loop where the model iterates on the previously generated
code based on test outputs. However, existing TDD-based methods fail
to leverage the rich information contained within \EBTs, which
includes crucial details such as target exception types and example
inputs that trigger exceptional behavior.

To this end, we present \textbf{Ex}ception \textbf{Coder} (\Tool), which, given the \tarmethod and
\EBTs, performs \conEng to provide \LLMs with useful context for retrofitting
\ERC. To identify all necessary information for our task, \Tool employs both
static and dynamic program analysis. Our static analysis process provides the
\LLM with contextual information, including \contextAS and \contextEC, which
supplies the context necessary for implementing syntactically correct \tss and
\conds. Dynamic analysis complements this approach by executing the provided
\EBTs and collecting runtime information, such as reachable code paths. This
information from execution gives \LLMs insights into how the exceptional
condition can be identified and where \tss can be inserted. We format this
contextual information into a prompt. This enables the \LLM to generate a new
method with \ERCs incorporated.

To limit the scope of this study, we implement and evaluate \Tool
only for \java projects. We construct a novel benchmark dataset derived from \java
repositories collected from GitHub. Our dataset specifically includes
\tarmethods that have \EBTs covering \tss in them. For selected
\tarmethods, we create ``stripped'' versions of them that maintain their
core functionality but with \ERCs removed.
We then use the corresponding \EBTs to evaluate whether \ERCs added by
our framework match the expected exceptional behavior specified in the
test suite.

To verify the effectiveness of the context provided by \Tool, we
compare against a baseline prompt that includes only the \tarmethod
and \EBTs, following the literature on
TDD~\cite{mathewsTestDrivenDevelopmentCode2024}.
This baseline excludes all context collected by our static and dynamic
analysis, which allows us to investigate how the context
provided by \Tool affects the \LLM's ability to complete the \task
task. Furthermore, to assess the generalizability of our approach
across diverse model capabilities, we evaluated \Tool on 5 distinct
models of varying sizes and architectures, including
\UseMacro{THead-llama3.1:8b-instruct-q8-0}~\cite{grattafioriLlama3Herd2024},
\UseMacro{THead-phi4:14b-q8-0}~\cite{abdinPhi4TechnicalReport2024},
\UseMacro{THead-qwen2.5-coder:7b-instruct-q8-0}~\cite{huiQwen25CoderTechnicalReport2024},
\UseMacro{THead-qwen2.5-coder:32b-instruct-q8-0}, and
\UseMacro{THead-gpt-5-mini}~\cite{OpenaiGPT52025}.
To evaluate the correctness of the generated \ERC, we use pass@k~\cite{chenEvaluatingLargeLanguage2021} metrics, which estimate the probability that at least one out of $k$ generated samples successfully passes the developer-provided tests. Additionally, to verify that each generated \ERC is equivalent to the ground truth, we run automatically generated tests produced by EvoSuite~\cite{FraserAndArcuri11EvoSuite} and Randoop~\cite{PachecoETAL07Randoop}, and manually inspect the generated code.

The results demonstrate that \Tool consistently outperforms the baseline across
all metrics for the selected models.
With \UseMacro{THead-qwen2.5-coder:32b-instruct-q8-0}, \Tool achieves
gains of \UseMacro{OverQwenLargePassAllOne} percentage points on
pass@1 (reaching
\UseMacro{res-real-mega-test-data-with-exception-with-project-with-gold-with-throw-run-all-pass-at-k-pass-at-1-llama_cpp-qwen2.5-coder:32b-instruct-q8_0-tuctn-all-info-multi_ebt}\%),
\UseMacro{OverQwenLargePassAllFive} p.p. on pass@5
(reaching
\UseMacro{res-real-mega-test-data-with-exception-with-project-with-gold-with-throw-run-all-pass-at-k-pass-at-5-llama_cpp-qwen2.5-coder:32b-instruct-q8_0-tuctn-all-info-multi_ebt}\%),
and \UseMacro{OverQwenLargePassAllTen} p.p. on pass@10
(reaching
\UseMacro{res-real-mega-test-data-with-exception-with-project-with-gold-with-throw-run-all-pass-at-k-pass-at-10-llama_cpp-qwen2.5-coder:32b-instruct-q8_0-tuctn-all-info-multi_ebt}\%)
when evaluated on developer-written test suites.
Similarly, with
\UseMacro{THead-gpt-5-mini}, \Tool achieves gains of
\UseMacro{OverGPTPassAllOne} p.p. on pass@1 (reaching
\UseMacro{res-real-mega-test-data-with-exception-with-project-with-gold-with-throw-run-all-pass-at-k-pass-at-1-azure-gpt-5-mini-tuctn-all-info-multi_ebt}\%),
\UseMacro{OverGPTPassAllFive} p.p. on pass@5 (reaching
\UseMacro{res-real-mega-test-data-with-exception-with-project-with-gold-with-throw-run-all-pass-at-k-pass-at-5-azure-gpt-5-mini-tuctn-all-info-multi_ebt}\%),
and \UseMacro{OverGPTPassAllTen} p.p. on pass@10
(reaching
\UseMacro{res-real-mega-test-data-with-exception-with-project-with-gold-with-throw-run-all-pass-at-k-pass-at-10-azure-gpt-5-mini-tuctn-all-info-multi_ebt}\%).
These improvements highlight the effectiveness of \Tool's
context-engineering approach in enabling language models to generate
better \ERCs.
Our manual inspection of the generated code further reveals limitations of \Tool,
pointing to directions for future work.

\vspace{5pt}
\noindent
Our work makes the following contributions:
\begin{itemize}[topsep=3pt,itemsep=3pt,partopsep=0ex,parsep=0ex,leftmargin=*]
\item We introduce and formalize the problem of retrofitting
existing code with \ERCs given \EBTs and code under test.
\item We present \Tool, which leverages static and dynamic program
analysis to perform \conEng, enabling large language models to
automatically generate \ERCs (at
appropriate locations in source code).
\item We construct a novel benchmark dataset derived from methods in
Java repositories collected from GitHub, specifically methods with
existing \EBTs for throw statements, which can evaluate generated \ERCs.
\item We evaluate the effectiveness of \Tool, and show that it
outperforms the baseline by enabling \LLMs to generate more
methods with \ERCs that compile and pass tests.
\end{itemize}

\vspace{5pt}
\noindent

\noindent
Our code, experimental scripts, and dataset are publicly available at
\url{https://github.com/EngineeringSoftware/excoder}.

\section{Task Definition}
\label{sec:task-definition}

In this section, we describe our task of retrofitting existing code
with \ERCs.

Given a \tarmethod ($M$) and a set of \EBTs
$\mathcal{E} = \{e_1, e_2, \ldots, e_n\}$, our goal is to automatically
retrofit \ERCs into $M$ such that the updated method ($M'$) satisfies
all tests in $\mathcal{E}$ while preserving the original functionality
of $M$ (passing all the existing non-exceptional tests). In this
context, \ERC refers to the combination of \tss and \conds, which are
the control structures surrounding \tss that determine when an
exception should be thrown. \Conds may take various forms, such as
\ifstates that detect exceptional program states, or \trycatchs that
convert non-specific exceptions (\eg\ \CodeIn{java.lang.RuntimeException}) into application-specific ones.

We assume any exception that the user wants to be thrown from $M$
originates directly from \tss within $M$ itself.\footnote{Note, if an exception thrown from a method that $M$ invokes is caught within $M$ and thrown again, we still consider it as originating directly from a \ts within $M$.} Also, at least one
\EBT is defined on $M$. From our observations, such cases are common
as they cover \DirectThrowPercentage of the methods with \EBTs in the
collected data (Section~\ref{sec:dataset-collection}).

An example \tarmethod selected from our dataset is shown in
Figure~\ref{fig:example}, from which the \ERCs have
been removed.  An \EBT (shown in Figure~\ref{fig:example-ebt}) is
defined for this method and specifies that a \exampleException must
be thrown when the object is initialized twice.
Given $M$, $\mathcal{E}$, and the repository where the \tarmethod is
located, an \task system should generate $M'$, as shown in
Figure~\ref{fig:example}, which incorporates \ERCs
that include an \cond (line \ref{example:guard-expression}) to detect
whether the object has been initialized and a \ts (line
\ref{example:throw-statement}) that throws \exampleException.

In this paper, we focus on performing \conEng to help \LLMs retrofit
\ERC.  Specifically, our goal is to identify and collect all useful
context from $M$, $\mathcal{E}$, and the repository containing $M$
that enables an \LLM to generate $M'$. We consider the task successful
when the generated $M'$ passes all \EBTs in $\mathcal{E}$ by throwing
the correct exception and also passes any
existing non-exceptional test covering the original \tarmethod ($M$),
which ensures that retrofitted \ERCs do not interfere with normal execution
paths.

\section{\Tool}

\begin{figure}[t]
\centering \includegraphics[width=\columnwidth,trim=0 9 0 4,clip]{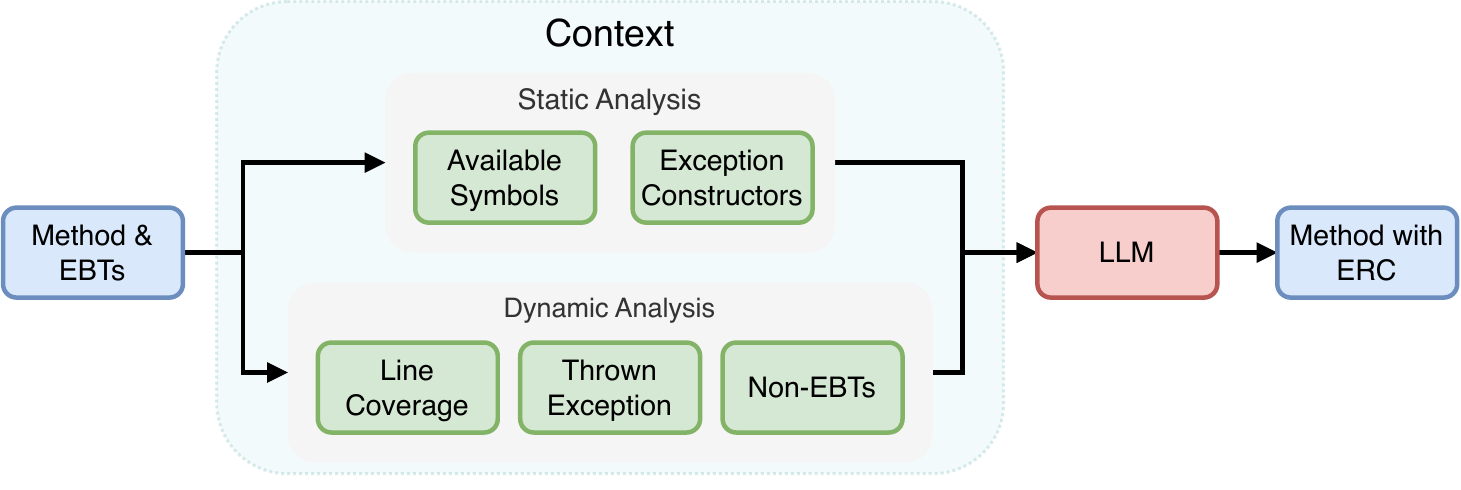}
\caption{Overview of \Tool.\label{fig:overview}}
\end{figure}

Figure~\ref{fig:overview} illustrates \Tool's workflow for
performing \conEng to assist the \LLM in completing the \task task. Given \tarmethods and
a corresponding set of \EBTs, \Tool employs static analysis to extract
contextual information including: \contextEC (\ShortEC) and \contextAS (\ShortAS). Additionally, it leverages dynamic analysis to capture runtime
information including: \contextTE (\ShortTE), \contextLC (\ShortLC), and
\contextNEBT
(\ShortNEBT). This contextual information generated by \Tool is integrated and
formulated into a structured prompt along with the \tarmethod and all \EBTs,
then provided to an \LLM to generate a modified method with \ERC. Here, we
ask the \LLM to add \ERC that can pass all given \EBTs. Compared to
resolving a single \EBT at a time, we believe this setup helps reduce the
overhead cost of prompting the \LLM multiple times and helps the \LLM better plan for
the structure of the \conds.
In the following paragraphs, we provide a detailed description of the static and
dynamic analysis techniques employed by \Tool and explain how they facilitate
\task.

\subsection{Static Analysis}

First, we introduce our static analysis process and the collected contextual
information. This information enables the \LLM to generate syntactically correct
code and encourages the \LLM to use existing methods, classes, and variables to
construct effective \ERC.

\MyPara{\ContextEC (\ShortEC)} To know how to throw the appropriate exception,
the \LLM must first know how to construct the exception object specified by the \EBTs.
Since the required parameters of the exception's constructors may fall outside
of the \LLM's knowledge, \Tool provides the constructor signatures to the
\LLM. Given an exception type, \Tool checks whether a corresponding definition
for its constructors exists in the repository by searching for Java files with a name that matches the exception type. If so, it will collect the
signatures and documentation for all constructors defined for that exception
type. In practice, we observe that signatures and documentation alone already provide
enough information for the \LLM to use the constructors correctly. When an
exception is not defined in the repository, \Tool will produce a
constructor signature by inspecting the class files in the project's dependencies,
identifying the exception constructor's parameter types and their order, and
finally composing the signature in the form of \CodeIn{public\ ExceptionName(PType0\
arg0,\ PType1\ arg1 ...)}.

\MyPara{\ContextAS(\ShortAS)} To enable the \LLM to produce syntactically correct \conds capable of inspecting input parameters and object states for exceptional behaviors, it is critical that the \LLM has knowledge of the methods and variables that can be potentially useful, such as the \CodeIn{initialized} variable used in the \cond in Figure~\ref{fig:example} line \ref{example:guard-expression}.
To this end, \Tool visits and collects all public method nodes and field nodes in the AST of the user-defined classes of the \tarmethod's input parameters, as well as all method (private and public) and field nodes defined within the class of the \tarmethod.
Following the same approach used for processing \contextEC, \Tool collects only the signature and documentation comment for each identified method and variable.

\subsection{Dynamic Analysis}

We now describe the context collected in \Tool's dynamic analysis
step.
Our dynamic analysis framework collects execution information that
helps \LLMs to identify the correct \conds and appropriate locations
for \ERC, so that the generated method can pass the given set of
\EBTs while preserving the target method's original functionality.

\MyPara{\ContextTE(\ShortTE)}
Oftentimes, when an exceptional behavior occurs, an exception is already thrown
by methods called within the target method, but the exception type differs from
the desired target exception type specified by the \EBTs. \Tool collects such
exceptions by executing each \EBT and logging exceptions thrown inside the
target method. Prior to execution, \Tool preprocesses the target method with JavaParser by
encapsulating the entire method body within a \CodeIn{try} block and appending a
\CodeIn{catch} block that handles all exception types.
Moreover, we not only add logging to the newly added \CodeIn{catch}
block, but we also log exceptions caught in existing \CodeIn{catch} blocks
inside the \tarmethod, which also provide useful information about the behavior
of the target method under the given inputs.
This context is essential for understanding exception handling.
Note that even though an exception is already thrown inside the
\tarmethod here, a developer may want \ERC that throws a custom, more
informative exception.

\MyPara{\ContextLC(\ShortLC)}
We use Java instrumentation to add logging for the line number of each
line inside the target method that is executed during \EBT execution
to capture the code coverage information. By analyzing each line of
the \tarmethod executed under exceptional inputs, \Tool can restrict
potential placement locations for \ERC to sections of the \tarmethod
that are reachable given the exceptional inputs specified by the
\EBTs. This ensures that the generated \ERC is always positioned
within a code branch that can actually be triggered by the exceptional
inputs. The \contextLC information also helps \LLMs determine the
value of each branching condition. This provides insight into the
program's state at decision points and helps \LLMs to infer the states
of the program that led to exceptional behaviors. By understanding
these states, \LLMs can better formulate the \cond that detects the
situation where an exception needs to be signaled.

\MyPara{\ContextNEBT(\ShortNEBT)}
\NEBTs cover the target method but do not verify throw
statements. They can help the \LLM compare exceptional inputs with normal
inputs, which helps with the generation of correct \conds. \NEBTs also
provide information on how the \tarmethod should execute normally. This
helps \LLMs implement \ERC without altering the target method's
expected behavior. To identify \nEBTs that test each target method, we
execute all tests in the project test suite and log each method that
is invoked during the execution through instrumentation. In this way, we can match the target
method with all \nEBTs that execute it.

\subsection{\LLM Generation}
\label{sec:llm-generation}

Finally, all context collected from both static analysis (\ShortEC and
\ShortAS) and dynamic analysis (\ShortTE, \ShortLC, and \ShortNEBT)
along with all the \EBTs ($\mathcal{E}$) given by the user and the
\tarmethod ($M$) are combined into a structured prompt. The prompt
construction can be represented as:
$$f_{\text{com}}(f_{\text{com}}(M, \FormulaTE), \FormulaLC)
\oplus \mathcal{E} \oplus \FormulaEC \oplus \FormulaAS \oplus \FormulaNEBT,$$ where
$\oplus$ denotes concatenation, and $f_{\text{com}}$ denotes the
function that annotates the given method $M$ with the given context as
inline comments.

Specifically, $f_{\text{com}}(\cdot, \FormulaTE)$ adds inline
comments at each line where an exception is thrown during test
execution, indicating the specific exception type that occurs at that
location. Similarly, $f_{\text{com}}(\cdot, \FormulaLC)$ inserts
comments at each line covered by an \EBT, specifying which particular
\EBT executed that line.

This prompt provides the \LLM with the essential context to reason about and implement appropriate \ERC, while also guiding the model to identify the locations within the method where they could be inserted.

\section{Dataset}

\begin{table}[t]
\begin{small}
\begin{center}
\caption{\UseMacro{TCap-dataset-stats}\label{tab:dataset-stats}}
\begin{tabular}{l |cccc}
\toprule
Split & \UseMacro{THead-data-stats-num-projects} & \UseMacro{THead-data-stats-num-methods} & \UseMacro{THead-data-stats-etest-sum} & \UseMacro{THead-data-stats-exception-type}
\\
\midrule
\UseMacro{THead-data-stats-all-data}
& \UseMacro{stat-num-projects-all-data}
& \UseMacro{stat-num-methods-all-data}
& \UseMacro{stat-etest-sum-all-data}
& \UseMacro{stat-exception-type-all-data}
\\
\UseMacro{THead-data-stats-mega-val-data-with-exception-with-project-with-gold-with-throw}
& \UseMacro{stat-num-projects-mega-val-data-with-exception-with-project-with-gold-with-throw}
& \UseMacro{stat-num-methods-mega-val-data-with-exception-with-project-with-gold-with-throw}
& \UseMacro{stat-etest-sum-mega-val-data-with-exception-with-project-with-gold-with-throw}
& \UseMacro{stat-exception-type-mega-val-data-with-exception-with-project-with-gold-with-throw}
\\
\UseMacro{THead-data-stats-real-mega-test-data-with-exception-with-project-with-gold-with-throw}
& \UseMacro{stat-num-projects-real-mega-test-data-with-exception-with-project-with-gold-with-throw}
& \UseMacro{stat-num-methods-real-mega-test-data-with-exception-with-project-with-gold-with-throw}
& \UseMacro{stat-etest-sum-real-mega-test-data-with-exception-with-project-with-gold-with-throw}
& \UseMacro{stat-exception-type-real-mega-test-data-with-exception-with-project-with-gold-with-throw}
\\
\bottomrule
\end{tabular}
\end{center}
\end{small}
\end{table}

To evaluate the usefulness of the context generated by \Tool on the novel \task task, we construct a new dataset comprised of \java methods collected from GitHub repositories. In this section, we describe our raw data collection process (Section~\ref{sec:dataset-collection}), our \ERC removal process (Section~\ref{sec:dataset-throw-remove}), the added tool-generated tests (Section~\ref{sec:dataset-tool-tests}), and the statistics of our dataset (Section~\ref{sec:dataset-statistics}).

\subsection{Raw Data Collection}
\label{sec:dataset-collection}

Following prior work~\cite{zhangExLongGeneratingExceptional2024}, we collect data from \java projects from
CodeSearchNet~\cite{CodeSearchNet} that are available on GitHub and satisfy the
following criteria: (1)~use the Maven build system; (2)~compile
successfully; and (3)~have a license that permits the use of their data.
For each project, we identify methods that have at least one \EBT
defined on them. Here, we categorize a test method as an \EBT if it
follows one of the widely used patterns identified by developers~\cite{MarcilioFuria21HowJavaProgrammersTestExceptionalBehavior}:
`try/catch', `expect test', `expect rule', and `assert throws'.

To ensure coverage of the most up-to-date Java code in our evaluation set,
we collect the most recent commit from each project with a cutoff date
of January 1, 2026. This initial collection from \NumCollectedProjects
projects yields \NumCollectedMethods methods. We then filter to
methods where the exceptions specified by the \EBTs originate from
\tss inside the target method, leaving \NumDirectThrowMethods methods,
and discard methods where exception types specified by the \EBTs do
not match the exception types in the \tss, leaving
\UseMacro{stat-num-methods-real-mega-test-data-with-exception}
methods. Next, we discard methods that do not pass the given \EBTs
even with their original \ERC implemented by the developers, to ensure
that the \task is achievable, reducing the set to
\UseMacro{stat-num-methods-real-mega-test-data-with-exception-with-project-with-gold}
methods. We then exclude methods that throw the correct exception even
after all \ERCs inside them are removed (\eg when the \CodeIn{catch}
block propagates the same exception originating from a function call
in the \CodeIn{try} block), as these cases do not require \task,
leaving
\UseMacro{stat-num-methods-real-mega-test-data-with-exception-with-project-with-gold-with-throw}
methods for our evaluation set.

\subsection{Throw Statement Removal}
\label{sec:dataset-throw-remove}

To automatically remove all \ERCs for each selected method, we
implemented the following rules.

When a \ts is
within a \CodeIn{catch} block, we remove the entire \CodeIn{catch}
block from the method. In cases where all \CodeIn{catch} blocks
associated with a \CodeIn{try} statement contain \tss, we remove the
entire \CodeIn{try/catch} construct and move all statements from the
\CodeIn{try} body to the outer scope. We took this approach
because the \CodeIn{try} block can be viewed as an \cond that
identifies the occurrence of exceptional behaviors. Conversely, when
one or more \CodeIn{catch} blocks do not contain \tss, we keep the
\CodeIn{try/catch} structure with only the \CodeIn{catch} blocks
without any \ts to preserve the original control flow structure.

For \CodeIn{if/else} statements, when a \CodeIn{then} block contains
only a \ts, we remove the entire \ifstate and move all statements from
the \CodeIn{else} block (if exists) to the outer scope, provided the
\CodeIn{else} block does not contain only a \ts. Conversely, when an
\CodeIn{else} block contains only a \ts, we remove the \CodeIn{else}
block and retain the \CodeIn{if} statement, provided the
\CodeIn{then} block does not contain only a \ts. If both blocks
contain only a \ts, we remove the entire \ifstate. This approach
also handles \CodeIn{else\CodeDash if} constructs, since they are
essentially \CodeIn{if} statements nested within \CodeIn{else}
blocks.

For \CodeIn{switch} statements, we remove all \CodeIn{case} blocks that contain \tss. If no \CodeIn{case} blocks remain after removal, we delete the entire \CodeIn{switch} construct. If only one \CodeIn{case} block remains, we move all statements from that case to the outer scope and remove the \CodeIn{switch} construct. If multiple \CodeIn{case} blocks remain, we keep the \CodeIn{switch} statement with the remaining cases.

Finally, we remove any \ts left in place by the rules above. This
can happen when a \ts appears alongside other statements in a
\CodeIn{then} or \CodeIn{else} block, the target method is defined just for throwing an
exception, or the \ts is defined at the end of the target method to
signal an exception when all other branches fail to return a value. In these cases, we believe it is most natural to simply remove
only the \ts.

Applying the rules above may leave a \tarmethod that no longer
compiles. \NumRmMissingReturnMethods of our \tarmethods fail with a
``missing return statement'' error, since the \ERC removal left one
execution branch without a \CodeIn{return} statement. The remaining
\NumRmUnreportedMethods also fail with an ``unreported exception''
error, since removing a \CodeIn{catch} block left the checked
exception raised in the corresponding \CodeIn{try} block neither
caught nor declared. We keep these data points since it is reasonable
to assume the user may intentionally provide such a \tarmethod to use
\Tool to add \tss to the method. In total, we have
\NumRmCompFailMethods (\RmCompFailPercentage) such data points.

\subsection{Tool-Generated Tests}
\label{sec:dataset-tool-tests}

To evaluate whether an \ERC generated from user-defined \EBTs is
semantically equivalent to the ground truth, we augment our evaluation
with additional \EBTs automatically generated by two popular Java test
generators: Randoop~\cite{PachecoETAL07Randoop} and EvoSuite~\cite{FraserAndArcuri11EvoSuite}. We generate tests per
project using each tool against the ground truth implementation. From the
generated tests, we keep only those that trigger an exception inside a
\tarmethod in our dataset and use the runtime stack trace to match
each such test to its \tarmethod. To bound the evaluation cost, we cap
the combined number of tool tests executed per \tarmethod at
\UseMacro{tool-test-cap}. Because EvoSuite generates significantly
fewer exception-triggering tests per \tarmethod than Randoop, Randoop
tests are counted first and EvoSuite tests fill the remaining budget.
In the end, our evaluation set has
\UseMacro{stat-tool-tests-avg-real-mega-test-data-with-exception-with-project-with-gold-with-throw}
tool-generated tests per \tarmethod on average.

\subsection{Dataset Statistics}

\begin{table}[t]
\begin{small}
\begin{center}
\caption{\UseMacro{TCap-dataset-throw-count}\label{tab:dataset-throw-count}}
\begin{tabular}{l |ccccc}
\toprule
Split & \UseMacro{THead-data-stats-catch-throw} & \UseMacro{THead-data-stats-if-throw} & \UseMacro{THead-data-stats-switch-throw} & \UseMacro{THead-data-stats-rest-throw} & \UseMacro{THead-data-stats-total-throw}
\\
\midrule
\UseMacro{THead-data-stats-all-data}
& \UseMacro{stat-catch-throw-all-data}
& \UseMacro{stat-if-throw-all-data}
& \UseMacro{stat-switch-throw-all-data}
& \UseMacro{stat-rest-throw-all-data}
& \UseMacro{stat-total-throw-all-data}
\\
\UseMacro{THead-data-stats-mega-val-data-with-exception-with-project-with-gold-with-throw}
& \UseMacro{stat-catch-throw-mega-val-data-with-exception-with-project-with-gold-with-throw}
& \UseMacro{stat-if-throw-mega-val-data-with-exception-with-project-with-gold-with-throw}
& \UseMacro{stat-switch-throw-mega-val-data-with-exception-with-project-with-gold-with-throw}
& \UseMacro{stat-rest-throw-mega-val-data-with-exception-with-project-with-gold-with-throw}
& \UseMacro{stat-total-throw-mega-val-data-with-exception-with-project-with-gold-with-throw}
\\
\UseMacro{THead-data-stats-real-mega-test-data-with-exception-with-project-with-gold-with-throw}
& \UseMacro{stat-catch-throw-real-mega-test-data-with-exception-with-project-with-gold-with-throw}
& \UseMacro{stat-if-throw-real-mega-test-data-with-exception-with-project-with-gold-with-throw}
& \UseMacro{stat-switch-throw-real-mega-test-data-with-exception-with-project-with-gold-with-throw}
& \UseMacro{stat-rest-throw-real-mega-test-data-with-exception-with-project-with-gold-with-throw}
& \UseMacro{stat-total-throw-real-mega-test-data-with-exception-with-project-with-gold-with-throw}
\\
\bottomrule
\end{tabular}
\end{center}
\end{small}
\end{table}

\label{sec:dataset-statistics}

We present the statistics of our collected dataset in
Table~\ref{tab:dataset-stats}. In our full dataset (including validation and evaluation sets), we collected
\UseMacro{stat-num-methods-all-data} methods and
\UseMacro{stat-etest-sum-all-data} \EBTs from
\UseMacro{stat-num-projects-all-data} eligible projects. The collected \EBTs
cover a range of \UseMacro{stat-exception-type-all-data} different
exception types.

Additionally, in Table~\ref{tab:dataset-throw-count}, we show the
number of each case of \ERC. Our full dataset includes
\UseMacro{stat-total-throw-all-data} throw statements in total. Among
them, \UseMacro{stat-catch-throw-all-data} \tss are in
\CodeIn{try\CodeDash catch} blocks, \UseMacro{stat-if-throw-all-data} \tss
are in
\CodeIn{if} statements (including both \CodeIn{then} blocks and
\CodeIn{else} blocks), \UseMacro{stat-switch-throw-all-data} \tss
are in \CodeIn{switch} statements, and \UseMacro{stat-rest-throw-all-data} are not inside any of the above 3 types of structures.

We randomly select half of the \java projects listed in CodeSearchNet to collect
data in the validation set (\UseMacro{THead-data-stats-val-no-switch-clean-v3}) and use
the remaining projects for evaluation set (\UseMacro{THead-data-stats-test-data}).
We use the validation set to guide our design decisions for \Tool, and the
evaluation set is used for evaluating the performance of \Tool and baselines.

\section{Evaluation Design}

\begin{table*}[t]
\begin{small}
\begin{center}
\caption{\UseMacro{TCap-res-model-comp-multi-ebt}\label{tab:model-comp-results}}
\setlength{\tabcolsep}{4.8pt}
\begin{tabular}{l | c |cccccccccccc}
\toprule
\multirow{2}{*}{\UseMacro{THead-model}}
& \multirow{2}{*}{\UseMacro{THead-prompt}}
& \multicolumn{3}{c}{\UseMacro{THead-compiled-at-k}}
& \multicolumn{3}{c}{\UseMacro{THead-ebts-pass-at-k}}
& \multicolumn{3}{c}{\UseMacro{THead-all-pass-at-k}}
& \multicolumn{3}{c}{\UseMacro{THead-allntools-pass-at-k}}
\\
&
& \UseMacro{THead-run-ebts-pass-at-k-compiled-at-1-small}
& \UseMacro{THead-run-ebts-pass-at-k-compiled-at-5-small}
& \UseMacro{THead-run-ebts-pass-at-k-compiled-at-10-small}
& \UseMacro{THead-run-ebts-pass-at-k-pass-at-1-small}
& \UseMacro{THead-run-ebts-pass-at-k-pass-at-5-small}
& \UseMacro{THead-run-ebts-pass-at-k-pass-at-10-small}
& \UseMacro{THead-run-all-pass-at-k-pass-at-1-small}
& \UseMacro{THead-run-all-pass-at-k-pass-at-5-small}
& \UseMacro{THead-run-all-pass-at-k-pass-at-10-small}
& \UseMacro{THead-run-all-with-tools-pass-at-k-pass-at-1-small}
& \UseMacro{THead-run-all-with-tools-pass-at-k-pass-at-5-small}
& \UseMacro{THead-run-all-with-tools-pass-at-k-pass-at-10-small}
\\
\midrule
\multirow{2}{*}{\UseMacro{THead-llama3.1:8b-instruct-q8-0}}
&\UseMacro{THead-base}
& \UseMacro{res-real-mega-test-data-with-exception-with-project-with-gold-with-throw-run-ebts-pass-at-k-compiled-at-1-llama_cpp-llama3.1:8b-instruct-q8_0-base-multi_ebt}
& \UseMacro{res-real-mega-test-data-with-exception-with-project-with-gold-with-throw-run-ebts-pass-at-k-compiled-at-5-llama_cpp-llama3.1:8b-instruct-q8_0-base-multi_ebt}
& \UseMacro{res-real-mega-test-data-with-exception-with-project-with-gold-with-throw-run-ebts-pass-at-k-compiled-at-10-llama_cpp-llama3.1:8b-instruct-q8_0-base-multi_ebt}
& \UseMacro{res-real-mega-test-data-with-exception-with-project-with-gold-with-throw-run-ebts-pass-at-k-pass-at-1-llama_cpp-llama3.1:8b-instruct-q8_0-base-multi_ebt}
& \UseMacro{res-real-mega-test-data-with-exception-with-project-with-gold-with-throw-run-ebts-pass-at-k-pass-at-5-llama_cpp-llama3.1:8b-instruct-q8_0-base-multi_ebt}
& \UseMacro{res-real-mega-test-data-with-exception-with-project-with-gold-with-throw-run-ebts-pass-at-k-pass-at-10-llama_cpp-llama3.1:8b-instruct-q8_0-base-multi_ebt}
& \UseMacro{res-real-mega-test-data-with-exception-with-project-with-gold-with-throw-run-all-pass-at-k-pass-at-1-llama_cpp-llama3.1:8b-instruct-q8_0-base-multi_ebt}
& \UseMacro{res-real-mega-test-data-with-exception-with-project-with-gold-with-throw-run-all-pass-at-k-pass-at-5-llama_cpp-llama3.1:8b-instruct-q8_0-base-multi_ebt}
& \UseMacro{res-real-mega-test-data-with-exception-with-project-with-gold-with-throw-run-all-pass-at-k-pass-at-10-llama_cpp-llama3.1:8b-instruct-q8_0-base-multi_ebt}
& \UseMacro{res-real-mega-test-data-with-exception-with-project-with-gold-with-throw-run-all-with-tools-pass-at-k-pass-at-1-llama_cpp-llama3.1:8b-instruct-q8_0-base-multi_ebt}
& \UseMacro{res-real-mega-test-data-with-exception-with-project-with-gold-with-throw-run-all-with-tools-pass-at-k-pass-at-5-llama_cpp-llama3.1:8b-instruct-q8_0-base-multi_ebt}
& \UseMacro{res-real-mega-test-data-with-exception-with-project-with-gold-with-throw-run-all-with-tools-pass-at-k-pass-at-10-llama_cpp-llama3.1:8b-instruct-q8_0-base-multi_ebt}
\\
&\UseMacro{THead-tuctn-all-info}
& \textbf{\UseMacro{res-real-mega-test-data-with-exception-with-project-with-gold-with-throw-run-ebts-pass-at-k-compiled-at-1-llama_cpp-llama3.1:8b-instruct-q8_0-tuctn-all-info-multi_ebt}}
& \textbf{\UseMacro{res-real-mega-test-data-with-exception-with-project-with-gold-with-throw-run-ebts-pass-at-k-compiled-at-5-llama_cpp-llama3.1:8b-instruct-q8_0-tuctn-all-info-multi_ebt}}
& \textbf{\UseMacro{res-real-mega-test-data-with-exception-with-project-with-gold-with-throw-run-ebts-pass-at-k-compiled-at-10-llama_cpp-llama3.1:8b-instruct-q8_0-tuctn-all-info-multi_ebt}}
& \textbf{\UseMacro{res-real-mega-test-data-with-exception-with-project-with-gold-with-throw-run-ebts-pass-at-k-pass-at-1-llama_cpp-llama3.1:8b-instruct-q8_0-tuctn-all-info-multi_ebt}}
& \textbf{\UseMacro{res-real-mega-test-data-with-exception-with-project-with-gold-with-throw-run-ebts-pass-at-k-pass-at-5-llama_cpp-llama3.1:8b-instruct-q8_0-tuctn-all-info-multi_ebt}}
& \textbf{\UseMacro{res-real-mega-test-data-with-exception-with-project-with-gold-with-throw-run-ebts-pass-at-k-pass-at-10-llama_cpp-llama3.1:8b-instruct-q8_0-tuctn-all-info-multi_ebt}}
& \textbf{\UseMacro{res-real-mega-test-data-with-exception-with-project-with-gold-with-throw-run-all-pass-at-k-pass-at-1-llama_cpp-llama3.1:8b-instruct-q8_0-tuctn-all-info-multi_ebt}}
& \textbf{\UseMacro{res-real-mega-test-data-with-exception-with-project-with-gold-with-throw-run-all-pass-at-k-pass-at-5-llama_cpp-llama3.1:8b-instruct-q8_0-tuctn-all-info-multi_ebt}}
& \textbf{\UseMacro{res-real-mega-test-data-with-exception-with-project-with-gold-with-throw-run-all-pass-at-k-pass-at-10-llama_cpp-llama3.1:8b-instruct-q8_0-tuctn-all-info-multi_ebt}}
& \textbf{\UseMacro{res-real-mega-test-data-with-exception-with-project-with-gold-with-throw-run-all-with-tools-pass-at-k-pass-at-1-llama_cpp-llama3.1:8b-instruct-q8_0-tuctn-all-info-multi_ebt}}
& \textbf{\UseMacro{res-real-mega-test-data-with-exception-with-project-with-gold-with-throw-run-all-with-tools-pass-at-k-pass-at-5-llama_cpp-llama3.1:8b-instruct-q8_0-tuctn-all-info-multi_ebt}}
& \textbf{\UseMacro{res-real-mega-test-data-with-exception-with-project-with-gold-with-throw-run-all-with-tools-pass-at-k-pass-at-10-llama_cpp-llama3.1:8b-instruct-q8_0-tuctn-all-info-multi_ebt}}
\\
\midrule
\multirow{2}{*}{\UseMacro{THead-phi4:14b-q8-0}}
&\UseMacro{THead-base}
& \UseMacro{res-real-mega-test-data-with-exception-with-project-with-gold-with-throw-run-ebts-pass-at-k-compiled-at-1-llama_cpp-phi4:14b-q8_0-base-multi_ebt}
& \UseMacro{res-real-mega-test-data-with-exception-with-project-with-gold-with-throw-run-ebts-pass-at-k-compiled-at-5-llama_cpp-phi4:14b-q8_0-base-multi_ebt}
& \UseMacro{res-real-mega-test-data-with-exception-with-project-with-gold-with-throw-run-ebts-pass-at-k-compiled-at-10-llama_cpp-phi4:14b-q8_0-base-multi_ebt}
& \UseMacro{res-real-mega-test-data-with-exception-with-project-with-gold-with-throw-run-ebts-pass-at-k-pass-at-1-llama_cpp-phi4:14b-q8_0-base-multi_ebt}
& \UseMacro{res-real-mega-test-data-with-exception-with-project-with-gold-with-throw-run-ebts-pass-at-k-pass-at-5-llama_cpp-phi4:14b-q8_0-base-multi_ebt}
& \UseMacro{res-real-mega-test-data-with-exception-with-project-with-gold-with-throw-run-ebts-pass-at-k-pass-at-10-llama_cpp-phi4:14b-q8_0-base-multi_ebt}
& \UseMacro{res-real-mega-test-data-with-exception-with-project-with-gold-with-throw-run-all-pass-at-k-pass-at-1-llama_cpp-phi4:14b-q8_0-base-multi_ebt}
& \UseMacro{res-real-mega-test-data-with-exception-with-project-with-gold-with-throw-run-all-pass-at-k-pass-at-5-llama_cpp-phi4:14b-q8_0-base-multi_ebt}
& \UseMacro{res-real-mega-test-data-with-exception-with-project-with-gold-with-throw-run-all-pass-at-k-pass-at-10-llama_cpp-phi4:14b-q8_0-base-multi_ebt}
& \UseMacro{res-real-mega-test-data-with-exception-with-project-with-gold-with-throw-run-all-with-tools-pass-at-k-pass-at-1-llama_cpp-phi4:14b-q8_0-base-multi_ebt}
& \UseMacro{res-real-mega-test-data-with-exception-with-project-with-gold-with-throw-run-all-with-tools-pass-at-k-pass-at-5-llama_cpp-phi4:14b-q8_0-base-multi_ebt}
& \UseMacro{res-real-mega-test-data-with-exception-with-project-with-gold-with-throw-run-all-with-tools-pass-at-k-pass-at-10-llama_cpp-phi4:14b-q8_0-base-multi_ebt}
\\
&\UseMacro{THead-tuctn-all-info}
& \textbf{\UseMacro{res-real-mega-test-data-with-exception-with-project-with-gold-with-throw-run-ebts-pass-at-k-compiled-at-1-llama_cpp-phi4:14b-q8_0-tuctn-all-info-multi_ebt}}
& \textbf{\UseMacro{res-real-mega-test-data-with-exception-with-project-with-gold-with-throw-run-ebts-pass-at-k-compiled-at-5-llama_cpp-phi4:14b-q8_0-tuctn-all-info-multi_ebt}}
& \textbf{\UseMacro{res-real-mega-test-data-with-exception-with-project-with-gold-with-throw-run-ebts-pass-at-k-compiled-at-10-llama_cpp-phi4:14b-q8_0-tuctn-all-info-multi_ebt}}
& \textbf{\UseMacro{res-real-mega-test-data-with-exception-with-project-with-gold-with-throw-run-ebts-pass-at-k-pass-at-1-llama_cpp-phi4:14b-q8_0-tuctn-all-info-multi_ebt}}
& \textbf{\UseMacro{res-real-mega-test-data-with-exception-with-project-with-gold-with-throw-run-ebts-pass-at-k-pass-at-5-llama_cpp-phi4:14b-q8_0-tuctn-all-info-multi_ebt}}
& \textbf{\UseMacro{res-real-mega-test-data-with-exception-with-project-with-gold-with-throw-run-ebts-pass-at-k-pass-at-10-llama_cpp-phi4:14b-q8_0-tuctn-all-info-multi_ebt}}
& \textbf{\UseMacro{res-real-mega-test-data-with-exception-with-project-with-gold-with-throw-run-all-pass-at-k-pass-at-1-llama_cpp-phi4:14b-q8_0-tuctn-all-info-multi_ebt}}
& \textbf{\UseMacro{res-real-mega-test-data-with-exception-with-project-with-gold-with-throw-run-all-pass-at-k-pass-at-5-llama_cpp-phi4:14b-q8_0-tuctn-all-info-multi_ebt}}
& \textbf{\UseMacro{res-real-mega-test-data-with-exception-with-project-with-gold-with-throw-run-all-pass-at-k-pass-at-10-llama_cpp-phi4:14b-q8_0-tuctn-all-info-multi_ebt}}
& \textbf{\UseMacro{res-real-mega-test-data-with-exception-with-project-with-gold-with-throw-run-all-with-tools-pass-at-k-pass-at-1-llama_cpp-phi4:14b-q8_0-tuctn-all-info-multi_ebt}}
& \textbf{\UseMacro{res-real-mega-test-data-with-exception-with-project-with-gold-with-throw-run-all-with-tools-pass-at-k-pass-at-5-llama_cpp-phi4:14b-q8_0-tuctn-all-info-multi_ebt}}
& \textbf{\UseMacro{res-real-mega-test-data-with-exception-with-project-with-gold-with-throw-run-all-with-tools-pass-at-k-pass-at-10-llama_cpp-phi4:14b-q8_0-tuctn-all-info-multi_ebt}}
\\
\midrule
\multirow{2}{*}{\UseMacro{THead-qwen2.5-coder:7b-instruct-q8-0}}
&\UseMacro{THead-base}
& \UseMacro{res-real-mega-test-data-with-exception-with-project-with-gold-with-throw-run-ebts-pass-at-k-compiled-at-1-llama_cpp-qwen2.5-coder:7b-instruct-q8_0-base-multi_ebt}
& \UseMacro{res-real-mega-test-data-with-exception-with-project-with-gold-with-throw-run-ebts-pass-at-k-compiled-at-5-llama_cpp-qwen2.5-coder:7b-instruct-q8_0-base-multi_ebt}
& \UseMacro{res-real-mega-test-data-with-exception-with-project-with-gold-with-throw-run-ebts-pass-at-k-compiled-at-10-llama_cpp-qwen2.5-coder:7b-instruct-q8_0-base-multi_ebt}
& \UseMacro{res-real-mega-test-data-with-exception-with-project-with-gold-with-throw-run-ebts-pass-at-k-pass-at-1-llama_cpp-qwen2.5-coder:7b-instruct-q8_0-base-multi_ebt}
& \UseMacro{res-real-mega-test-data-with-exception-with-project-with-gold-with-throw-run-ebts-pass-at-k-pass-at-5-llama_cpp-qwen2.5-coder:7b-instruct-q8_0-base-multi_ebt}
& \UseMacro{res-real-mega-test-data-with-exception-with-project-with-gold-with-throw-run-ebts-pass-at-k-pass-at-10-llama_cpp-qwen2.5-coder:7b-instruct-q8_0-base-multi_ebt}
& \UseMacro{res-real-mega-test-data-with-exception-with-project-with-gold-with-throw-run-all-pass-at-k-pass-at-1-llama_cpp-qwen2.5-coder:7b-instruct-q8_0-base-multi_ebt}
& \UseMacro{res-real-mega-test-data-with-exception-with-project-with-gold-with-throw-run-all-pass-at-k-pass-at-5-llama_cpp-qwen2.5-coder:7b-instruct-q8_0-base-multi_ebt}
& \UseMacro{res-real-mega-test-data-with-exception-with-project-with-gold-with-throw-run-all-pass-at-k-pass-at-10-llama_cpp-qwen2.5-coder:7b-instruct-q8_0-base-multi_ebt}
& \UseMacro{res-real-mega-test-data-with-exception-with-project-with-gold-with-throw-run-all-with-tools-pass-at-k-pass-at-1-llama_cpp-qwen2.5-coder:7b-instruct-q8_0-base-multi_ebt}
& \UseMacro{res-real-mega-test-data-with-exception-with-project-with-gold-with-throw-run-all-with-tools-pass-at-k-pass-at-5-llama_cpp-qwen2.5-coder:7b-instruct-q8_0-base-multi_ebt}
& \UseMacro{res-real-mega-test-data-with-exception-with-project-with-gold-with-throw-run-all-with-tools-pass-at-k-pass-at-10-llama_cpp-qwen2.5-coder:7b-instruct-q8_0-base-multi_ebt}
\\
&\UseMacro{THead-tuctn-all-info}
& \textbf{\UseMacro{res-real-mega-test-data-with-exception-with-project-with-gold-with-throw-run-ebts-pass-at-k-compiled-at-1-llama_cpp-qwen2.5-coder:7b-instruct-q8_0-tuctn-all-info-multi_ebt}}
& \textbf{\UseMacro{res-real-mega-test-data-with-exception-with-project-with-gold-with-throw-run-ebts-pass-at-k-compiled-at-5-llama_cpp-qwen2.5-coder:7b-instruct-q8_0-tuctn-all-info-multi_ebt}}
& \textbf{\UseMacro{res-real-mega-test-data-with-exception-with-project-with-gold-with-throw-run-ebts-pass-at-k-compiled-at-10-llama_cpp-qwen2.5-coder:7b-instruct-q8_0-tuctn-all-info-multi_ebt}}
& \textbf{\UseMacro{res-real-mega-test-data-with-exception-with-project-with-gold-with-throw-run-ebts-pass-at-k-pass-at-1-llama_cpp-qwen2.5-coder:7b-instruct-q8_0-tuctn-all-info-multi_ebt}}
& \textbf{\UseMacro{res-real-mega-test-data-with-exception-with-project-with-gold-with-throw-run-ebts-pass-at-k-pass-at-5-llama_cpp-qwen2.5-coder:7b-instruct-q8_0-tuctn-all-info-multi_ebt}}
& \textbf{\UseMacro{res-real-mega-test-data-with-exception-with-project-with-gold-with-throw-run-ebts-pass-at-k-pass-at-10-llama_cpp-qwen2.5-coder:7b-instruct-q8_0-tuctn-all-info-multi_ebt}}
& \textbf{\UseMacro{res-real-mega-test-data-with-exception-with-project-with-gold-with-throw-run-all-pass-at-k-pass-at-1-llama_cpp-qwen2.5-coder:7b-instruct-q8_0-tuctn-all-info-multi_ebt}}
& \textbf{\UseMacro{res-real-mega-test-data-with-exception-with-project-with-gold-with-throw-run-all-pass-at-k-pass-at-5-llama_cpp-qwen2.5-coder:7b-instruct-q8_0-tuctn-all-info-multi_ebt}}
& \textbf{\UseMacro{res-real-mega-test-data-with-exception-with-project-with-gold-with-throw-run-all-pass-at-k-pass-at-10-llama_cpp-qwen2.5-coder:7b-instruct-q8_0-tuctn-all-info-multi_ebt}}
& \textbf{\UseMacro{res-real-mega-test-data-with-exception-with-project-with-gold-with-throw-run-all-with-tools-pass-at-k-pass-at-1-llama_cpp-qwen2.5-coder:7b-instruct-q8_0-tuctn-all-info-multi_ebt}}
& \textbf{\UseMacro{res-real-mega-test-data-with-exception-with-project-with-gold-with-throw-run-all-with-tools-pass-at-k-pass-at-5-llama_cpp-qwen2.5-coder:7b-instruct-q8_0-tuctn-all-info-multi_ebt}}
& \textbf{\UseMacro{res-real-mega-test-data-with-exception-with-project-with-gold-with-throw-run-all-with-tools-pass-at-k-pass-at-10-llama_cpp-qwen2.5-coder:7b-instruct-q8_0-tuctn-all-info-multi_ebt}}
\\
\midrule
\multirow{2}{*}{\UseMacro{THead-qwen2.5-coder:32b-instruct-q8-0}}
&\UseMacro{THead-base}
& \UseMacro{res-real-mega-test-data-with-exception-with-project-with-gold-with-throw-run-ebts-pass-at-k-compiled-at-1-llama_cpp-qwen2.5-coder:32b-instruct-q8_0-base-multi_ebt}
& \UseMacro{res-real-mega-test-data-with-exception-with-project-with-gold-with-throw-run-ebts-pass-at-k-compiled-at-5-llama_cpp-qwen2.5-coder:32b-instruct-q8_0-base-multi_ebt}
& \UseMacro{res-real-mega-test-data-with-exception-with-project-with-gold-with-throw-run-ebts-pass-at-k-compiled-at-10-llama_cpp-qwen2.5-coder:32b-instruct-q8_0-base-multi_ebt}
& \UseMacro{res-real-mega-test-data-with-exception-with-project-with-gold-with-throw-run-ebts-pass-at-k-pass-at-1-llama_cpp-qwen2.5-coder:32b-instruct-q8_0-base-multi_ebt}
& \UseMacro{res-real-mega-test-data-with-exception-with-project-with-gold-with-throw-run-ebts-pass-at-k-pass-at-5-llama_cpp-qwen2.5-coder:32b-instruct-q8_0-base-multi_ebt}
& \UseMacro{res-real-mega-test-data-with-exception-with-project-with-gold-with-throw-run-ebts-pass-at-k-pass-at-10-llama_cpp-qwen2.5-coder:32b-instruct-q8_0-base-multi_ebt}
& \UseMacro{res-real-mega-test-data-with-exception-with-project-with-gold-with-throw-run-all-pass-at-k-pass-at-1-llama_cpp-qwen2.5-coder:32b-instruct-q8_0-base-multi_ebt}
& \UseMacro{res-real-mega-test-data-with-exception-with-project-with-gold-with-throw-run-all-pass-at-k-pass-at-5-llama_cpp-qwen2.5-coder:32b-instruct-q8_0-base-multi_ebt}
& \UseMacro{res-real-mega-test-data-with-exception-with-project-with-gold-with-throw-run-all-pass-at-k-pass-at-10-llama_cpp-qwen2.5-coder:32b-instruct-q8_0-base-multi_ebt}
& \UseMacro{res-real-mega-test-data-with-exception-with-project-with-gold-with-throw-run-all-with-tools-pass-at-k-pass-at-1-llama_cpp-qwen2.5-coder:32b-instruct-q8_0-base-multi_ebt}
& \UseMacro{res-real-mega-test-data-with-exception-with-project-with-gold-with-throw-run-all-with-tools-pass-at-k-pass-at-5-llama_cpp-qwen2.5-coder:32b-instruct-q8_0-base-multi_ebt}
& \UseMacro{res-real-mega-test-data-with-exception-with-project-with-gold-with-throw-run-all-with-tools-pass-at-k-pass-at-10-llama_cpp-qwen2.5-coder:32b-instruct-q8_0-base-multi_ebt}
\\
&\UseMacro{THead-tuctn-all-info}
& \textbf{\UseMacro{res-real-mega-test-data-with-exception-with-project-with-gold-with-throw-run-ebts-pass-at-k-compiled-at-1-llama_cpp-qwen2.5-coder:32b-instruct-q8_0-tuctn-all-info-multi_ebt}}
& \textbf{\UseMacro{res-real-mega-test-data-with-exception-with-project-with-gold-with-throw-run-ebts-pass-at-k-compiled-at-5-llama_cpp-qwen2.5-coder:32b-instruct-q8_0-tuctn-all-info-multi_ebt}}
& \textbf{\UseMacro{res-real-mega-test-data-with-exception-with-project-with-gold-with-throw-run-ebts-pass-at-k-compiled-at-10-llama_cpp-qwen2.5-coder:32b-instruct-q8_0-tuctn-all-info-multi_ebt}}
& \textbf{\UseMacro{res-real-mega-test-data-with-exception-with-project-with-gold-with-throw-run-ebts-pass-at-k-pass-at-1-llama_cpp-qwen2.5-coder:32b-instruct-q8_0-tuctn-all-info-multi_ebt}}
& \textbf{\UseMacro{res-real-mega-test-data-with-exception-with-project-with-gold-with-throw-run-ebts-pass-at-k-pass-at-5-llama_cpp-qwen2.5-coder:32b-instruct-q8_0-tuctn-all-info-multi_ebt}}
& \textbf{\UseMacro{res-real-mega-test-data-with-exception-with-project-with-gold-with-throw-run-ebts-pass-at-k-pass-at-10-llama_cpp-qwen2.5-coder:32b-instruct-q8_0-tuctn-all-info-multi_ebt}}
& \textbf{\UseMacro{res-real-mega-test-data-with-exception-with-project-with-gold-with-throw-run-all-pass-at-k-pass-at-1-llama_cpp-qwen2.5-coder:32b-instruct-q8_0-tuctn-all-info-multi_ebt}}
& \textbf{\UseMacro{res-real-mega-test-data-with-exception-with-project-with-gold-with-throw-run-all-pass-at-k-pass-at-5-llama_cpp-qwen2.5-coder:32b-instruct-q8_0-tuctn-all-info-multi_ebt}}
& \textbf{\UseMacro{res-real-mega-test-data-with-exception-with-project-with-gold-with-throw-run-all-pass-at-k-pass-at-10-llama_cpp-qwen2.5-coder:32b-instruct-q8_0-tuctn-all-info-multi_ebt}}
& \textbf{\UseMacro{res-real-mega-test-data-with-exception-with-project-with-gold-with-throw-run-all-with-tools-pass-at-k-pass-at-1-llama_cpp-qwen2.5-coder:32b-instruct-q8_0-tuctn-all-info-multi_ebt}}
& \textbf{\UseMacro{res-real-mega-test-data-with-exception-with-project-with-gold-with-throw-run-all-with-tools-pass-at-k-pass-at-5-llama_cpp-qwen2.5-coder:32b-instruct-q8_0-tuctn-all-info-multi_ebt}}
& \textbf{\UseMacro{res-real-mega-test-data-with-exception-with-project-with-gold-with-throw-run-all-with-tools-pass-at-k-pass-at-10-llama_cpp-qwen2.5-coder:32b-instruct-q8_0-tuctn-all-info-multi_ebt}}
\\
\midrule
\multirow{2}{*}{\UseMacro{THead-gpt-5-mini}}
&\UseMacro{THead-base}
& \UseMacro{res-real-mega-test-data-with-exception-with-project-with-gold-with-throw-run-ebts-pass-at-k-compiled-at-1-azure-gpt-5-mini-base-multi_ebt}
& \UseMacro{res-real-mega-test-data-with-exception-with-project-with-gold-with-throw-run-ebts-pass-at-k-compiled-at-5-azure-gpt-5-mini-base-multi_ebt}
& \UseMacro{res-real-mega-test-data-with-exception-with-project-with-gold-with-throw-run-ebts-pass-at-k-compiled-at-10-azure-gpt-5-mini-base-multi_ebt}
& \UseMacro{res-real-mega-test-data-with-exception-with-project-with-gold-with-throw-run-ebts-pass-at-k-pass-at-1-azure-gpt-5-mini-base-multi_ebt}
& \UseMacro{res-real-mega-test-data-with-exception-with-project-with-gold-with-throw-run-ebts-pass-at-k-pass-at-5-azure-gpt-5-mini-base-multi_ebt}
& \UseMacro{res-real-mega-test-data-with-exception-with-project-with-gold-with-throw-run-ebts-pass-at-k-pass-at-10-azure-gpt-5-mini-base-multi_ebt}
& \UseMacro{res-real-mega-test-data-with-exception-with-project-with-gold-with-throw-run-all-pass-at-k-pass-at-1-azure-gpt-5-mini-base-multi_ebt}
& \UseMacro{res-real-mega-test-data-with-exception-with-project-with-gold-with-throw-run-all-pass-at-k-pass-at-5-azure-gpt-5-mini-base-multi_ebt}
& \UseMacro{res-real-mega-test-data-with-exception-with-project-with-gold-with-throw-run-all-pass-at-k-pass-at-10-azure-gpt-5-mini-base-multi_ebt}
& \UseMacro{res-real-mega-test-data-with-exception-with-project-with-gold-with-throw-run-all-with-tools-pass-at-k-pass-at-1-azure-gpt-5-mini-base-multi_ebt}
& \UseMacro{res-real-mega-test-data-with-exception-with-project-with-gold-with-throw-run-all-with-tools-pass-at-k-pass-at-5-azure-gpt-5-mini-base-multi_ebt}
& \UseMacro{res-real-mega-test-data-with-exception-with-project-with-gold-with-throw-run-all-with-tools-pass-at-k-pass-at-10-azure-gpt-5-mini-base-multi_ebt}
\\
&\UseMacro{THead-tuctn-all-info}
& \textbf{\UseMacro{res-real-mega-test-data-with-exception-with-project-with-gold-with-throw-run-ebts-pass-at-k-compiled-at-1-azure-gpt-5-mini-tuctn-all-info-multi_ebt}}
& \textbf{\UseMacro{res-real-mega-test-data-with-exception-with-project-with-gold-with-throw-run-ebts-pass-at-k-compiled-at-5-azure-gpt-5-mini-tuctn-all-info-multi_ebt}}
& \textbf{\UseMacro{res-real-mega-test-data-with-exception-with-project-with-gold-with-throw-run-ebts-pass-at-k-compiled-at-10-azure-gpt-5-mini-tuctn-all-info-multi_ebt}}
& \textbf{\UseMacro{res-real-mega-test-data-with-exception-with-project-with-gold-with-throw-run-ebts-pass-at-k-pass-at-1-azure-gpt-5-mini-tuctn-all-info-multi_ebt}}
& \textbf{\UseMacro{res-real-mega-test-data-with-exception-with-project-with-gold-with-throw-run-ebts-pass-at-k-pass-at-5-azure-gpt-5-mini-tuctn-all-info-multi_ebt}}
& \textbf{\UseMacro{res-real-mega-test-data-with-exception-with-project-with-gold-with-throw-run-ebts-pass-at-k-pass-at-10-azure-gpt-5-mini-tuctn-all-info-multi_ebt}}
& \textbf{\UseMacro{res-real-mega-test-data-with-exception-with-project-with-gold-with-throw-run-all-pass-at-k-pass-at-1-azure-gpt-5-mini-tuctn-all-info-multi_ebt}}
& \textbf{\UseMacro{res-real-mega-test-data-with-exception-with-project-with-gold-with-throw-run-all-pass-at-k-pass-at-5-azure-gpt-5-mini-tuctn-all-info-multi_ebt}}
& \textbf{\UseMacro{res-real-mega-test-data-with-exception-with-project-with-gold-with-throw-run-all-pass-at-k-pass-at-10-azure-gpt-5-mini-tuctn-all-info-multi_ebt}}
& \textbf{\UseMacro{res-real-mega-test-data-with-exception-with-project-with-gold-with-throw-run-all-with-tools-pass-at-k-pass-at-1-azure-gpt-5-mini-tuctn-all-info-multi_ebt}}
& \textbf{\UseMacro{res-real-mega-test-data-with-exception-with-project-with-gold-with-throw-run-all-with-tools-pass-at-k-pass-at-5-azure-gpt-5-mini-tuctn-all-info-multi_ebt}}
& \textbf{\UseMacro{res-real-mega-test-data-with-exception-with-project-with-gold-with-throw-run-all-with-tools-pass-at-k-pass-at-10-azure-gpt-5-mini-tuctn-all-info-multi_ebt}}
\\
\bottomrule
\end{tabular}
\end{center}
\end{small}
\end{table*}

We assess the effectiveness of \Tool by answering the following research questions:

\DefMacro{rq-modelComp}{RQ1\xspace}
\DefMacro{rq-promptComp}{RQ2\xspace}
\DefMacro{rq-lenComp}{RQ3\xspace}
\DefMacro{rq-iter}{RQ4\xspace}

\MyParaOnly{\UseMacro{rq-modelComp}}: How does \Tool help \LLMs of different architectures and parameter scales perform the \task task?

\MyParaOnly{\UseMacro{rq-promptComp}}: How much does each of our prompt components help \Tool generate \ERCs?

\MyParaOnly{\UseMacro{rq-lenComp}}: How does the complexity of the target method influence \Tool's performance?

\MyParaOnly{\UseMacro{rq-iter}}: How does \Tool perform when combined with \LLM self-repair?

\subsection{Baselines}
\label{sec:baselines}

We evaluate \Tool against a baseline approach (\Base) that provides an
\LLM with only the target method that requires \task and the
corresponding \EBTs that define the expected exceptional behavior,
excluding all context collected by our static and dynamic analysis,
similar to prior TDD-based code generation work~\cite{mathewsTestDrivenDevelopmentCode2024}. Since \task is a
novel task, no existing technique directly applies, which makes it
hard to establish a comparable baseline. We therefore additionally
treat the ablation variants in \UseMacro{rq-promptComp}, where the
models are provided with an additional source of context, as
baselines. Likewise, Iter-\Base in \UseMacro{rq-iter}, which augments
\Base with \LLM self-repair driven by compiler and test feedback,
serves as a self-repair baseline.

\subsection{Evaluation Metrics}
\label{sec:eval-metrics}

Following prior work~\cite{zhangExLongGeneratingExceptional2024, NieETAL23TeCo, RaoETAL23CAT}, we use four @$k$-based~\cite{chenEvaluatingLargeLanguage2021} metrics, each of which estimates the probability that at least one out of $k$ generated samples satisfies a given success criterion. For each task, @$k$ is evaluated by generating $n \geq k$ samples per task (in this paper we use $n=10$ and $k \leq 10$), and using the following unbiased estimator to evaluate:
$$ \text{@}k := \mathbb{E}_{\text{Problems}}\left[1 - \frac{\binom{n-\sum_{s\in \text{samples}}c(s)}{k}}{\binom{n}{k}}\right]$$
Here, $c(s)$ is a binary indicator function, returning $1$ if sample $s$ satisfies a predefined success criterion and $0$ otherwise. The definition of $c(s)$ varies for each specific metric and is detailed below. We report the average metrics over all tasks in the evaluation dataset.

\MyParaOnly{\UseMacro{THead-compiled-at-k}:} This metric estimates the probability that at least one out of $k$ generated samples can be compiled without errors. Here, $c(s) = 1$ if sample $s$ successfully compiles.

\MyParaOnly{\UseMacro{THead-ebts-pass-at-k}:} This metric checks whether the \ERC completes the task given by the user, by estimating the probability that at least one out of $k$ generated samples passes all associated Exceptional Behavior Tests (\EBTs). Here, $c(s) = 1$ if sample $s$ passes all \EBTs.

\MyParaOnly{\UseMacro{THead-all-pass-at-k}:} This metric checks whether the \ERC completes the task given by the user without breaking the original functionality of the \tarmethod, by estimating the probability that at least one out of $k$ generated samples passes all user-defined tests on the \tarmethod, covering both Exceptional Behavior Tests (\EBTs) and Non-Exceptional Behavior Tests (\NEBTs). Accordingly, $c(s) = 1$ if sample $s$ passes both \EBTs and \nEBTs.

\MyParaOnly{\UseMacro{THead-allntools-pass-at-k}:} This metric adds behavioral checks beyond the user-written tests, which can underspecify true developer intent, by extending \UseMacro{THead-all-pass-at-k} with tests automatically generated by EvoSuite~\cite{FraserAndArcuri11EvoSuite} and Randoop~\cite{PachecoETAL07Randoop} on the \tarmethod. Accordingly, $c(s) = 1$ if sample $s$ passes all \EBTs and \nEBTs, as well as \EBTs generated by EvoSuite and Randoop. These generated tests are additional test oracles, not ground truth for developer intent (Section~\ref{sec:limitations}).

\subsection{Implementation Details}
\label{sec:imple-details}

For \UseMacro{rq-modelComp} we evaluate \Tool on 4 different
open-source models (\UseMacro{THead-llama3.1:8b-instruct-q8-0}~\cite{grattafioriLlama3Herd2024},
\UseMacro{THead-phi4:14b-q8-0}~\cite{abdinPhi4TechnicalReport2024},
\UseMacro{THead-qwen2.5-coder:7b-instruct-q8-0}~\cite{huiQwen25CoderTechnicalReport2024}, and
\UseMacro{THead-qwen2.5-coder:32b-instruct-q8-0}) and one
closed-source model (\UseMacro{THead-gpt-5-mini}~\cite{OpenaiGPT52025}). We selected
these models based on three criteria: (1)~\textbf{parameter size},
ranging from 7b to 32b parameters to assess how model capacity affects
performance; (2)~\textbf{training objective}, covering both
coding-specialized (\UseMacro{THead-qwen2.5-coder:7b-instruct-q8-0}
and \UseMacro{THead-qwen2.5-coder:32b-instruct-q8-0}) and
general-purpose \LLMs (\UseMacro{THead-llama3.1:8b-instruct-q8-0},
\UseMacro{THead-phi4:14b-q8-0}); and (3)~\textbf{community adoption},
selecting among the most widely used models available at the time of
our study. For cost-efficient inference, we run an 8-bit quantized
version of each open-source model using the ``Q8\_0'' quantization method
implemented in the llama.cpp library~\cite{GgmlorgLlamacpp2025}.

\subsection{Hardware}

We run all evaluations for \Tool and baseline on a machine with Intel
Xeon w5-3433 @ 4.2 GHz (16 cores, 32 threads) CPU, 130 GB of RAM, 2
NVIDIA RTX 5000 Ada Generation GPUs, Ubuntu 24.04, Python 3.11, Java
8, and Maven 3.8.6.

\section{Results}
\label{sec:result}

\begin{table*}[t]
\begin{small}
\begin{center}
\caption{\UseMacro{TCap-res-prompt-comp-multi-ebt}\label{tab:prompt-comp-results}}
\setlength{\tabcolsep}{8.1pt}
\begin{tabular}{l |cccccccccccc}
\toprule
\multirow{2}{*}{\UseMacro{THead-prompt}}
& \multicolumn{3}{c}{\UseMacro{THead-compiled-at-k}}
& \multicolumn{3}{c}{\UseMacro{THead-ebts-pass-at-k}}
& \multicolumn{3}{c}{\UseMacro{THead-all-pass-at-k}}
& \multicolumn{3}{c}{\UseMacro{THead-allntools-pass-at-k}}
\\
& \UseMacro{THead-run-ebts-pass-at-k-compiled-at-1-small}
& \UseMacro{THead-run-ebts-pass-at-k-compiled-at-5-small}
& \UseMacro{THead-run-ebts-pass-at-k-compiled-at-10-small}
& \UseMacro{THead-run-ebts-pass-at-k-pass-at-1-small}
& \UseMacro{THead-run-ebts-pass-at-k-pass-at-5-small}
& \UseMacro{THead-run-ebts-pass-at-k-pass-at-10-small}
& \UseMacro{THead-run-all-pass-at-k-pass-at-1-small}
& \UseMacro{THead-run-all-pass-at-k-pass-at-5-small}
& \UseMacro{THead-run-all-pass-at-k-pass-at-10-small}
& \UseMacro{THead-run-all-with-tools-pass-at-k-pass-at-1-small}
& \UseMacro{THead-run-all-with-tools-pass-at-k-pass-at-5-small}
& \UseMacro{THead-run-all-with-tools-pass-at-k-pass-at-10-small}
\\
\midrule
\UseMacro{THead-tuctn-all-info}
& \textbf{\UseMacro{res-real-mega-test-data-with-exception-with-project-with-gold-with-throw-run-ebts-pass-at-k-compiled-at-1-llama_cpp-qwen2.5-coder:32b-instruct-q8_0-tuctn-all-info-multi_ebt}}
& \textbf{\UseMacro{res-real-mega-test-data-with-exception-with-project-with-gold-with-throw-run-ebts-pass-at-k-compiled-at-5-llama_cpp-qwen2.5-coder:32b-instruct-q8_0-tuctn-all-info-multi_ebt}}
& \textbf{\UseMacro{res-real-mega-test-data-with-exception-with-project-with-gold-with-throw-run-ebts-pass-at-k-compiled-at-10-llama_cpp-qwen2.5-coder:32b-instruct-q8_0-tuctn-all-info-multi_ebt}}
& \textbf{\UseMacro{res-real-mega-test-data-with-exception-with-project-with-gold-with-throw-run-ebts-pass-at-k-pass-at-1-llama_cpp-qwen2.5-coder:32b-instruct-q8_0-tuctn-all-info-multi_ebt}}
& \textbf{\UseMacro{res-real-mega-test-data-with-exception-with-project-with-gold-with-throw-run-ebts-pass-at-k-pass-at-5-llama_cpp-qwen2.5-coder:32b-instruct-q8_0-tuctn-all-info-multi_ebt}}
& \textbf{\UseMacro{res-real-mega-test-data-with-exception-with-project-with-gold-with-throw-run-ebts-pass-at-k-pass-at-10-llama_cpp-qwen2.5-coder:32b-instruct-q8_0-tuctn-all-info-multi_ebt}}
& \textbf{\UseMacro{res-real-mega-test-data-with-exception-with-project-with-gold-with-throw-run-all-pass-at-k-pass-at-1-llama_cpp-qwen2.5-coder:32b-instruct-q8_0-tuctn-all-info-multi_ebt}}
& \textbf{\UseMacro{res-real-mega-test-data-with-exception-with-project-with-gold-with-throw-run-all-pass-at-k-pass-at-5-llama_cpp-qwen2.5-coder:32b-instruct-q8_0-tuctn-all-info-multi_ebt}}
& \textbf{\UseMacro{res-real-mega-test-data-with-exception-with-project-with-gold-with-throw-run-all-pass-at-k-pass-at-10-llama_cpp-qwen2.5-coder:32b-instruct-q8_0-tuctn-all-info-multi_ebt}}
& \textbf{\UseMacro{res-real-mega-test-data-with-exception-with-project-with-gold-with-throw-run-all-with-tools-pass-at-k-pass-at-1-llama_cpp-qwen2.5-coder:32b-instruct-q8_0-tuctn-all-info-multi_ebt}}
& \textbf{\UseMacro{res-real-mega-test-data-with-exception-with-project-with-gold-with-throw-run-all-with-tools-pass-at-k-pass-at-5-llama_cpp-qwen2.5-coder:32b-instruct-q8_0-tuctn-all-info-multi_ebt}}
& \textbf{\UseMacro{res-real-mega-test-data-with-exception-with-project-with-gold-with-throw-run-all-with-tools-pass-at-k-pass-at-10-llama_cpp-qwen2.5-coder:32b-instruct-q8_0-tuctn-all-info-multi_ebt}}
\\
\midrule
\UseMacro{THead-only-avsym}
& \UseMacro{res-real-mega-test-data-with-exception-with-project-with-gold-with-throw-run-ebts-pass-at-k-compiled-at-1-llama_cpp-qwen2.5-coder:32b-instruct-q8_0-only-avsym-multi_ebt}
& \UseMacro{res-real-mega-test-data-with-exception-with-project-with-gold-with-throw-run-ebts-pass-at-k-compiled-at-5-llama_cpp-qwen2.5-coder:32b-instruct-q8_0-only-avsym-multi_ebt}
& \UseMacro{res-real-mega-test-data-with-exception-with-project-with-gold-with-throw-run-ebts-pass-at-k-compiled-at-10-llama_cpp-qwen2.5-coder:32b-instruct-q8_0-only-avsym-multi_ebt}
& \UseMacro{res-real-mega-test-data-with-exception-with-project-with-gold-with-throw-run-ebts-pass-at-k-pass-at-1-llama_cpp-qwen2.5-coder:32b-instruct-q8_0-only-avsym-multi_ebt}
& \UseMacro{res-real-mega-test-data-with-exception-with-project-with-gold-with-throw-run-ebts-pass-at-k-pass-at-5-llama_cpp-qwen2.5-coder:32b-instruct-q8_0-only-avsym-multi_ebt}
& \UseMacro{res-real-mega-test-data-with-exception-with-project-with-gold-with-throw-run-ebts-pass-at-k-pass-at-10-llama_cpp-qwen2.5-coder:32b-instruct-q8_0-only-avsym-multi_ebt}
& \UseMacro{res-real-mega-test-data-with-exception-with-project-with-gold-with-throw-run-all-pass-at-k-pass-at-1-llama_cpp-qwen2.5-coder:32b-instruct-q8_0-only-avsym-multi_ebt}
& \UseMacro{res-real-mega-test-data-with-exception-with-project-with-gold-with-throw-run-all-pass-at-k-pass-at-5-llama_cpp-qwen2.5-coder:32b-instruct-q8_0-only-avsym-multi_ebt}
& \UseMacro{res-real-mega-test-data-with-exception-with-project-with-gold-with-throw-run-all-pass-at-k-pass-at-10-llama_cpp-qwen2.5-coder:32b-instruct-q8_0-only-avsym-multi_ebt}
& \UseMacro{res-real-mega-test-data-with-exception-with-project-with-gold-with-throw-run-all-with-tools-pass-at-k-pass-at-1-llama_cpp-qwen2.5-coder:32b-instruct-q8_0-only-avsym-multi_ebt}
& \UseMacro{res-real-mega-test-data-with-exception-with-project-with-gold-with-throw-run-all-with-tools-pass-at-k-pass-at-5-llama_cpp-qwen2.5-coder:32b-instruct-q8_0-only-avsym-multi_ebt}
& \UseMacro{res-real-mega-test-data-with-exception-with-project-with-gold-with-throw-run-all-with-tools-pass-at-k-pass-at-10-llama_cpp-qwen2.5-coder:32b-instruct-q8_0-only-avsym-multi_ebt}
\\
\UseMacro{THead-only-lcov}
& \UseMacro{res-real-mega-test-data-with-exception-with-project-with-gold-with-throw-run-ebts-pass-at-k-compiled-at-1-llama_cpp-qwen2.5-coder:32b-instruct-q8_0-only-lcov-multi_ebt}
& \UseMacro{res-real-mega-test-data-with-exception-with-project-with-gold-with-throw-run-ebts-pass-at-k-compiled-at-5-llama_cpp-qwen2.5-coder:32b-instruct-q8_0-only-lcov-multi_ebt}
& \UseMacro{res-real-mega-test-data-with-exception-with-project-with-gold-with-throw-run-ebts-pass-at-k-compiled-at-10-llama_cpp-qwen2.5-coder:32b-instruct-q8_0-only-lcov-multi_ebt}
& \UseMacro{res-real-mega-test-data-with-exception-with-project-with-gold-with-throw-run-ebts-pass-at-k-pass-at-1-llama_cpp-qwen2.5-coder:32b-instruct-q8_0-only-lcov-multi_ebt}
& \UseMacro{res-real-mega-test-data-with-exception-with-project-with-gold-with-throw-run-ebts-pass-at-k-pass-at-5-llama_cpp-qwen2.5-coder:32b-instruct-q8_0-only-lcov-multi_ebt}
& \UseMacro{res-real-mega-test-data-with-exception-with-project-with-gold-with-throw-run-ebts-pass-at-k-pass-at-10-llama_cpp-qwen2.5-coder:32b-instruct-q8_0-only-lcov-multi_ebt}
& \UseMacro{res-real-mega-test-data-with-exception-with-project-with-gold-with-throw-run-all-pass-at-k-pass-at-1-llama_cpp-qwen2.5-coder:32b-instruct-q8_0-only-lcov-multi_ebt}
& \UseMacro{res-real-mega-test-data-with-exception-with-project-with-gold-with-throw-run-all-pass-at-k-pass-at-5-llama_cpp-qwen2.5-coder:32b-instruct-q8_0-only-lcov-multi_ebt}
& \UseMacro{res-real-mega-test-data-with-exception-with-project-with-gold-with-throw-run-all-pass-at-k-pass-at-10-llama_cpp-qwen2.5-coder:32b-instruct-q8_0-only-lcov-multi_ebt}
& \UseMacro{res-real-mega-test-data-with-exception-with-project-with-gold-with-throw-run-all-with-tools-pass-at-k-pass-at-1-llama_cpp-qwen2.5-coder:32b-instruct-q8_0-only-lcov-multi_ebt}
& \UseMacro{res-real-mega-test-data-with-exception-with-project-with-gold-with-throw-run-all-with-tools-pass-at-k-pass-at-5-llama_cpp-qwen2.5-coder:32b-instruct-q8_0-only-lcov-multi_ebt}
& \UseMacro{res-real-mega-test-data-with-exception-with-project-with-gold-with-throw-run-all-with-tools-pass-at-k-pass-at-10-llama_cpp-qwen2.5-coder:32b-instruct-q8_0-only-lcov-multi_ebt}
\\
\UseMacro{THead-only-nebt}
& \UseMacro{res-real-mega-test-data-with-exception-with-project-with-gold-with-throw-run-ebts-pass-at-k-compiled-at-1-llama_cpp-qwen2.5-coder:32b-instruct-q8_0-only-nebt-multi_ebt}
& \UseMacro{res-real-mega-test-data-with-exception-with-project-with-gold-with-throw-run-ebts-pass-at-k-compiled-at-5-llama_cpp-qwen2.5-coder:32b-instruct-q8_0-only-nebt-multi_ebt}
& \UseMacro{res-real-mega-test-data-with-exception-with-project-with-gold-with-throw-run-ebts-pass-at-k-compiled-at-10-llama_cpp-qwen2.5-coder:32b-instruct-q8_0-only-nebt-multi_ebt}
& \UseMacro{res-real-mega-test-data-with-exception-with-project-with-gold-with-throw-run-ebts-pass-at-k-pass-at-1-llama_cpp-qwen2.5-coder:32b-instruct-q8_0-only-nebt-multi_ebt}
& \UseMacro{res-real-mega-test-data-with-exception-with-project-with-gold-with-throw-run-ebts-pass-at-k-pass-at-5-llama_cpp-qwen2.5-coder:32b-instruct-q8_0-only-nebt-multi_ebt}
& \UseMacro{res-real-mega-test-data-with-exception-with-project-with-gold-with-throw-run-ebts-pass-at-k-pass-at-10-llama_cpp-qwen2.5-coder:32b-instruct-q8_0-only-nebt-multi_ebt}
& \UseMacro{res-real-mega-test-data-with-exception-with-project-with-gold-with-throw-run-all-pass-at-k-pass-at-1-llama_cpp-qwen2.5-coder:32b-instruct-q8_0-only-nebt-multi_ebt}
& \UseMacro{res-real-mega-test-data-with-exception-with-project-with-gold-with-throw-run-all-pass-at-k-pass-at-5-llama_cpp-qwen2.5-coder:32b-instruct-q8_0-only-nebt-multi_ebt}
& \UseMacro{res-real-mega-test-data-with-exception-with-project-with-gold-with-throw-run-all-pass-at-k-pass-at-10-llama_cpp-qwen2.5-coder:32b-instruct-q8_0-only-nebt-multi_ebt}
& \UseMacro{res-real-mega-test-data-with-exception-with-project-with-gold-with-throw-run-all-with-tools-pass-at-k-pass-at-1-llama_cpp-qwen2.5-coder:32b-instruct-q8_0-only-nebt-multi_ebt}
& \UseMacro{res-real-mega-test-data-with-exception-with-project-with-gold-with-throw-run-all-with-tools-pass-at-k-pass-at-5-llama_cpp-qwen2.5-coder:32b-instruct-q8_0-only-nebt-multi_ebt}
& \UseMacro{res-real-mega-test-data-with-exception-with-project-with-gold-with-throw-run-all-with-tools-pass-at-k-pass-at-10-llama_cpp-qwen2.5-coder:32b-instruct-q8_0-only-nebt-multi_ebt}
\\
\UseMacro{THead-only-threxc}
& \UseMacro{res-real-mega-test-data-with-exception-with-project-with-gold-with-throw-run-ebts-pass-at-k-compiled-at-1-llama_cpp-qwen2.5-coder:32b-instruct-q8_0-only-threxc-multi_ebt}
& \UseMacro{res-real-mega-test-data-with-exception-with-project-with-gold-with-throw-run-ebts-pass-at-k-compiled-at-5-llama_cpp-qwen2.5-coder:32b-instruct-q8_0-only-threxc-multi_ebt}
& \UseMacro{res-real-mega-test-data-with-exception-with-project-with-gold-with-throw-run-ebts-pass-at-k-compiled-at-10-llama_cpp-qwen2.5-coder:32b-instruct-q8_0-only-threxc-multi_ebt}
& \UseMacro{res-real-mega-test-data-with-exception-with-project-with-gold-with-throw-run-ebts-pass-at-k-pass-at-1-llama_cpp-qwen2.5-coder:32b-instruct-q8_0-only-threxc-multi_ebt}
& \UseMacro{res-real-mega-test-data-with-exception-with-project-with-gold-with-throw-run-ebts-pass-at-k-pass-at-5-llama_cpp-qwen2.5-coder:32b-instruct-q8_0-only-threxc-multi_ebt}
& \UseMacro{res-real-mega-test-data-with-exception-with-project-with-gold-with-throw-run-ebts-pass-at-k-pass-at-10-llama_cpp-qwen2.5-coder:32b-instruct-q8_0-only-threxc-multi_ebt}
& \UseMacro{res-real-mega-test-data-with-exception-with-project-with-gold-with-throw-run-all-pass-at-k-pass-at-1-llama_cpp-qwen2.5-coder:32b-instruct-q8_0-only-threxc-multi_ebt}
& \UseMacro{res-real-mega-test-data-with-exception-with-project-with-gold-with-throw-run-all-pass-at-k-pass-at-5-llama_cpp-qwen2.5-coder:32b-instruct-q8_0-only-threxc-multi_ebt}
& \UseMacro{res-real-mega-test-data-with-exception-with-project-with-gold-with-throw-run-all-pass-at-k-pass-at-10-llama_cpp-qwen2.5-coder:32b-instruct-q8_0-only-threxc-multi_ebt}
& \UseMacro{res-real-mega-test-data-with-exception-with-project-with-gold-with-throw-run-all-with-tools-pass-at-k-pass-at-1-llama_cpp-qwen2.5-coder:32b-instruct-q8_0-only-threxc-multi_ebt}
& \UseMacro{res-real-mega-test-data-with-exception-with-project-with-gold-with-throw-run-all-with-tools-pass-at-k-pass-at-5-llama_cpp-qwen2.5-coder:32b-instruct-q8_0-only-threxc-multi_ebt}
& \UseMacro{res-real-mega-test-data-with-exception-with-project-with-gold-with-throw-run-all-with-tools-pass-at-k-pass-at-10-llama_cpp-qwen2.5-coder:32b-instruct-q8_0-only-threxc-multi_ebt}
\\
\UseMacro{THead-cmtu}
& \UseMacro{res-real-mega-test-data-with-exception-with-project-with-gold-with-throw-run-ebts-pass-at-k-compiled-at-1-llama_cpp-qwen2.5-coder:32b-instruct-q8_0-cmtu-multi_ebt}
& \UseMacro{res-real-mega-test-data-with-exception-with-project-with-gold-with-throw-run-ebts-pass-at-k-compiled-at-5-llama_cpp-qwen2.5-coder:32b-instruct-q8_0-cmtu-multi_ebt}
& \UseMacro{res-real-mega-test-data-with-exception-with-project-with-gold-with-throw-run-ebts-pass-at-k-compiled-at-10-llama_cpp-qwen2.5-coder:32b-instruct-q8_0-cmtu-multi_ebt}
& \UseMacro{res-real-mega-test-data-with-exception-with-project-with-gold-with-throw-run-ebts-pass-at-k-pass-at-1-llama_cpp-qwen2.5-coder:32b-instruct-q8_0-cmtu-multi_ebt}
& \UseMacro{res-real-mega-test-data-with-exception-with-project-with-gold-with-throw-run-ebts-pass-at-k-pass-at-5-llama_cpp-qwen2.5-coder:32b-instruct-q8_0-cmtu-multi_ebt}
& \UseMacro{res-real-mega-test-data-with-exception-with-project-with-gold-with-throw-run-ebts-pass-at-k-pass-at-10-llama_cpp-qwen2.5-coder:32b-instruct-q8_0-cmtu-multi_ebt}
& \UseMacro{res-real-mega-test-data-with-exception-with-project-with-gold-with-throw-run-all-pass-at-k-pass-at-1-llama_cpp-qwen2.5-coder:32b-instruct-q8_0-cmtu-multi_ebt}
& \UseMacro{res-real-mega-test-data-with-exception-with-project-with-gold-with-throw-run-all-pass-at-k-pass-at-5-llama_cpp-qwen2.5-coder:32b-instruct-q8_0-cmtu-multi_ebt}
& \UseMacro{res-real-mega-test-data-with-exception-with-project-with-gold-with-throw-run-all-pass-at-k-pass-at-10-llama_cpp-qwen2.5-coder:32b-instruct-q8_0-cmtu-multi_ebt}
& \UseMacro{res-real-mega-test-data-with-exception-with-project-with-gold-with-throw-run-all-with-tools-pass-at-k-pass-at-1-llama_cpp-qwen2.5-coder:32b-instruct-q8_0-cmtu-multi_ebt}
& \UseMacro{res-real-mega-test-data-with-exception-with-project-with-gold-with-throw-run-all-with-tools-pass-at-k-pass-at-5-llama_cpp-qwen2.5-coder:32b-instruct-q8_0-cmtu-multi_ebt}
& \UseMacro{res-real-mega-test-data-with-exception-with-project-with-gold-with-throw-run-all-with-tools-pass-at-k-pass-at-10-llama_cpp-qwen2.5-coder:32b-instruct-q8_0-cmtu-multi_ebt}
\\
\UseMacro{THead-base}
& \UseMacro{res-real-mega-test-data-with-exception-with-project-with-gold-with-throw-run-ebts-pass-at-k-compiled-at-1-llama_cpp-qwen2.5-coder:32b-instruct-q8_0-base-multi_ebt}
& \UseMacro{res-real-mega-test-data-with-exception-with-project-with-gold-with-throw-run-ebts-pass-at-k-compiled-at-5-llama_cpp-qwen2.5-coder:32b-instruct-q8_0-base-multi_ebt}
& \UseMacro{res-real-mega-test-data-with-exception-with-project-with-gold-with-throw-run-ebts-pass-at-k-compiled-at-10-llama_cpp-qwen2.5-coder:32b-instruct-q8_0-base-multi_ebt}
& \UseMacro{res-real-mega-test-data-with-exception-with-project-with-gold-with-throw-run-ebts-pass-at-k-pass-at-1-llama_cpp-qwen2.5-coder:32b-instruct-q8_0-base-multi_ebt}
& \UseMacro{res-real-mega-test-data-with-exception-with-project-with-gold-with-throw-run-ebts-pass-at-k-pass-at-5-llama_cpp-qwen2.5-coder:32b-instruct-q8_0-base-multi_ebt}
& \UseMacro{res-real-mega-test-data-with-exception-with-project-with-gold-with-throw-run-ebts-pass-at-k-pass-at-10-llama_cpp-qwen2.5-coder:32b-instruct-q8_0-base-multi_ebt}
& \UseMacro{res-real-mega-test-data-with-exception-with-project-with-gold-with-throw-run-all-pass-at-k-pass-at-1-llama_cpp-qwen2.5-coder:32b-instruct-q8_0-base-multi_ebt}
& \UseMacro{res-real-mega-test-data-with-exception-with-project-with-gold-with-throw-run-all-pass-at-k-pass-at-5-llama_cpp-qwen2.5-coder:32b-instruct-q8_0-base-multi_ebt}
& \UseMacro{res-real-mega-test-data-with-exception-with-project-with-gold-with-throw-run-all-pass-at-k-pass-at-10-llama_cpp-qwen2.5-coder:32b-instruct-q8_0-base-multi_ebt}
& \UseMacro{res-real-mega-test-data-with-exception-with-project-with-gold-with-throw-run-all-with-tools-pass-at-k-pass-at-1-llama_cpp-qwen2.5-coder:32b-instruct-q8_0-base-multi_ebt}
& \UseMacro{res-real-mega-test-data-with-exception-with-project-with-gold-with-throw-run-all-with-tools-pass-at-k-pass-at-5-llama_cpp-qwen2.5-coder:32b-instruct-q8_0-base-multi_ebt}
& \UseMacro{res-real-mega-test-data-with-exception-with-project-with-gold-with-throw-run-all-with-tools-pass-at-k-pass-at-10-llama_cpp-qwen2.5-coder:32b-instruct-q8_0-base-multi_ebt}
\\
\bottomrule
\end{tabular}
\end{center}
\end{small}
\end{table*}

In the following sections we present our evaluation results and answer our research questions.

\subsection{\UseMacro{rq-modelComp}: Effectiveness Across Models}
\label{sec:res-model-comp}

The results on the effectiveness of \Tool on different models are presented in Table~\ref{tab:model-comp-results}.

\Tool consistently outperforms the baseline across all five models on
all metrics. Among open-source models, the most significant
performance gains are observed with
\UseMacro{THead-qwen2.5-coder:32b-instruct-q8-0}. Specifically, \Tool
achieves improvements of \UseMacro{OverQwenLargeCompFive} percentage
points on \UseMacro{THead-run-ebts-pass-at-k-compiled-at-5},
\UseMacro{OverQwenLargePassEBTFive} p.p. on
\UseMacro{THead-run-ebts-pass-at-k-pass-at-5},
\UseMacro{OverQwenLargePassAllFive} p.p. on
\UseMacro{THead-run-all-pass-at-k-pass-at-5}, and
\UseMacro{OverQwenLargePassAllToolsFive} p.p. on
\UseMacro{THead-run-all-with-tools-pass-at-k-pass-at-5}.
This performance gain likely stems from
\UseMacro{THead-qwen2.5-coder:32b-instruct-q8-0}'s capability to
interpret and leverage the contextual information provided by \Tool.
On the state-of-the-art closed-source model
\UseMacro{THead-gpt-5-mini}, \Tool also demonstrates notable
improvements over the baseline: \UseMacro{OverGPTCompFive} p.p. on
\UseMacro{THead-run-ebts-pass-at-k-compiled-at-5},
\UseMacro{OverGPTPassEBTFive} p.p. on
\UseMacro{THead-run-ebts-pass-at-k-pass-at-5},
\UseMacro{OverGPTPassAllFive} p.p. on
\UseMacro{THead-run-all-pass-at-k-pass-at-5}, and
\UseMacro{OverGPTPassAllToolsFive} p.p. on
\UseMacro{THead-run-all-with-tools-pass-at-k-pass-at-5}. These results
demonstrate that \Tool provides meaningful gains even on the most
capable models, and that the improvements persist when generated \ERCs
are checked against automatically generated tests. At the same time,
\UseMacro{THead-run-all-with-tools-pass-at-k-pass-at-5} remains
noticeably below \UseMacro{THead-run-all-pass-at-k-pass-at-5} for
every model, including \Tool itself, indicating that some \ERCs pass
all user-written tests yet still deviate from the ground truth. We
further examine this in Section~\ref{sec:qual-equiv}.

\begin{figure}[t!]
\centering
\begin{subfigure}[b]{0.49\columnwidth}
\centering
\includegraphics[width=\linewidth,trim=28 27 5 11,clip]{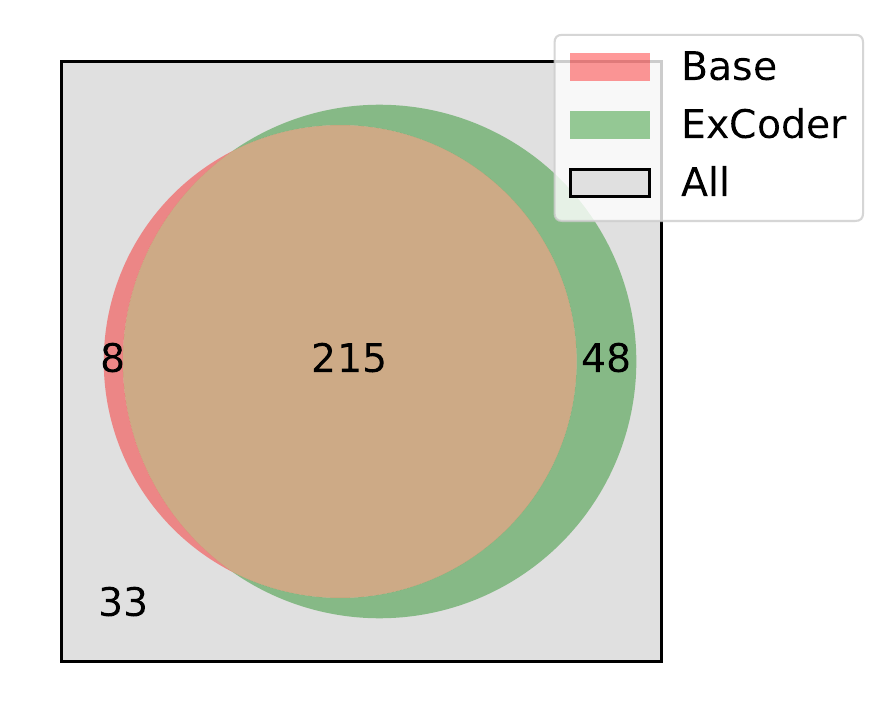}
\caption{Pass all user-written tests (\EBTs and \nEBTs)\newline}
\label{fig:run-ebts-venn}
\end{subfigure}
\hfill
\begin{subfigure}[b]{0.49\columnwidth}
\centering
\includegraphics[width=\linewidth,trim=28 27 5 11,clip]{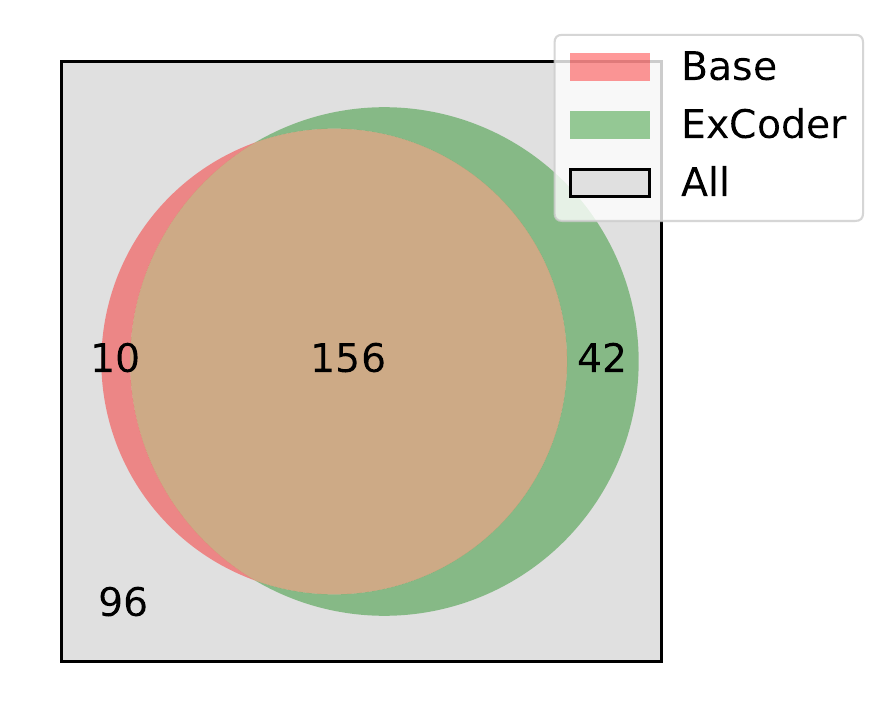}
\caption{Manually verified to be semantically equivalent to the ground truth}
\label{fig:qual-venn}
\end{subfigure}

\caption{\label{fig:venn}Overlap between \tarmethods
successfully handled by the baseline versus \Tool. Each
circle covers the \tarmethods for which at least one of the
10 generated samples meets the criterion.}
\end{figure}

Figure~\ref{fig:run-ebts-venn} illustrates the overlap and differences between the sets of
\tarmethods where the task was successfully completed (i.e., achieving a score
of 1.0 in \UseMacro{THead-run-all-pass-at-k-pass-at-10}) using the baseline
prompt versus \Tool with \UseMacro{THead-qwen2.5-coder:32b-instruct-q8-0}. \Tool
successfully solved all but
\UseMacro{overlap-real-mega-test-data-with-exception-with-project-with-gold-with-throw-run-all-llama_cpp-qwen2.5-coder:32b-instruct-q8_0-multi_ebt-only-base}
of the \task tasks that the baseline prompt solved, while additionally solving
\UseMacro{overlap-real-mega-test-data-with-exception-with-project-with-gold-with-throw-run-all-llama_cpp-qwen2.5-coder:32b-instruct-q8_0-multi_ebt-only-tuctn-all-info}
tasks where the baseline failed. This demonstrates that the contextual
information provided by \Tool meaningfully enhances the \LLM's \task capability
without significantly compromising its performance on cases where simpler
prompts suffice.
The
\UseMacro{overlap-real-mega-test-data-with-exception-with-project-with-gold-with-throw-run-all-llama_cpp-qwen2.5-coder:32b-instruct-q8_0-multi_ebt-only-base}
tasks where the baseline succeeded but \Tool did not can be attributed to cases
where the \task task was relatively straightforward, yet the additional context
introduced noise that made it harder for the \LLM to focus on the most relevant
information in the \tarmethod and \EBTs. Although these cases are rare, they
highlight a promising future direction of developing techniques to prioritize
the most relevant contextual information for the \LLM-based \task task. Overall, our
results suggest that \Tool's context engineering provides a robust net benefit,
substantially expanding the range of solvable tasks.

\subsection{\UseMacro{rq-promptComp}: Ablation Study of Prompt Components}

To evaluate the contribution of each component in \Tool, we conduct an ablation
study in which each variant augments the baseline prompt with exactly one
context component. Table~\ref{tab:prompt-comp-results} reports results on
\UseMacro{THead-qwen2.5-coder:32b-instruct-q8-0}. These variants also serve as
the enhanced versions of the baseline described in Section~\ref{sec:baselines}, and \Tool
outperforms all of them on every metric in the table.

Adding \contextAS or \contextEC alone yields large improvements across all
correctness metrics, demonstrating that these contexts are the easiest for
the \LLM to utilize: \contextEC supply the exact constructor signature
required to generate the exception object that the \EBT checks for, and
\contextAS enumerate the methods and fields that can legally appear in the
exceptional condition check. This information can be readily utilized by the
\LLM to generate correct \ERC with the right function calls and exception
construction.

The remaining components, \contextLC, \contextTE, and \contextNEBT,
yield noticeably smaller gains that appear only in pass@5 or
pass@10. This pattern indicates that these contexts do help the model
eventually generate correct \ERCs given enough samples, but the model
struggles to utilize them on every attempt. The underlying reason is
that these contexts describe program behavior rather than code that
needs to be generated. Using them correctly requires the \LLM to
reason about program states and control flow, which is a substantially
harder task~\cite{thimmaiah2026llms}. We leave as future work how
these richer signals can be more fully utilized, e.g., through more
targeted annotation or specialized prompting strategies.

\begin{figure*}[!tbp]
\centering
\begin{subfigure}[b]{0.32\textwidth}
\centering
\includegraphics[width=\linewidth]{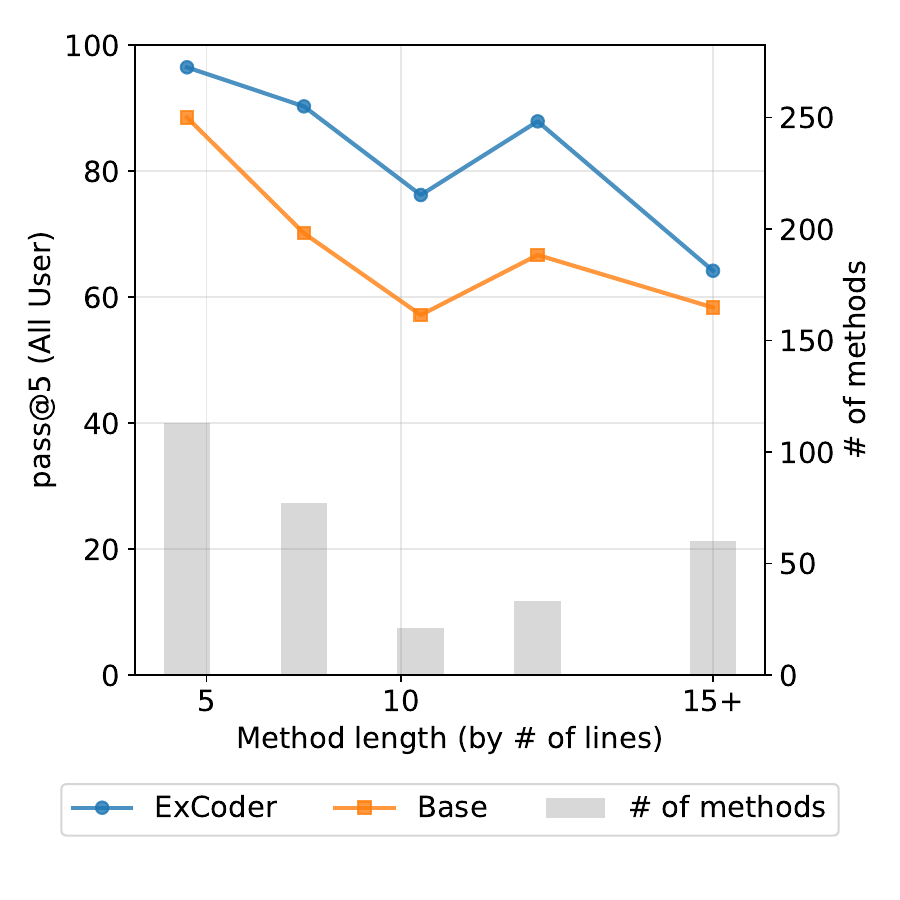}
\caption{Method length}
\label{fig:lencomp-len}
\end{subfigure}
\hfill
\begin{subfigure}[b]{0.32\textwidth}
\centering
\includegraphics[width=\linewidth]{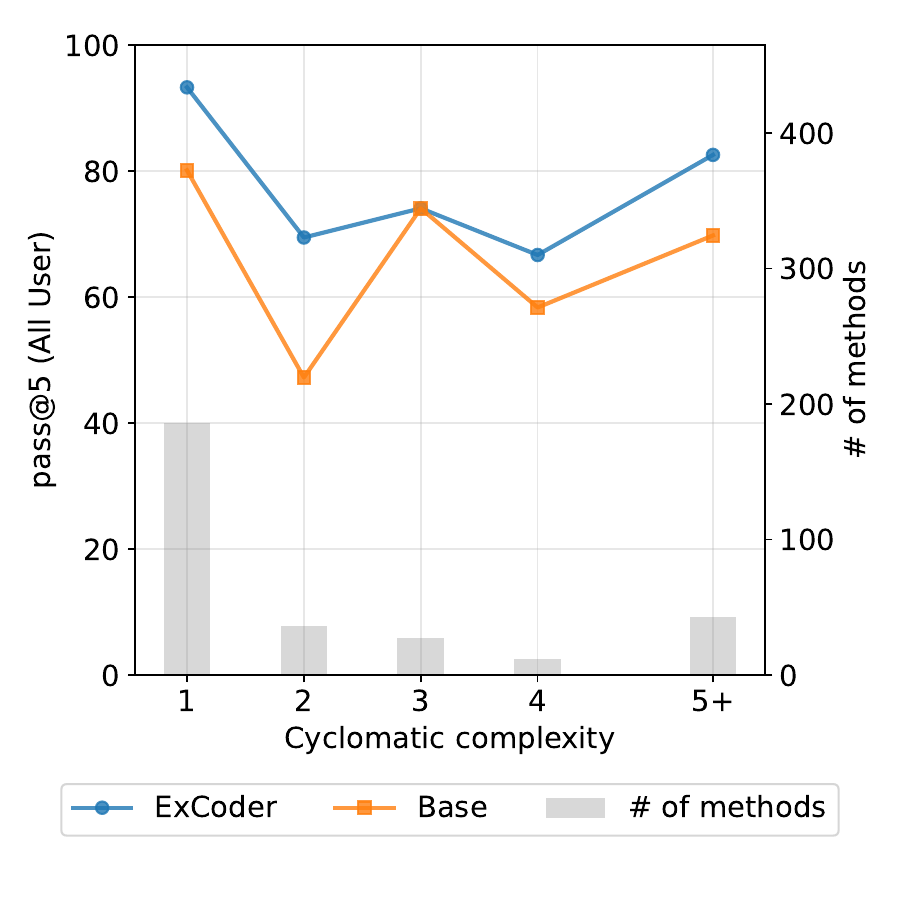}
\caption{Cyclomatic complexity}
\label{fig:lencomp-cc}
\end{subfigure}
\hfill
\begin{subfigure}[b]{0.32\textwidth}
\centering
\includegraphics[width=\linewidth]{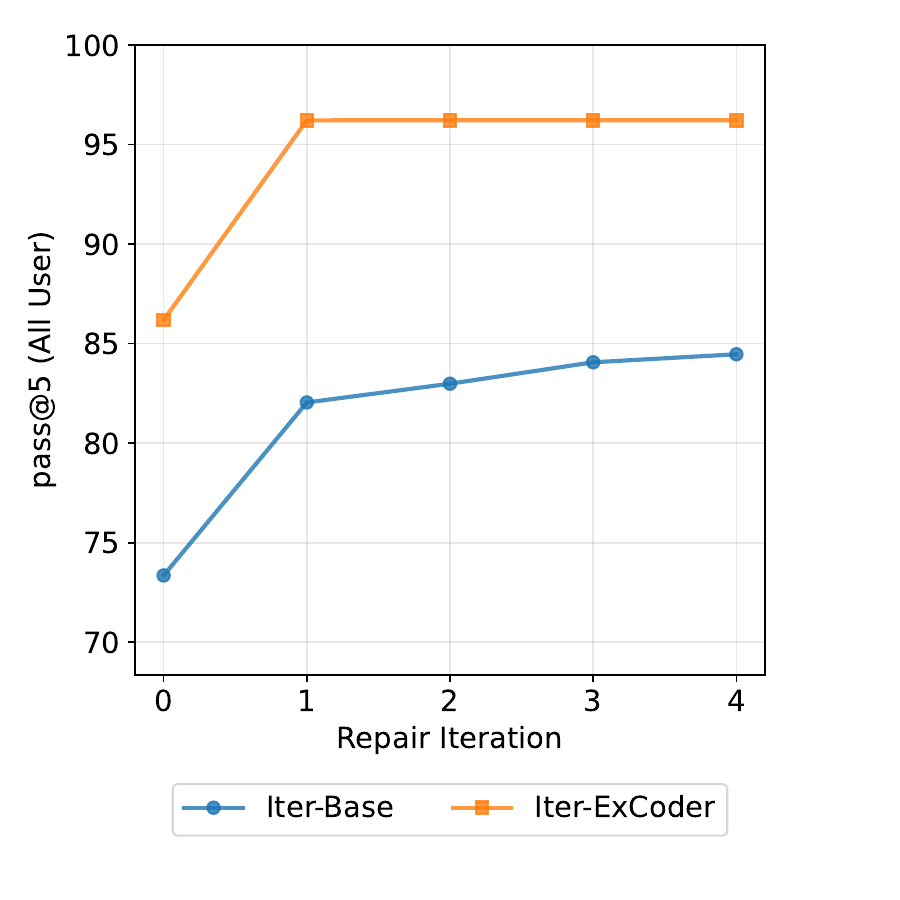}
\caption{Iterations of self-repair}
\label{fig:itercomp}
\end{subfigure}

\caption{Performance of \Tool compared with \Base on \tarmethods of
different (a)~length and (b)~cyclomatic complexity, where bars show the
number of \tarmethods in each bin, and (c)~performance of Iter-\Tool
compared with Iter-\Base at each iteration of the repair process.}
\label{fig:lencomp}
\end{figure*}

\subsection{\UseMacro{rq-lenComp}: Effect of Target Method Complexity}

Figures~\ref{fig:lencomp-len} and~\ref{fig:lencomp-cc} present our
analysis of how \tarmethod complexity influences \Tool's and the
baseline's effectiveness, evaluated on
\UseMacro{THead-qwen2.5-coder:32b-instruct-q8-0}. Here, we use the
number of lines of code and cyclomatic complexity as a proxy for
method complexity and use \UseMacro{THead-run-all-pass-at-k-pass-at-5}
to measure performance.

Methods exceeding \UseMacro{ComplexityLocClosedMax} lines  constitute only
\UseMacro{ComplexityLocOpenPercent} of the \tarmethods, and their lengths vary significantly,
ranging up to \UseMacro{ComplexityLocMax} lines. Similarly, methods with
cyclomatic complexity of \UseMacro{ComplexityCcOpenMin} or more constitute only
\UseMacro{ComplexityCcOpenPercent} of the \tarmethods, with complexity
ranging up to \UseMacro{ComplexityCcMax}. We therefore report them in single
open-ended ``\UseMacro{ComplexityLocOpenLabel}'' and
``\UseMacro{ComplexityCcOpenLabel}'' bins rather than excluding them, while
all other bins have equal width.

\Tool matches or outperforms \Base in every bin of both measures. We
observe that \Tool's advantage is largest for methods of
\UseMacro{ComplexityLocMidRangeLabel} lines and for methods with
cyclomatic complexity of two. We attribute this to the fact that
simpler methods typically have \conds that are easier to reason about,
such as checking whether an input is null, while the longest methods
provide more information in their method body, reducing the benefit of
the additional context. In contrast, methods in the middle of the
complexity range often involve complex conditions that could benefit
from the additional context provided by \Tool, while the methods
themselves do not provide enough information for \task on their own.
Moreover, for cyclomatic complexity, which counts branches rather than
lines, measuring behavioral complexity relatively independent of the
information contained in code, we see the advantage widen again to
\UseMacro{ComplexityCcOpenGapValue} p.p. at
\UseMacro{ComplexityCcOpenLabel}, so \Tool still helps when behavior
is hard to parse.

\begin{table*}[t]
\begin{footnotesize}
\begin{center}
\caption{\UseMacro{TCap-res-prompt-comp-repair}\label{tab:prompt-comp-results-repair}}
\begin{tabular}{l |cccccccccccc}
\toprule
\multirow{2}{*}{\UseMacro{THead-prompt}}
& \multicolumn{3}{c}{\UseMacro{THead-compiled-at-k}}
& \multicolumn{3}{c}{\UseMacro{THead-ebts-pass-at-k}}
& \multicolumn{3}{c}{\UseMacro{THead-all-pass-at-k}}
& \multicolumn{3}{c}{\UseMacro{THead-allntools-pass-at-k}}
\\
& \UseMacro{THead-run-ebts-pass-at-k-compiled-at-1-small}
& \UseMacro{THead-run-ebts-pass-at-k-compiled-at-5-small}
& \UseMacro{THead-run-ebts-pass-at-k-compiled-at-10-small}
& \UseMacro{THead-run-ebts-pass-at-k-pass-at-1-small}
& \UseMacro{THead-run-ebts-pass-at-k-pass-at-5-small}
& \UseMacro{THead-run-ebts-pass-at-k-pass-at-10-small}
& \UseMacro{THead-run-all-pass-at-k-pass-at-1-small}
& \UseMacro{THead-run-all-pass-at-k-pass-at-5-small}
& \UseMacro{THead-run-all-pass-at-k-pass-at-10-small}
& \UseMacro{THead-run-all-with-tools-pass-at-k-pass-at-1-small}
& \UseMacro{THead-run-all-with-tools-pass-at-k-pass-at-5-small}
& \UseMacro{THead-run-all-with-tools-pass-at-k-pass-at-10-small}
\\
\midrule
\UseMacro{THead-base}
& \UseMacro{res-real-mega-test-data-with-exception-with-project-with-gold-with-throw-run-ebts-pass-at-k-compiled-at-1-llama_cpp-qwen2.5-coder:32b-instruct-q8_0-base-multi_ebt}
& \UseMacro{res-real-mega-test-data-with-exception-with-project-with-gold-with-throw-run-ebts-pass-at-k-compiled-at-5-llama_cpp-qwen2.5-coder:32b-instruct-q8_0-base-multi_ebt}
& \UseMacro{res-real-mega-test-data-with-exception-with-project-with-gold-with-throw-run-ebts-pass-at-k-compiled-at-10-llama_cpp-qwen2.5-coder:32b-instruct-q8_0-base-multi_ebt}
& \UseMacro{res-real-mega-test-data-with-exception-with-project-with-gold-with-throw-run-ebts-pass-at-k-pass-at-1-llama_cpp-qwen2.5-coder:32b-instruct-q8_0-base-multi_ebt}
& \UseMacro{res-real-mega-test-data-with-exception-with-project-with-gold-with-throw-run-ebts-pass-at-k-pass-at-5-llama_cpp-qwen2.5-coder:32b-instruct-q8_0-base-multi_ebt}
& \UseMacro{res-real-mega-test-data-with-exception-with-project-with-gold-with-throw-run-ebts-pass-at-k-pass-at-10-llama_cpp-qwen2.5-coder:32b-instruct-q8_0-base-multi_ebt}
& \UseMacro{res-real-mega-test-data-with-exception-with-project-with-gold-with-throw-run-all-pass-at-k-pass-at-1-llama_cpp-qwen2.5-coder:32b-instruct-q8_0-base-multi_ebt}
& \UseMacro{res-real-mega-test-data-with-exception-with-project-with-gold-with-throw-run-all-pass-at-k-pass-at-5-llama_cpp-qwen2.5-coder:32b-instruct-q8_0-base-multi_ebt}
& \UseMacro{res-real-mega-test-data-with-exception-with-project-with-gold-with-throw-run-all-pass-at-k-pass-at-10-llama_cpp-qwen2.5-coder:32b-instruct-q8_0-base-multi_ebt}
& \UseMacro{res-real-mega-test-data-with-exception-with-project-with-gold-with-throw-run-all-with-tools-pass-at-k-pass-at-1-llama_cpp-qwen2.5-coder:32b-instruct-q8_0-base-multi_ebt}
& \UseMacro{res-real-mega-test-data-with-exception-with-project-with-gold-with-throw-run-all-with-tools-pass-at-k-pass-at-5-llama_cpp-qwen2.5-coder:32b-instruct-q8_0-base-multi_ebt}
& \UseMacro{res-real-mega-test-data-with-exception-with-project-with-gold-with-throw-run-all-with-tools-pass-at-k-pass-at-10-llama_cpp-qwen2.5-coder:32b-instruct-q8_0-base-multi_ebt}
\\
\UseMacro{THead-base-repair@4}
& \UseMacro{res-real-mega-test-data-with-exception-with-project-with-gold-with-throw-run-ebts-pass-at-k-compiled-at-1-llama_cpp-qwen2.5-coder:32b-instruct-q8_0-base-repair@4-repair}
& \UseMacro{res-real-mega-test-data-with-exception-with-project-with-gold-with-throw-run-ebts-pass-at-k-compiled-at-5-llama_cpp-qwen2.5-coder:32b-instruct-q8_0-base-repair@4-repair}
& \UseMacro{res-real-mega-test-data-with-exception-with-project-with-gold-with-throw-run-ebts-pass-at-k-compiled-at-10-llama_cpp-qwen2.5-coder:32b-instruct-q8_0-base-repair@4-repair}
& \UseMacro{res-real-mega-test-data-with-exception-with-project-with-gold-with-throw-run-ebts-pass-at-k-pass-at-1-llama_cpp-qwen2.5-coder:32b-instruct-q8_0-base-repair@4-repair}
& \UseMacro{res-real-mega-test-data-with-exception-with-project-with-gold-with-throw-run-ebts-pass-at-k-pass-at-5-llama_cpp-qwen2.5-coder:32b-instruct-q8_0-base-repair@4-repair}
& \UseMacro{res-real-mega-test-data-with-exception-with-project-with-gold-with-throw-run-ebts-pass-at-k-pass-at-10-llama_cpp-qwen2.5-coder:32b-instruct-q8_0-base-repair@4-repair}
& \UseMacro{res-real-mega-test-data-with-exception-with-project-with-gold-with-throw-run-all-pass-at-k-pass-at-1-llama_cpp-qwen2.5-coder:32b-instruct-q8_0-base-repair@4-repair}
& \UseMacro{res-real-mega-test-data-with-exception-with-project-with-gold-with-throw-run-all-pass-at-k-pass-at-5-llama_cpp-qwen2.5-coder:32b-instruct-q8_0-base-repair@4-repair}
& \UseMacro{res-real-mega-test-data-with-exception-with-project-with-gold-with-throw-run-all-pass-at-k-pass-at-10-llama_cpp-qwen2.5-coder:32b-instruct-q8_0-base-repair@4-repair}
& \UseMacro{res-real-mega-test-data-with-exception-with-project-with-gold-with-throw-run-all-with-tools-pass-at-k-pass-at-1-llama_cpp-qwen2.5-coder:32b-instruct-q8_0-base-repair@4-repair}
& \UseMacro{res-real-mega-test-data-with-exception-with-project-with-gold-with-throw-run-all-with-tools-pass-at-k-pass-at-5-llama_cpp-qwen2.5-coder:32b-instruct-q8_0-base-repair@4-repair}
& \UseMacro{res-real-mega-test-data-with-exception-with-project-with-gold-with-throw-run-all-with-tools-pass-at-k-pass-at-10-llama_cpp-qwen2.5-coder:32b-instruct-q8_0-base-repair@4-repair}
\\
\midrule
\UseMacro{THead-tuctn-all-info}
& \UseMacro{res-real-mega-test-data-with-exception-with-project-with-gold-with-throw-run-ebts-pass-at-k-compiled-at-1-llama_cpp-qwen2.5-coder:32b-instruct-q8_0-tuctn-all-info-multi_ebt}
& \UseMacro{res-real-mega-test-data-with-exception-with-project-with-gold-with-throw-run-ebts-pass-at-k-compiled-at-5-llama_cpp-qwen2.5-coder:32b-instruct-q8_0-tuctn-all-info-multi_ebt}
& \UseMacro{res-real-mega-test-data-with-exception-with-project-with-gold-with-throw-run-ebts-pass-at-k-compiled-at-10-llama_cpp-qwen2.5-coder:32b-instruct-q8_0-tuctn-all-info-multi_ebt}
& \UseMacro{res-real-mega-test-data-with-exception-with-project-with-gold-with-throw-run-ebts-pass-at-k-pass-at-1-llama_cpp-qwen2.5-coder:32b-instruct-q8_0-tuctn-all-info-multi_ebt}
& \UseMacro{res-real-mega-test-data-with-exception-with-project-with-gold-with-throw-run-ebts-pass-at-k-pass-at-5-llama_cpp-qwen2.5-coder:32b-instruct-q8_0-tuctn-all-info-multi_ebt}
& \UseMacro{res-real-mega-test-data-with-exception-with-project-with-gold-with-throw-run-ebts-pass-at-k-pass-at-10-llama_cpp-qwen2.5-coder:32b-instruct-q8_0-tuctn-all-info-multi_ebt}
& \UseMacro{res-real-mega-test-data-with-exception-with-project-with-gold-with-throw-run-all-pass-at-k-pass-at-1-llama_cpp-qwen2.5-coder:32b-instruct-q8_0-tuctn-all-info-multi_ebt}
& \UseMacro{res-real-mega-test-data-with-exception-with-project-with-gold-with-throw-run-all-pass-at-k-pass-at-5-llama_cpp-qwen2.5-coder:32b-instruct-q8_0-tuctn-all-info-multi_ebt}
& \UseMacro{res-real-mega-test-data-with-exception-with-project-with-gold-with-throw-run-all-pass-at-k-pass-at-10-llama_cpp-qwen2.5-coder:32b-instruct-q8_0-tuctn-all-info-multi_ebt}
& \UseMacro{res-real-mega-test-data-with-exception-with-project-with-gold-with-throw-run-all-with-tools-pass-at-k-pass-at-1-llama_cpp-qwen2.5-coder:32b-instruct-q8_0-tuctn-all-info-multi_ebt}
& \UseMacro{res-real-mega-test-data-with-exception-with-project-with-gold-with-throw-run-all-with-tools-pass-at-k-pass-at-5-llama_cpp-qwen2.5-coder:32b-instruct-q8_0-tuctn-all-info-multi_ebt}
& \UseMacro{res-real-mega-test-data-with-exception-with-project-with-gold-with-throw-run-all-with-tools-pass-at-k-pass-at-10-llama_cpp-qwen2.5-coder:32b-instruct-q8_0-tuctn-all-info-multi_ebt}
\\
\UseMacro{THead-tuctn-all-info-repair@4}
& \textbf{\UseMacro{res-real-mega-test-data-with-exception-with-project-with-gold-with-throw-run-ebts-pass-at-k-compiled-at-1-llama_cpp-qwen2.5-coder:32b-instruct-q8_0-tuctn-all-info-repair@4-repair}}
& \textbf{\UseMacro{res-real-mega-test-data-with-exception-with-project-with-gold-with-throw-run-ebts-pass-at-k-compiled-at-5-llama_cpp-qwen2.5-coder:32b-instruct-q8_0-tuctn-all-info-repair@4-repair}}
& \textbf{\UseMacro{res-real-mega-test-data-with-exception-with-project-with-gold-with-throw-run-ebts-pass-at-k-compiled-at-10-llama_cpp-qwen2.5-coder:32b-instruct-q8_0-tuctn-all-info-repair@4-repair}}
& \textbf{\UseMacro{res-real-mega-test-data-with-exception-with-project-with-gold-with-throw-run-ebts-pass-at-k-pass-at-1-llama_cpp-qwen2.5-coder:32b-instruct-q8_0-tuctn-all-info-repair@4-repair}}
& \textbf{\UseMacro{res-real-mega-test-data-with-exception-with-project-with-gold-with-throw-run-ebts-pass-at-k-pass-at-5-llama_cpp-qwen2.5-coder:32b-instruct-q8_0-tuctn-all-info-repair@4-repair}}
& \textbf{\UseMacro{res-real-mega-test-data-with-exception-with-project-with-gold-with-throw-run-ebts-pass-at-k-pass-at-10-llama_cpp-qwen2.5-coder:32b-instruct-q8_0-tuctn-all-info-repair@4-repair}}
& \textbf{\UseMacro{res-real-mega-test-data-with-exception-with-project-with-gold-with-throw-run-all-pass-at-k-pass-at-1-llama_cpp-qwen2.5-coder:32b-instruct-q8_0-tuctn-all-info-repair@4-repair}}
& \textbf{\UseMacro{res-real-mega-test-data-with-exception-with-project-with-gold-with-throw-run-all-pass-at-k-pass-at-5-llama_cpp-qwen2.5-coder:32b-instruct-q8_0-tuctn-all-info-repair@4-repair}}
& \textbf{\UseMacro{res-real-mega-test-data-with-exception-with-project-with-gold-with-throw-run-all-pass-at-k-pass-at-10-llama_cpp-qwen2.5-coder:32b-instruct-q8_0-tuctn-all-info-repair@4-repair}}
& \textbf{\UseMacro{res-real-mega-test-data-with-exception-with-project-with-gold-with-throw-run-all-with-tools-pass-at-k-pass-at-1-llama_cpp-qwen2.5-coder:32b-instruct-q8_0-tuctn-all-info-repair@4-repair}}
& \textbf{\UseMacro{res-real-mega-test-data-with-exception-with-project-with-gold-with-throw-run-all-with-tools-pass-at-k-pass-at-5-llama_cpp-qwen2.5-coder:32b-instruct-q8_0-tuctn-all-info-repair@4-repair}}
& \textbf{\UseMacro{res-real-mega-test-data-with-exception-with-project-with-gold-with-throw-run-all-with-tools-pass-at-k-pass-at-10-llama_cpp-qwen2.5-coder:32b-instruct-q8_0-tuctn-all-info-repair@4-repair}}
\\
\bottomrule
\end{tabular}
\end{center}
\end{footnotesize}
\end{table*}

\subsection{\UseMacro{rq-iter}: Combination with Self-Repair}

Prior TDD-based code generation work~\cite{mathewsTestDrivenDevelopmentCode2024}
employs an \LLM self-repair approach, where compilation and execution output is
fed back to the model to fix issues in the generated code. This approach is
natural for TDD-based code generation, as input tests can provide direct feedback
that guides the \LLM toward a correct solution.

To investigate whether \Tool remains beneficial within this iterative framework,
we implemented two repair-based variants: \UseMacro{THead-base-repair@4}, which
uses only the baseline prompt, and \UseMacro{THead-tuctn-all-info-repair@4},
which retains all context provided by \Tool. Both variants use
\UseMacro{THead-qwen2.5-coder:32b-instruct-q8-0} and perform up to 4 iterations
of repair, where the \LLM receives compilation output and test errors after each
failed attempt.

The results, shown in Table~\ref{tab:prompt-comp-results-repair} and
Figure~\ref{fig:itercomp}, demonstrate that iterative repair improves
performance for both prompts. With 4 rounds of repair,
\UseMacro{THead-base-repair@4} achieves
\UseMacro{res-real-mega-test-data-with-exception-with-project-with-gold-with-throw-run-all-pass-at-k-pass-at-5-llama_cpp-qwen2.5-coder:32b-instruct-q8_0-base-repair@4-repair}\%
on \UseMacro{THead-run-all-pass-at-k-pass-at-5}, while
\UseMacro{THead-tuctn-all-info-repair@4} reaches
\UseMacro{res-real-mega-test-data-with-exception-with-project-with-gold-with-throw-run-all-pass-at-k-pass-at-5-llama_cpp-qwen2.5-coder:32b-instruct-q8_0-tuctn-all-info-repair@4-repair}\%.
As shown in Figure~\ref{fig:itercomp}, most tasks are successfully resolved in
the first repair iteration, with only a few additional tasks solved in
subsequent rounds. This shows that while compiler and test feedback can help \LLMs identify surface-level errors in the repair iteration, such feedback becomes insufficient when the underlying issue stems from a lack of context, which leads to diminishing returns in subsequent rounds. Furthermore, even after self-repair steps,
\UseMacro{THead-base-repair@4} still trails behind
\UseMacro{THead-tuctn-all-info-repair@4} by
\UseMacro{OverQwenLargeRepairPassAllFive} p.p. on
\UseMacro{THead-run-all-pass-at-k-pass-at-5} and by
\UseMacro{OverQwenLargeRepairPassAllToolsFive} p.p. on
\UseMacro{THead-run-all-with-tools-pass-at-k-pass-at-5}. These findings indicate that
self-repair and \Tool complement each other effectively, and that self-repair
alone cannot substitute for the contextual information provided by \Tool in
\task.

\section{Qualitative Analysis}

\subsection{Cause of Failure Cases}
\label{sec:qual-failure-causes}

We first investigate the individual cases in which the generated \ERC
fails, and identify what causes each failure. These failures fall into two kinds. In the first, the \LLM
knows what to check but cannot write code that correctly implements
it. In the second, the \ERC compiles but the \cond checks the
wrong property. Due to space limits, we show an example only for the
first kind.
We present an example of the first kind
of failure in Figure~\ref{fig:fail-vocab-example}, where the \ERC has a compile error even though the \LLM correctly infers the underlying exceptional condition. The \tarmethod issues an HTTP request and must raise
the library's \CodeIn{TempoDBException}, carrying the server's message
and status code, when the request fails. The ground truth reads that
payload with \CodeIn{result.getMessage()} and
\CodeIn{result.getCode()} (Figure~\ref{fig:fail-vocab-example-gt}).
The generated \ERC places the \cond correctly and calls the right
constructor (Figure~\ref{fig:fail-vocab-example-tool}), but it uses
\CodeIn{State.ERROR}, \CodeIn{result.getErrorMessage()}, and
\CodeIn{result.getErrorCode()}, none of which is actually defined in
the project, so the \ERC does not compile. The \LLM hallucinates these
symbols, and the root cause is a gap in
\contextAS, which does not include the methods and fields defined on
the classes of local variables such as \CodeIn{result}. We omit these
because our early experiments showed that including them lowers performance
overall. The cases in which the \LLM strictly needs those symbols are
rare, and the extra context crowds out the information the \LLM does
use.

\subsection{Equivalence to Ground Truth}
\label{sec:qual-equiv}

\begin{figure}[!tbp]
\begin{subfigure}{\columnwidth}
\begin{lstlisting}[language=java-diff]
public Iterator<SingleValue> iterator() {
    Result<SingleValueSegment> result = client.execute(request, SingleValueSegment.class);
    ...
    if (result.getState() == State.SUCCESS) {
        ...
+   } else if (result.getState() == State.ERROR) {
+       throw new TempoDBException(result.getErrorMessage(), result.getErrorCode());
    }
    return iterator;
}
\end{lstlisting}
\caption{\ERC generated with \Tool.}
\label{fig:fail-vocab-example-tool}
\end{subfigure}
\begin{subfigure}{\columnwidth}
\begin{lstlisting}[language=java-diff]
public Iterator<SingleValue> iterator() {
    Result<SingleValueSegment> result = client.execute(request, SingleValueSegment.class);
    ...
    if (result.getState() == State.SUCCESS) {
        ...
+   } else {
+       throw new TempoDBException(result.getMessage(), result.getCode());
    }
    return iterator;
}
\end{lstlisting}
\caption{The original implementation from the ground truth method.}
\label{fig:fail-vocab-example-gt}
\end{subfigure}
\vspace{-5pt}
\caption{\label{fig:fail-vocab-example}A failure caused by
model hallucination, from the \CodeIn{SingleValueCursor} class in
\CodeIn{tempodb/tempodb\CodeDash java}.}
\vspace{-10pt}
\end{figure}

In practice, \ERCs can pass all user-written tests yet still deviate
from the ground truth, which represents the real user intention. To
investigate the reasons behind this deviation, we manually inspect
each \ERC generated by
\UseMacro{THead-qwen2.5-coder:32b-instruct-q8-0} with the baseline prompt and \Tool that passes all \EBTs
and \nEBTs for each \tarmethod. If there are multiple different
\LLM outputs for one \tarmethod that pass all tests, we only check the
first one. To collect these manual inspection data, we ask
\UseMacro{qual-human-rater-count} inspectors with extensive Java
programming experience, and \UseMacro{qual-agent-model}
\UseMacro{qual-agent-harness-version}
with \UseMacro{qual-agent-model-version}, to compare
the generated \ERC and the ground truth \ERC and then decide whether
the two pieces of code are functionally equivalent. The first inspector
and \UseMacro{qual-agent-model} judge generations with both the
baseline prompt and \Tool, while the second inspector only judges the
\ERCs generated with \Tool. Pairwise Cohen's $\kappa$ between inspectors ranges from
\UseMacro{agree-kappa-min} to \UseMacro{agree-kappa-max}. We then
reconcile every \ERC on which the inspectors disagree by re-reading
the code and checking \UseMacro{qual-agent-model}'s justification, and
we report the reconciled labels.

\begin{figure}[!tbp]
\begin{subfigure}{\columnwidth}
\begin{lstlisting}[language=java-diff]
public void setData(File file) {
+   if (this.embed != null && !this.embed.isEmpty()) {
+       throw new IllegalArgumentException("Embed code is already set.");
+   }
    this.data = file;
}
\end{lstlisting}
\caption{\ERC generated with \Tool.}
\label{fig:ebt-qual-example-tool}
\end{subfigure}
\begin{subfigure}{\columnwidth}
\begin{lstlisting}[language=java-diff]
public void setData(File file) {
+   if (embed != null) {
+       throw new IllegalArgumentException("Cannot supply both embed & data");
+   }
    this.data = file;
}
\end{lstlisting}
\caption{The original implementation from the ground truth method.}
\label{fig:ebt-qual-example-gt}
\end{subfigure}
\begin{subfigure}{\columnwidth}
\begin{lstlisting}[language=java-pretty]
@Test(expected = IllegalArgumentException.class)
public void setEmbedCodeWithData() {
    post.setEmbedCode("something");
    post.setData(new File("some_path"));
}
\end{lstlisting}
\caption{An \EBT defined to test the target method.}
\label{fig:ebt-qual-example-ebt}
\end{subfigure}
\vspace{-10pt}
\caption{\label{fig:ebt-qual-example}Example of a
\UseMacro{THead-qual-too-lenient} false positive, from the
\CodeIn{VideoPost} class in \CodeIn{tumblr/jumblr}.}
\end{figure}

In Figure~\ref{fig:qual-venn}, we show the result of our equivalence
labeling. We found that the generated \ERC is functionally equivalent to
the ground truth for
\UseMacro{qual-base-semantic-match-yes-of-methods-pct}\%
of all \tarmethods (\UseMacro{qual-base-semantic-match-yes-count} out of
\UseMacro{stat-num-methods-real-mega-test-data-with-exception-with-project-with-gold-with-throw})
in the evaluation set with the baseline and
\UseMacro{qual-tuctn-all-info-semantic-match-yes-of-methods-pct}\%
(\UseMacro{qual-tuctn-all-info-semantic-match-yes-count} out of
\UseMacro{stat-num-methods-real-mega-test-data-with-exception-with-project-with-gold-with-throw})
with \Tool.

To understand the sources of false positives, the inspectors additionally classify every false positive into four categories, and reconcile labels through
the same process as described above. We define the categories as follows, where
$T$ is the set of inputs on which the generated \ERC throws an exception and $G$ is the set
on which the ground truth \ERC throws:
(1)~\CateTL: $T$ is a strict subset of $G$;
(2)~\CateTS: $T$ is a strict superset of $G$;
(3)~\CateCD: the generated code drops or rewrites a statement of the
\tarmethod the \LLM was given, rather than only inserting an \ERC into it;
(4)~\CateWH: everything else, including cases where neither $T$ nor
$G$ contains the other, or where the generated \ERC throws at a
different location than the ground truth, causing
the two to throw with different side effects already applied.

\begin{table}[!tbp]
\vspace{10pt}
\begin{small}
\begin{center}
\caption{\UseMacro{TCap-qualitative}\label{tab:qualitative}}
\vspace{3pt}
\begin{tabular}{l |cc}
\toprule
(\%) & \UseMacro{THead-base} & \UseMacro{THead-tuctn-all-info}
\\
\midrule
\UseMacro{THead-qual-semantic-match-no} & \UseMacro{qual-base-semantic-match-no-of-total-pct} & \UseMacro{qual-tuctn-all-info-semantic-match-no-of-total-pct}
\\
\midrule
\UseMacro{THead-qual-too-lenient} & \UseMacro{qual-base-too-lenient-pct} & \UseMacro{qual-tuctn-all-info-too-lenient-pct}
\\
\UseMacro{THead-qual-too-strict} & \UseMacro{qual-base-too-strict-pct} & \UseMacro{qual-tuctn-all-info-too-strict-pct}
\\
\UseMacro{THead-qual-wrong-handling} & \UseMacro{qual-base-wrong-handling-pct} & \UseMacro{qual-tuctn-all-info-wrong-handling-pct}
\\
\UseMacro{THead-qual-destroyed-code} & \UseMacro{qual-base-destroyed-code-pct} & \UseMacro{qual-tuctn-all-info-destroyed-code-pct}
\\
\bottomrule
\end{tabular}
\end{center}
\end{small}
\end{table}

In Table~\ref{tab:qualitative}, we show, for each of the baseline prompt and
\Tool, the share of false positives among the \tarmethods for which the prompt
generates a passing \ERC
(\UseMacro{qual-base-total-samples-count} \tarmethods for the baseline and
\UseMacro{qual-tuctn-all-info-total-samples-count} for \Tool), and the
distribution of those false positives across the categories. The two prompts' false positive rates differ by
\UseMacro{qual-fp-rate-gap-pct} points, and no category share differs
by more than \UseMacro{qual-max-share-gap-pct} points. This suggests
that \Tool can increase the number of outputs that pass the given \EBTs,
but it does not reduce the rate at which a test-passing \ERC is a
false positive. Notably, we see that a large portion of the false
positives is due to \catetl,
\UseMacro{qual-base-too-lenient-pct}\% for the baseline and
\UseMacro{qual-tuctn-all-info-too-lenient-pct}\% for \Tool. We
believe this is because the ability to generate \ERC that aligns with
developer intent is largely limited by the coverage of the \EBTs
provided. A weaker \cond can still pass every \EBT if no \EBT includes
an example input on which the weaker \cond and the developer's
intention disagree. We leave the question of how to write \EBTs to
communicate the user's intended condition for future work.

To provide a deeper understanding of false positives, we look at an
example shown in Figure~\ref{fig:ebt-qual-example}. The example is selected from
the category of \catetl, as it is the most common type of false positive. The
original implementation, shown in Figure~\ref{fig:ebt-qual-example-gt}, is taken from the \CodeIn{VideoPost} class in
\CodeIn{tumblr/jumblr}, a Java wrapper for the Tumblr API. In a
\CodeIn{VideoPost}, the video source can be specified either as an HTML embed
code or as a video file, but not both. The ground truth \ERC throws an
exception when \CodeIn{setData} is called while \CodeIn{embed} is non-null,
enforcing mutual exclusivity between the two fields. The provided \EBT, in Figure~\ref{fig:ebt-qual-example-ebt}, verifies
this behavior by first setting the embed code to ``something'' and then calling
\CodeIn{setData} to trigger the exception, covering the target \cond. In Figure~\ref{fig:ebt-qual-example-tool}, we show the \Tool-generated implementation (using \UseMacro{THead-qwen2.5-coder:32b-instruct-q8-0}). It checks that
\CodeIn{embed} is neither null nor an empty string before throwing an exception. This
difference means that when \CodeIn{embed} is set to an empty string, the ground
truth would reject a subsequent \CodeIn{setData} call, while the generated code
would incorrectly allow it. The consequence is that a \CodeIn{VideoPost} could
be constructed with both \CodeIn{embed} and \CodeIn{data} set, leading to
undefined behavior when the malformed data is sent to the backend server. We
believe this is because there are three common patterns
for checking an unset string that the \LLM learns during its training: checking for
null, for empty (string with length 0), and for blank (string with only whitespace). Setting the embed code to ``something'' is
equally consistent with all three, so the \EBT does not tell the \LLM which one is the user's intended check.

\FloatBarrier

\section{Limitations}
\label{sec:limitations}

\MyPara{LLM randomness} LLMs are probabilistic, so their outputs may
vary across runs. To limit the effect of this randomness, we generate
\UseMacro{eval-num-samples} outputs per task. For the open-source
models, we use temperature \UseMacro{eval-temperature}.
\UseMacro{THead-gpt-5-mini} uses the provider's decoding settings.

\MyPara{Benchmark and evaluation} (1)~Removing developer-written \ERC
from existing methods approximates, but does not reproduce, naturally
missing \ERC. (2)~pass@k measures whether a sampled candidate passes a
test oracle, not whether it is semantically equivalent or useful in
practice. (3)~While EvoSuite and Randoop add behavioral checks that
strengthen the test oracle beyond user-written tests, they may still
miss relevant cases and do not fully establish developer intent.

\MyPara{Scope} \Tool targets exceptions that originate from \tss
within the \tarmethod (Section~\ref{sec:task-definition}), allowing us to develop a deeper
technical solution for this specific task. We leave exceptions that
propagate from callees or are wrapped along call chains as future
work. Our dataset is drawn from CodeSearchNet and includes only
compilable Java projects that use Maven. \Tool's \conEng design is
language-agnostic. However, some generalization gap may remain for
other projects, languages, or build systems.

\section{Related Work}

There has been a lot of prior work on code
generation~\cite{roziereCodeLlamaOpen2023,
liCompetitionLevelCodeGeneration2022,
chenEvaluatingLargeLanguage2021, zhang-etal-2024-codeagent,
olausson2024repair, MaSWEGPT2025}, fault
localization~\cite{wongSurveySoftwareFault2016,renieresFaultLocalizationNearest2003,jonesVisualizationTestInformation,GroupingBasedStrategyImprove,liuNotAllExceptions2025},
and test-driven
development~\cite{pancurImpactTestdrivenDevelopment2011,pancurEmpiricalEvaluationTestdriven2003,huangEmpiricalInvestigationEffectiveness2009,liCompetitionLevelCodeGeneration2022,leCodeRLMasteringCode2022,mathewsTestDrivenDevelopmentCode2024}.

\MyPara{Machine learning for generating exception related code}
ThEx~\cite{zhongWhichExceptionShall2022} helps developers decide which exception should be thrown given a code
snippet.
FuzzyCatch~\cite{nguyenRecommendingExceptionHandling2019} uses fuzzy logic to predict if a runtime exception would occur in a given code
snippet and recommends code to handle that exception.
Neurex~\cite{caiProgrammingAssistantException2024} bases its exception handling recommender on CodeBERT~\cite{fengCodeBERTPreTrainedModel2020}. It
determines if a \trycatch is needed, identifies statements for the try
block, and specifies exception types for the catch clause.
exLong~\cite{zhangExLongGeneratingExceptional2024} automatically generates exceptional behavior tests to help developers check that their code detects unwanted events and throws appropriate exceptions.

With large language models~\cite{vaswaniAttentionAllYou2023,roziereCodeLlamaOpen2023, brownLanguageModelsAre2020}, many models can now generate the complete exception handling code for a given method. Knowledge-driven Prompt Chaining (KPC)~\cite{renMisuseMasteryEnhancing2023} generates exception handling code using iterative check-rewrite steps with fine-grained, knowledge-driven prompts.

In contrast to this prior work, which modifies code to avoid unhandled
exceptions, we automatically retrofit \ERC so that a method throws the
correct exception under an exceptional condition rather than handling it.
Our approach leverages test-driven development of exceptional behavior
instead of analyzing general code patterns to handle runtime errors.

\MyPara{Fault localization}
Finding the appropriate locations for \ERC is a crucial part of \task. The similar problem of localizing exceptional behaviors within software systems has been extensively studied in the field of fault localization~\cite{wongSurveySoftwareFault2016}.
Tarantula~\cite{jonesVisualizationTestInformation}, a widely-used spectrum-based technique, computes the suspiciousness of each statement based on coverage information. Coverage of the tests is also used in \Tool as one of the contexts collected in dynamic analysis. ABEL~\cite{liuNotAllExceptions2025} leverages attention-based models to automatically rank the most suspicious exceptions, helping developers focus on resolving the most relevant ones.
However, our approach differs fundamentally from fault localization. Unlike fault localization techniques that identify the locations of bugs, \Tool focuses on identifying any location where \ERC can be effectively implemented within the given program context.

\MyPara{Test-driven development}
Test-driven development (TDD) is a software development approach where tests are written before production code. Many studies~\cite{pancurImpactTestdrivenDevelopment2011,pancurEmpiricalEvaluationTestdriven2003,huangEmpiricalInvestigationEffectiveness2009} have been conducted to affirm its effectiveness.

Recent advances in LLMs have enabled new applications of \TDD principles in automated code generation. AlphaCode~\cite{liCompetitionLevelCodeGeneration2022} leverages test cases from competitive programming problems to filter and validate sampled implementations. CodeRL~\cite{leCodeRLMasteringCode2022} frames code generation as a reinforcement learning problem, using unit test execution results as reward signals to guide iterative code improvement. TGen~\cite{mathewsTestDrivenDevelopmentCode2024} employs a multi-agent approach with a coder agent for initial code generation and a remediation agent that iteratively fixes code based on test feedback until all tests pass.

While these TDD-based approaches demonstrate effectiveness for general code generation tasks, our work targets a specific aspect of software development: generation of \ERC based on exceptional behavior tests. This focused approach differs from existing work by specifically leveraging the contextual information provided by \EBTs to guide \task rather than addressing general functional requirements.

\section{Conclusion}

We present the first work on automatically retrofitting \ERC into
\tarmethods given exceptional behavior tests (\EBTs) using large
language models (\LLMs). We introduce \Tool, which performs \conEng
and constructs targeted prompts by extracting necessary context from
target methods and their corresponding \EBTs through a combination of
static and dynamic program analysis techniques. To evaluate \Tool, we
develop a novel dataset comprising methods without \ERC paired with
\EBTs that specify the required exceptional behavior to be
implemented. Our evaluation demonstrates that \Tool consistently
improves the performance of \LLMs on \task across different model sizes
and architectures.

\section{Acknowledgment}

We thank Cheng Ding, Ivan Grigorik, Tong-Nong Lin, Aditya Thimmaiah,
and anonymous reviewers for helpful feedback and discussions.
This work was supported in part by the U.S. National Science
Foundation (NSF) Nos.~CCF-2217696, CCF-2313027, CCF-2403036; and a
gift by Cisco Research.
Any opinions, findings, and conclusions or recommendations expressed
in this material are those of the authors and do not necessarily reflect
the views of the NSF or Cisco.

\bibliographystyle{IEEEtran}
\bibliography{bib}

\end{document}